%% file: zak_proc_v3_arxiv.tex
\documentclass[11pt]{article}

\usepackage{graphicx}              
\usepackage{amsmath}               
\usepackage{amsfonts}              
\usepackage{amsthm}                
\usepackage{amscd}
\usepackage{amssymb}
\usepackage[top=2cm,bottom=3cm, left=3cm, right=3cm]{geometry}

\newcommand{\be}{\begin{equation}}
\newcommand{\ee}{\end{equation}}
\newcommand{\bea}{\begin{eqnarray}}
\newcommand{\eea}{\end{eqnarray}}
\newcommand{\nn}{\nonumber}

\newcommand{\Hkin}{{\cal H}_{\rm kin}}
\newcommand{\Hphys}{{\cal H}_{\rm phys}}
\newcommand{\Dkin}{{\cal D}_{\rm kin}}
\newcommand{\Dphys}{{\cal D}_{\rm phys}}

\newcommand{\R}{\mathbb{R}}                  

\newcommand{\C}{\mathbb{C}}

\newtheorem{prop}{Proposition}   

\newcommand{\scpr}[2]{\langle#1\, \vert \, #2 \rangle}
\newcommand{\sscpr}[3]{\langle#1\, \vert \, #2 \, \vert \, #3\rangle} 
\newcommand{\norm}[1]{\left\lVert #1 \right\rVert}
\newcommand{\betr}[1]{\left\lvert #1 \right\rvert}
\newcommand{\ket}[1]{\lvert\, #1\rangle}

\newcommand{\um}{\mathbb{I}}
\newcommand{\id}{\text{id}}
\newcommand{\abar}{\overline{\mathcal{A}}}
\newcommand{\bg}[1]{#1^{(0)}}

\DeclareMathOperator{\cyl}{Cyl}
\DeclareMathOperator{\tr}{tr}

\newcommand{\flux}{\hat{E}_j(S)} 
\newcommand{\snf}{T_{\gamma,\vec{j},\vec{I}}}  
\newcommand{\rivf}{R^j_e}     
\newcommand{\livf}{L^j_e}  
\newcommand{\Hdiff}{{\cal H}_{\rm diff}} 

\begin{document}
\title{From Classical To Quantum Gravity: Introduction to Loop Quantum Gravity}


\author{Kristina Giesel\\
{\small \textit{Friedrich-Alexander-University of Erlangen-N\"urnberg, Department of Physics,}}\\
{\small \textit{Institute for Quantum Gravity, Erlangen (Germany)}}
\\
{\small E-mail: kristina.giesel@gravity.fau.de}\\[1cm]
Hanno Sahlmann\\
{\small \textit{Asia Pacific Center for Theoretical Physics, Pohang (South Korea)}}\\
{\small \textit{Physics Department, Pohang University of Science and Technology, Pohang (South Korea)}}\\
{\small E-mail: sahlmann@apctp.org}}

\maketitle

\abstract{We present an introduction to the canonical quantization of gravity performed in loop quantum gravity, based on lectures held at the 3rd quantum geometry and quantum gravity school in Zakopane in 2011. A special feature of this introduction is the inclusion of new proposals for coupling matter to gravity that can be used to deparametrize the theory, thus making its dynamics more tractable. The classical and quantum aspects of these new proposals are explained alongside the standard quantization of vacuum general relativity in loop quantum gravity.}


\section{Introduction}
\label{se:intro}
\input{introd}

\section{Classical theory}
\label{se:class}
This section deals with the classical setup for loop quantum gravity. We will start with a derivation of the Ashtekar variables for general relativity, that are the elementary phase space variables used in the classical theory in section \ref{suse:canon}. Afterwards in section \ref{suse:Cldynamics} we present a discussion on the dynamics of general relativity with a foucs on those properties of the classical dynamics that are relevant for the quantization later on. Finally, we present the classical formulation of two models  that have been recently introduced for the dynamics of loop quantum gravity in section \ref{suse:bkmod} and \ref{suse:scalar}.

\subsection{Canonical connection variables}
\label{suse:canon}
\input{canon}
\subsection{Dynamics in the classical theory}
\label{suse:Cldynamics}
\input{dynamics}
\subsubsection{Brown-Kuchar model}
\label{suse:bkmod}
\input{BKModCl}

\subsubsection{Scalar field model}
\label{suse:scalar}
\input{ScalarMod}

\section{Quantum kinematics}
\label{se:qukin}
This section includes a introductory discussion on the quantum kinematics of loop quantum gravity corresponding to the sector of the theory where the constraints have not been solved yet. For the reason that the kinematical representation is rather different from the usual Fock representation we introduce it in section \ref{suse:al} and discuss its properties. Afterwards in section \ref{suse:geo} we discuss geometric operators in the framework of loop quantum gravity. These are operators that correspond to classical geometrical objects like length, area and volume. Section \ref{suse:gdcons} deals with the question how the classical Gauss and diffeomorphism constraints can be quantized and how solutions to the constraint operators can be constructed. Finally, in section \ref{suse:reps} we present more details on the kinematical representation and explain how it can be generalized.

\subsection{Ashtekar-Lewandowski representation}
\label{suse:al}
\input{al}

\subsection{Geometric operators}
\label{suse:geo}
\input{area_operator}

\subsection{Solving Gauss and diffeomorphism constraints}
\label{suse:gdcons}
In accordance with Dirac's program for constraint quantization an implementation of the classical constraints as operators on the kinematical Hilbert space is needed. In section \ref{se:gauss} we will discuss this quantization for the Gauss constraint and explain how solutions to the constraint can be constructed. The quantization and the construction of solutions to the spatial diffeomorphism constraint requires a framework called refined algebraic quantization that will be introduced in section \ref{se:RAQ} and applied to the diffeomorphism constraint in section \ref{se:Diffeo}.
\subsubsection{Quantization and solution of the Gauss constraint}
\input{gauss.tex}
\label{se:gauss}
\subsubsection{Refined algebraic quantization}
\label{se:RAQ}
\input{RAQ}

\subsubsection{Quantization and solutions of the diffeomorphism constraint}
\label{se:Diffeo}
\input{diffeo}

\subsection{More on representations of $\mathfrak{A}$}
\label{suse:reps}
\input{reps}

\section{Quantum dynamics}
\label{se:qudyn}
In this section we discuss the quantum dynamics of loop quantum gravity. In the case that we use pure Dirac quantization the dynamics is encoded in the Hamiltonian constraint and its quantization is explained in section \ref{suse:ham}. When we consider the Brown-Kuchar or the scalar field model earlier introduced the dynamics is encoded in a so called physical Hamiltonian. However, the quantization of the latter relies in both models on techniques that have been used for the Hamiltonian constraint. The details of the quantization of the Brown-Kuchar model are discussed in section \ref{suse:qbkmod}, whereas the quantum theory of the scalar field model is presented in \ref{suse:qscalar}.

\subsection{The quantum Hamilton constraint}
\label{suse:ham}
\input{ham}

\subsection{The quantum Brown-Kuchar model}
\label{suse:qbkmod}
\input{qbkmod}

\subsection{The quantum scalar-field model}
\label{suse:qscalar}
\input{qscalar}


\section{Summary and open problems}
\label{se:sum}
\input{concl}

\section*{Acknowledgments}
The authors would like to thank the organizers of the 3rd quantum geometry and quantum gravity school in Zakopane for their kind invitation. The authors would also like to thank Jerzy Lewandowski for carefully reading our review and giving valuable feedback that improved the current version of the review.
This work was supported by the European Science Foundation through its network Quantum Geometry and Quantum Gravity.

\end{document}

%% file: introd.tex
Loop quantum gravity aims to formulate a quantum theory for Einstein's classical theory of general relativity. As in the case of quantum mechanics one takes a given classical theory, in our case general relativity, as the starting point and then tries to develop the corresponding quantum theory. However, when we aim to quantize general relativity, we face two aspects that are particular to general relativity. At first, in contrast to the gauge theories that have been quantized in the context of the standard model of particle physics, in general relativity the metric itself becomes a dynamical object. For this reason we are no longer in the situation that we can consider matter put on a fixed background geometry and then describe how matter evolves and interacts with each other on a given fixed spacetime. In general relativity matter and geometry interact, and this highly non-linear interaction is described by Einstein's equations. Secondly, in general relativity, due to the requirement of general covariance, spatial and temporal coordinates do not posses any physical significance. This is also reflected in the fact that general relativity can be understood as a gauge theory with spacetime diffeomorphisms as the gauge group. Thus, observables in general relativity, as these are gauge invariant quantities, are usually more difficult to construct than in other gauge theories due to the complexity of the gauge group of spacetime diffeomorphisms. In the context of loop quantum gravity (LQG) these two points are taking as a guiding principle for formulating the quantum theory. The first point mentioned above explains why a non-perturbative and thus background independent quantization is chosen. A more detailed discussion about this motivation can be also found in the article of Ashtekar in these proceedings. Although techniques from ordinary lattice gauge theory are adopted for LQG, the representation we end up with  is rather different from the Fock representation used in ordinary quantum field theories and again the role of diffeomorphisms enters crucially here. This representation allows to formulate a quantum analogue of Einstein's equation, also called Wheeler-deWitt equations,  and we will discuss those quantum Einstein equations and their derivation in this review.
The construction of observables and therefore the description of the  physically relevant sector of loop quantum gravity can be performed in different ways. One possibility is to extract the gauge invariant sector of the theory at classical level and then quantize. Another is to follow Dirac's idea and derive the gauge invariant part of the theory at the quantum level. Also, we can combine those ideas and only partially reduce the system at the classical level. In these lectures we will briefly discuss how one proceeds in those cases and what are the properties of the resulting quantum theories. One particular aspecty of our introduction is the inclusion of new proposals for coupling matter to gravity that can be used to deparametrize the theory, thus making its dynamics more tractable. We discuss the classical and quantum aspects of these new proposals alongside the standard quantization of vacuum general relativity in loop quantyum gravity. We hope that this parallel treatment makes them more acessible and allows a clear comparison to the standard approach.  
\\
\\
Of course we will not be able to present a complete introduction to loop quantum gravity but the idea of the article is to provide a rather brief introduction to the techniques used in LQG. For instance we will only focus on the canonical part of the theory in this article. A path integral approach to LQG in the context of spin foams will be explained in an article by Rovelli in these proceedings. We will only briefly discuss the connection this has to the canonical approach reviewed here in section \ref{se:qudyn} on quantum dynamics. 
Introductions to LQG that contain a much more detailed description can be found in Rovelli's \cite{Rovelli:2004tv} and Thiemann's book \cite{Thiemann:2007zz}. A book on LQG that is particularly addressed to undergraduate students has been published by Gambini and Pullin \cite{GambiniPullin}. Furthermore, Bojowald's book \cite{Bojowald} presents an introduction to LQG with a focus on applications of the theory. Other already existing lecture notes on loop quantum gravity can for instance be found in
 \cite{Ashtekar:2004eh,Perez:2006dm,Perez:2009zz,Dona:2010hm,Sahlmann:2010zf}. 
 \\
 \\
 We have structured this article into four main parts. The first part in section \ref{se:class} introduces the classical framework, that is needed for general relativity in oder to take it as a starting point for loop quantum gravity. In section \ref{se:qukin} we discuss the quantization of general relativity and explain how the kinematics of loop quantum gravity can be formulated. Afterwards we discuss the quantum dynamics in section \ref{se:qudyn} and finally summarize and conclude in section \ref{se:sum}.

%% file: canon.tex
In these lectures we describe a canonical quantization of gravity. Therefore we need to start with a \emph{canonical description} of the classical theory, that is, a description in terms of a phase space, canonical variables, and a Hamiltonian. The canonical variables are just coordinates in the phase space, so strictly speaking they are irrelevant in the classical description. Their change is just a canonical transformation, and hence does not change the physics. But in our context, we have to pay close attention to the choice of canonical variables, as it affects the quantum description. Canonical transformations rarely correspond to unitary maps in the quantum theory:
\begin{equation}
\begin{CD}
\text{Canonical formulation 1} @>>> \text{Quantum theory 1}\\
@VV{\text{canon. trafo}}V @VV{\ncong}V\\
\text{Canonical formulation 2} @>>> \text{Quantum theory 2}
\end{CD}
\end{equation}
So the canonical variables have to be chosen with care. How to do this? While any choice will ultimately have to be justified by the results, one can be guided by other factors. In our case, this is the following fact \cite{Ashtekar:1986yd,Barbero:1994ap,Immirzi:1996di}: 
\begin{quote}
There is a formulation of general relativity in which the phase space is is precisely that of SU(2) Yang-Mills theory.   
\end{quote}
Before we describe this formulation, let us consider the canonical formulation of Maxwell theory as a warm-up. The action is 
\begin{equation}
S[A]=-\frac{1}{4}\int_\mathbb{M} F_{\mu\nu}F^{\mu\nu}\, \text{d}^4x,  
\end{equation}
with $A$ the 4-potential and $F$ its curvature. Now we chose an equal-time surface $\Sigma_t$ relative to some inertial observer time $t$. The canonical coordinates are the 4-potential $A_\mu$ and  
and the conjugate momenta $P^\mu=\delta S/\delta \dot{A}_\mu$. We find that the spatial momenta are given by the electrical field, $P^a=-E^a$. Moreover, $P^0=0$ which implies that $A_0$ is non-dynamical in our description. We can then rewrite the action as
\begin{equation}
S=\int\text{d}t\int\text{d}^3x\, -E^a\dot A_a - \frac{1}{2}(E^2+B^2) +A_0\nabla \cdot E. 
\end{equation}
From this form, one can conclude the following: 
\begin{enumerate}
\item $A_a,E^b$ are the canonical coordinates. 
\item $(E^2+B^2)/2$ is the Hamiltonian of the system. 
\item $A_0$ is a Lagrange multiplier. It enforces the \emph{constraint} $\nabla \cdot E=0$, which is just Gauss' law in the absence of charges.   
\end{enumerate}
Now we will go through the exact same steps for general relativity. To be able to do this, one needs to start from a formulation in terms of a connection and a (co-) frame. What is a \emph{frame}? It is simply a basis $e_I\equiv e_I^{\mu}\partial_\mu$ of the tangent space at each point of space-time. Equivalently, it can be viewed as a point-dependent map $\R^4\rightarrow T_p M$. If there is an \emph{internal} metric, that is, a metric $\eta$ on $\R^4$, the frame defines a space-time metric $g$, by declaring the components of the frame to be orthogonal:
\begin{equation}
g^{\mu\nu}:= e^\mu_Ie^\nu_J\eta^{IJ}.  
\end{equation} 
Vice versa, a space-time metric $g$ defines an orthonormal frame, but only up to SO(3,1) rotations in the internal space $\R^4$. A co-frame is the same as a frame, but with respect to the co-tangent bundle, $e^I\equiv e^I_\mu dx^\mu$. A frame defines a co-frame, and vice versa, by simply pulling indices with the corresponding internal and external metric, or, equivalently, inverting the matrix of components.          

With that said, we can write the action of general relativity in terms of a co-frame $e^I$ and an so(3,1) connection $\omega$:
\begin{equation}
S[\omega,e]=S_\text{P}[\omega,e]+S_\text{H}[\omega,e], 
\end{equation} 
with \cite{Holst:1995pc}
\begin{align}
S_\text{P}[\omega,e]&=\frac{1}{2\kappa}\int \text{d}^4x\,\, \epsilon_{IJKL}e^I\wedge e^J\wedge F(\omega)^{KL}\\
S_\text{H}[\omega,e]&= -\frac{1}{\kappa\beta}\int \text{d}^4x\,\, e^I\wedge e^J\wedge F(\omega)_{IJ}
\end{align}
Here $\kappa=16\pi G$ is the coupling constant of gravity, and $\beta$ is an additional parameter, the \emph{Barbero-Immirzi parameter} to be discussed below. 
The \emph{Palatini action} $S_\text{P}$ was long known as a viable action for general relativity. The \emph{Holst action}  $S_\text{H}[\omega,e]$ does not change that: The equations $\delta S/\delta \omega$ are equivalent to $D^{(\omega)}e=0$ \cite{Holst:1995pc}, which can be solved for $\omega\equiv\omega(e)$. Re-inserting $\omega(e)$ into the action gives 
\begin{equation}
S_\text{H}[\omega(e),e]= 0,\qquad S_\text{P}[\omega(e),e]= S_{\text{EH}}[g(e)],
\end{equation} 
where $S_{\text{EH}}$ is the Einstein-Hilbert action. Thus the above action principle leads to the equations of motion of general relativity, irrespective of the value of $\beta$.%
\footnote{$\beta$ \emph{does} acquire physical significance in the case that spinor matter is coupled to gravity, see \cite{Perez:2005pm} for details.}
The values $\beta=\pm i$ are, however, special in the sense that the resulting canonical formulation has special properties \cite{Ashtekar:1986yd}, as we will see shortly.

To go over to the Hamiltonian formulation, we first have to chose a \emph{time function} $t$ on the space-time manifold, whose level surfaces $\Sigma_t$ give a foliation of the space time into spatial slices. Additionally we pick a \emph{time co-vector field} $t^\alpha$ with $t^\alpha\partial_\alpha t=1$, and decompose it into tangential and normal components with respect to $\Sigma_t$, 
\begin{equation}
t^\alpha=Nn^\alpha+N^\alpha, 
\end{equation}  
where $n^\alpha$ is the unit normal, and the \emph{shift vector} $N^\alpha$ is tangential. 
Next, we partially fix the gauge freedom in the co-frame by linking it with the normal one-form $n_\alpha$, 
\begin{equation}
e^0_\mu \overset{!}{=} n_\mu.
\label{eq:timegauge}
\end{equation}
This gauge is called \emph{time gauge}. Since $n_\mu$ is time-like, only SO(3) remains as gauge group. The covariant fields can now be decomposed accordingly, and adapted coordinates be chosen. The co-frame assumes the structure 
\begin{equation}
(e_\mu^I)=\left(
\begin{array}{c|ccc}
N&\quad &N^i&\quad\\
\hline
0&&&\\
0&&e^i_a&\\
0&&&	
\end{array}\right)
\end{equation}
where we let $I$ run horizontally, and $\mu$ vertically. The index $i$ now runs from one to three, and $a$ is a tangent space index for $\Sigma_t$. Analogously, $\omega$ can be decomposed into SO(3) connections $\Gamma^i_a:=\epsilon^{i0}{}_{KL}\omega^{KL}_a$, $K^i_a:=\omega^{i0}_a$ and the rest, i.e., the components of $\omega_0$. The action can now be expressed in terms of the decomposed fields, and it organizes precisely as in the case of the electromagnetic fields \cite{Ashtekar:2004eh}: 
\begin{equation}
S=\frac{2}{\kappa\beta}\int \text{d}t \int_{\Sigma_t} E^{a}_i\dot{A}^i_a-\underbrace{\left(\omega^i_0G_i+N^aC'_a +NC'\right)}_{=:h_{\rm can}} 
\label{eq:AshAction}
\end{equation}
with 
\begin{equation}
A^i_a=\Gamma^i_a+\beta K^i_a, \qquad E^a_i=\sqrt{\det q}\, e^a_i, \qquad q_{ab}=e^i_ae^j_b\delta_{ij}.
\end{equation}
$q$ denotes the metric on $\Sigma_t$ and the dot is the time derivative $t^\alpha\partial_\alpha$. We see that $A$ and $E$ are conjugate canonical variables, 
\begin{equation}
\left\{A^i_a(x),E^b_j(y)\right\}=\frac{\kappa}{2}\beta\delta_a^b\delta_j^i\delta(x,y).
\end{equation}
where $\kappa:=16\pi G$ with $G$ being Newton's constant.
The Hamiltonian density $h_{\rm can}$ is a sum of constraints,  
and the constraints
\begin{equation}
\begin{split}
G_i=D^{(A)}_aE^a_i, \qquad 
C'_a=E^b_iF^i_{ab} +(\ldots)^i_a G_i,\\
C'=\frac{\beta}{2}\frac{E^a_iE^b_j}{\det E}\left[\epsilon^{ij}{}_kF^k_{ab}-2(1+\beta^2)K_{[a}^iK_{b]}^j\right] +(\ldots)^i_a G_i. 
\end{split}
\label{eq:Constraints}
\end{equation}
By subtracting multiples of $G_i$ from $G'_a$ and $C'$, we obtain a set of independent constraints  $G_i, G_a, C$. The constraint equations $G_i= G_a= C=0$, together with the evolution equations
\begin{equation}
\left\{A(x),h_\text{can}\right\}=\dot{A}(x), \qquad \left\{E(x),h_\text{can}\right\}=\dot{E}(x)
\end{equation}
are completely equivalent to Einsteins equations. But as it is the case for all reparametrization invariant systems, time evolution is a kind of gauge transformation. Concretely, 
\begin{align*}
\{A_a,G(\Lambda)\}&=-D^{(A)}_a\Lambda=\left.\frac{\text{d}}{\text{d}\epsilon}\right\rvert_{\epsilon=0}g_\epsilon A g^{-1}_\epsilon+g_\epsilon\text{d}g^{-1}_\epsilon,\\
\{E^a,G(\Lambda)\}&=[\Lambda,E^a]=\left.\frac{\text{d}}{\text{d}\epsilon}\right\rvert_{\epsilon=0}g_\epsilon E^ag^{-1}_\epsilon
\end{align*}
with $g_\epsilon=\exp(\epsilon\Lambda)$ and $G(\Lambda)=\int G_i\Lambda^i$, so $G$ generates gauge transformations. Moreover, 
\begin{equation}
\{A,\vec{C}(\vec{N})\}=\mathcal{L}_{\vec{N}}A, \qquad \{E,\vec{C}(\vec{N})\}=\mathcal{L}_{\vec{N}}E,  
\end{equation}
so $\vec{C}$ generates spatial diffeomorphisms. Lastly $C$ is related to $\mathcal{L}_{Nn^\alpha}$, i.e., to the diffeomorphisms in a direction normal to $\Sigma$. 
The constraints form an algebra, the \emph{Dirac algebra},
\begin{equation}
\{G(\Lambda),G(\Lambda')\}=G([\Lambda',\Lambda])\qquad 
\{G(\Lambda),\vec{C}(\vec{N})\}= -G(\mathcal{L}_{\vec{N}}\Lambda)\qquad
\{\vec{C}(\vec{N}),\vec{C}(\vec{N}')\}=\vec{C}([\vec{N},\vec{N}']).
\label{eq:dirac1}
\end{equation}
The Hamiltonian constraint $C$ is gauge invariant and transforms under diffeomorphisms in the expected way, 
\begin{equation}
\{C(N),G(\lambda)\}=0, \qquad \{C(N),\vec{C}(\vec{N})\}=C(\mathcal{L}_{\vec{N}}N).
\label{eq:dirac2}
\end{equation}
Up to here, the structure is that of an infinite dimensional Lie algebra. But the bracket of two Hamiltonian constraints is more complicated, 
\begin{equation}
\{C(N),C(M)\}=-\frac{\kappa^2\beta^2}{4} \vec{C}(\vec{S}), \qquad 
S^a=\frac{E^aE^b}{\det q}(N\partial_b M-M\partial_b N).
\label{eq:dirac3}
\end{equation}
It contains a function of the phase space point on the right hand side, hence the structure is not that of a Lie algebra anymore. 

Before we finish this section, some remarks are in order: 
\begin{enumerate}
	\item We have just described a canonical formulation of Einstein gravity in $D+1=4$ dimensions in terms of a phase spaces that is identical to that of SO(D) Yang-Mills theory. This formalism relies on a coincidence that only happens for $D=3$: An SO($D$) connection has $D(D-1)/2$ components, whereas a spatial frame has $D$. If they are to be canonically conjugate variables, they have to have the same number of components, which restricts to $D=3$. But this does not mean that there are no other ways to formulate canonical GR in gauge theory variables. For example, it has long been known that GR in D+1=3 has such a formulation in terms of SO(2,1) connections and their conjugate momenta, see for example \cite{Carlip:1998uc} for an introduction. More recently it has been shown that gravity in $D+1$ dimensions can indeed be formulated in terms of SO($D+1$) Yang-Mills theory variables \cite{TNv}, but there are additional constraints that have to be implemented in order to bring the number of degrees of freedom in line with that of GR.  

\item In the following, we will go over from a formulation in terms of SO(3) to one in terms of its covering group SU(2). This must be done in order to couple fermions to gravity, but it does not change classical or quantum theory much. The biggest change is that also representations with half-integer spin will be allowed in the quantum theory. 

\item As remarked before, in the presence of fermions, physical predictions do become dependent on the value of $\beta$ \cite{Perez:2005pm}. Solving $\delta S/\delta\omega=0$ and re-inserting into the action gives an effective 4-fermion interaction with $\beta$-dependent strength. The effect is however suppressed by the gravitational coupling constant, and is thus extremely small. 

\item We have seen that the constraints commute on the constraint surface $G=\vec{C}=C=0$. This means that they form a \emph{first class system}, and thus the constraints can be imposed in the quantum theory as operator equations. This is the result of imposing the time gauge. Without it, the situation is more complicated, but also very interesting \cite{Alexandrov:2000jw,Alexandrov:2001wt,Alexandrov:2002br,Livine:2002ak,Cianfrani:2008zv,Cianfrani:2009wi,Cianfrani:2009sz,Geiller:2011cv,Geiller:2011bh}.          
\end{enumerate}

%% file: dynamics.tex
We saw in the last section that we need to choose the classical variables in which  we would like to formulate our classical theory with care. Whatever classical choice we make will also have an influence on the quantum theory that we obtain. In this section we will briefly discuss some aspects of the classical dynamics of general relativity and see that also here  one needs to think carefully how to formulate the classical dynamics because each choice will enter crucially in the properties of the quantum dynamics that we obtain later on and discuss in section 
\ref{se:qukin}.
\\
\\
The dynamics of classical general relativity is encoded in the canonical Hamiltonian introduced in the last section. As we saw the Hamiltonian consists of a linear combination of constraints only and the smeared version of the constraints  satisfy the Dirac algebra shown in equations (\ref{eq:dirac1}), (\ref{eq:dirac2}) and (\ref{eq:dirac3}).
The constraint $\vec{C}$ generates spatial diffeomorphisms within the spatial hypersurface, the Gauss constraint generates SU(2) gauge transformations and the Hamiltonian constraint generates diffeomorphisms  orthogonal to the hypersurfaces. Note that the latter  is only true on shell, that is when the equation of motion are satisfied. From the explicit form of the constraints as well as from the Dirac algebra we see that the most complicated among them is the Hamiltonian constraint. In particular it is only the algebra of the Hamiltonian constraints that involves structure functions instead of structure constants and thus their algebra is no true Lie algebra anymore. As a consequence the corresponding quantum operators need to satisfy a complicated quantum algebra in order to be implemented without anomalies. A further complication is that in the representation used in LQG the infinitesimal operators for the spatial diffeomorphisms cannot be implemented as operators for the reason that their finite counterparts are not weakly continuous (see also section \ref{se:Diffeo} for more details). Hence, strictly speaking one is not even able to check whether the algebra of the quantum Hamiltonian constraints is correctly implemented. However, in \cite{TQSD} an anomaly-free quantization of the Hamiltonian constraint has been introduced and its anomaly freeness has been achieved by quantizing the operator in such a way that the commutator annihilates spatially diffeomorphism invariant states. This is what we would expect from an operator version of $\vec{C}$ if it exists.  In order to obtain the physical sector of the quantum theory later on using the Dirac quantization procedure we look for solutions of the constraints in the quantum theory. Also here the complicated algebra of the Hamiltonian constraint operators makes our life more difficult and although general solutions to the Gauss and diffeomorphism constraint can be constructed one has not been successful in the case of the Hamiltonian constraint. However, this is not too surprising since otherwise we would be able to write down the general solution of quantum gravity and this is not even possible for classical general relativity.  During the last years a couple of new proposals have been introduced for describing the (canonical) dynamics of LQG. Their aim is  to reformulate the dynamics in a  technically different but physically equivalent way in order to extract the physical sector out of LQG. One of the first proposals in this direction is the so called master constraint program \cite{TMCP}. The idea of the master constraint is to reformulate the dynamics of the Hamiltonian constraints in such a way that the algebra simplifies. For this purpose one introduces at the classical level a master constraint involving the squares of the Hamiltonian constraints
\be
\label{Master}
{\bf M}=\int\limits_{\Sigma}d^3x\frac{[C(x)]^2}{\sqrt{\det{q}}}
\ee
Classically due to the square of the Hamiltonian ${\bf M}=0$ is equivalent to $C(x)=0$ for all $x\in\Sigma$. The $1/\sqrt{q}$ has been introduced because then ${\bf M}$ is spatially diffeomorphism invariant and thus Poisson commutes with $\vec{C}$ and also because then $\bf M$ has density weight one which will be useful for the quantization of the master constraint. The master constraint satisfies the trivial algebra $\{{\bf M},{\bf M}\}=0$ and this carries also over to quantum theory where the corresponding operator is required to satisfy this trivial algebra as well. Therefore as far as the algebra is concerned the Master constraint simplifies the situation. However, classically as well as in the quantum theory at the end we are not only interested in solutions to the constraints but we also would like to have so called weak Dirac observables. These are quantities that are gauge invariant, meaning that they commute with all constraints of the theory on the constraint surface. However, on the constraint surface for the master constraints we have $\{f,{\bf M}\}=0$ for all phase space functions $f$. Therefore it seems that using the master constraint we loose the ability to detect weak Dirac observables with respect to the Hamiltonian constraint. However, this is not the case because instead of using the condition that weak Dirac observables need to commute weakly with the constraint we can also require for a weak observable $O$
\be
\label{OM}
\{O,\{O,{\bf M}\}\}_{{\bf M}=0}=0
\ee
This condition is no longer linear in $O$ as before and hence we get a system of non linear partial differential equations for the observables.  Formally, it seems that the price we have to pay for a simpler constraint algebra is a more complicated equation for weak Dirac observables. However, whether the condition in (\ref{OM}) is indeed harder to satisfy needs to be checked in applications of the master constraint program. As far as the quantum theory is concerned the master constraint program has advantages with respect to using the Hamiltonian constraint as discussed in \cite{TMCP}. The master constraint program has been tested in a variety  models in a series of papers \cite{DTMCP}. 
As introduced in \cite{TMCP} we can also define a so called extended master constraints that involves also the diffeomorphism and Gauss constraint 
\be
{\bf M}=\int\limits_{\Sigma}d^3x\frac{[C(x)]^2+q^{ab}C_aC_b + \delta^{jk}G_jG_k}{\sqrt{\det{q}}}
\ee
In this case the constraint algebra trivializes completely and condition (\ref{OM}) selects weak Dirac observables for all constraints at once. The extended master constraint has for instance been used in \cite{Giesel:2006uj} where the semiclassical properties of the quantum dynamics have been analyzed in the context of Algebraic Quantum Gravity (AQG).
\\
\\
Apart from solutions to the constraints and observables at the end of the day we also would like to describe the evolution of those observables in order to be able to describe dynamics in the physical sector of the theory. Certainly, their evolution cannot be described by the canonical Hamiltonian for the reason that observables, by definition, need to Poisson commute with all constraints. That evolution for observables is frozen when one considers the canonical Hamiltonian as the generator of their dynamics is known as the \emph{problem of time} in general relativity. Therefore the question arises whether we can reformulate the classical dynamics of general relativity in a way that we can improve our situation twofold:
\begin{itemize}
 \item
  we would like to get a less complicated algebra for the constraints and
\item
 we would like to work in a formalism where the evolution of Dirac observables is naturally implemented
\end{itemize}
The second point can be addressed in the context of the relational formalism introduced in \cite{RRF} and mathematically improved in \cite{DObs}. The idea of this framework is to introduce reference fields that are used to construct Dirac observables and with respect to which the dynamics of the remaining degrees of freedom can be described. In the case of GR this means to introduce reference fields that label points in space and time when we want to construct observables with respect to the Hamiltonian and diffeomorphism constraint. It turns out that by a suitable choice of reference fields we can also make progress regarding the first point and simplify the constraint algebra. In the following of this section we will discuss the classical theory of two particular models where this philosophy has been used and the quantization has been performed using LQG techniques. The quantization of these models will be discussed in section \ref{suse:qbkmod} and \ref{suse:qscalar} respectively.
\\
\\
Before going into the details of those models we will briefly comment on two different strategies to quantize systems with constraints. Given a classical system with constraints we have the option to solve the constraints classically and work with the so called reduced (or physical) phase space or we can follow a procedure introduced by Dirac and quantize the kinematical phase space and then implement the classical constraints as operators. Following the latter we obtain as an intermediate step a kinematical Hilbert space $\mathcal{H}_{kin}$ and require for physical states that they are annihilated by all constraint operators, which are the quantum Einstein equations in the case of general relativity. Solutions to the quantum Einstein Equations live in the physical Hilbert space $\mathcal{H}_{phys}$. In the case of reduced phase space quantization one needs to quantize not the kinematical algebra but the algebra of observables and obtains directly the physical Hilbert space $\mathcal{H}_{phys}$. The dynamics, as we will see in the following two sections, is described by a so called physical Hamiltonian at the classical level, whose explicit form depends on the chosen model. The Hamiltonian is called physical here for the reason that in contrast to the canonical Hamiltonian it is not vanishing on the constraint surface. In the reduced case the quantum dynamics is encoded in the operator corresponding to the physical Hamiltonian and the quantum Einstein equations are given by the quantum evolution of the quantized observables generated by the physical Hamiltonian operator. More details and concrete applications will follow in section \ref{suse:qbkmod} and section \ref{suse:qscalar}. Whether one chooses reduced phase space or Dirac quantization is rather model dependent. In general relativity the constraints are complicated and general solutions to the constraints are difficult to construct. That is why often Dirac quantization is favored. Furthermore, in general the algebra of observables has a more complicated structure than the kinematical one and hence to find representations of it can become a challenging task.  On the other hand finding solutions that are annihilated by all constraints operators and construct the inner product for the physical Hilbert space can be very difficult as well and without the physical Hilbert space the physical relevance of the results is rather hard to evaluate. The models discussed in the next sections have both the properties that the algebra of the observables is isomorphic to the corresponding kinematical algebra and thus representations can be easily found. They also share the property that not all constraints are reduced at the classical level but part of them are solved in the quantum theory. Therefore in both models  a combination of Dirac and reduced phase space quantization is used which is also a choice one can make for quantizing systems with constraints. 

%% file: BKModCl.tex
The idea of the Brown-Kuchar Model \cite{BK} is to introduce additional matter dust fields that can serve as a reference system for general relativity. In the language of the relational formalism we need one reference field for each constraint that we would like to reduce at the classical level. The action introduced in \cite{BK} has the following form
\begin{equation}
\label{dustAction}
S_{\rm dust}=-\frac{1}{2}\int_{M} d^4X\sqrt{|\det(g)|}\rho(g^{\mu\nu}U_{\mu}U_{\nu}+1)
\end{equation}
Here $M$ denotes the spacetime manifold, $g$ the spacetime metric, $\rho$ is the dust energy density and $U_{\mu}=-T_{,\mu}+W_jS^j_{,\mu}$ denotes the dust four velocity, that is itself expressed in terms of 7 scalar fields $T,S^j, W_j$ where $j$ runs from 1 to 3. The action is understood as a functional of the metric $g$ and the eight scalar field $\rho,T,S^j, W_j$  that are  considered in addition to the gravitational and possible other standard model matter degrees of freedom. Here we will restrict our discussion to gravity and dust only, but the generalisation to additional standard model matter is straightforward.
That the action in (\ref{dustAction}) is associated with dust is justified by looking at its physical interpretation. The energy momentum tensor $T_{\mu\nu}$ is that of pressureless dust. The Euler-Lagrange equations for the scalar fields yield that 
$U^{\mu}=g^{\mu\nu}U_{\nu}$ is a geodesic congruence and the fields $W_j,S^j$ are constant along those geodesics. Furthermore $T$ defines proper time along each geodesic. The canonical analysis of the action above shows that the system including the dust involves second class constraints. Introducing the associated Dirac bracket and solving those second class constraints strongly yields that neither $\rho$ nor $W_j$ are independent variables but can be expressed in terms of the remaining phase space variables. Moreover, as long as only functions of $T,S^j, A$ and their conjugate momenta $P,P_j, E$ are considered, the Dirac bracket reduces to the Poisson bracket, and it is those functions that we are interested in the following. For the reason that the fields $S^j$ are constant  and  the field $T$ defines proper time along each geodesics they are a natural choice as the reference fields for space and time respectively. The field $T$ will be used to construct observables with respect to the Hamiltonian constraint $C$ and the three fields $S^j$ will serve as reference fields for the spatial diffeomorphism constraints $C_a$. The choice of reference fields in the relational formalism is completely arbitrary and those fields do not even need to be additional matter fields. For instance one could also choose gravitational degrees of freedom. However, what will be influenced by that choice is the form of the algebra of the constructed observables as well as the form of the physical Hamiltonian. Since we are interested in quantizing the reduced theory later on we would like to choose those reference fields that lead to a simple observable algebra as well as to a physical Hamiltonian, which can be implemented as an operator later on. As we will discuss now the dust fields $T,S^j$ satisfy both requirements and can be interpreted as a free falling observer that is dynamically coupled to the system. The particular choice of $T$ and $S^j$ should not be understood as the only convenient clock and rods for general relativity but a possible choice that has been made in this model and that allows, as we will see in section \ref{suse:qbkmod},  to complete the quantization program. Likewise to the situation when one chooses a particular gauge fixing it might be necessary to consider more than one choice of reference fields in order to consider dynamics in a large region of spacetime. By considering the phase space whose coordinates are given by $T,S^j,A$ and their conjugate momenta we work in an extended phase space picture where time and space are themselves configuration variables.
The (first class) constraints of the system gravity plus dust have the following form:
\bea
\label{eq:ConsDust1}
C^{\rm tot}&=&C+C^{\rm dust}, \quad\rm{with}\quad C^{\rm dust}=-\sqrt{P^2+q^{ab}(PT_{,a}+P_jS^j_{,a})(PT_{,b}+P_jS^j_{,b})}\nonumber\\
C_a^{\rm tot}&=&C_a+C_a^{\rm dust}\quad{\rm with}\quad C_a^{\rm dust}=PT_{,a}+P_jS^j_{,a}
\eea  
where $C,C_a$ denote the gravitational parts of the Hamiltonian and diffeomorphism constraints shown in (\ref{eq:Constraints}) and the Gauss constraint is  unaltered the one in (\ref{eq:Constraints}). What Brown and Kuchar observed in their seminal article was that (i) the fields $P,T$ enter into $C^{\rm tot}$ only in the combination that also occurs in $C_a^{\rm dust}$, and using $C^{\rm tot}_a=0$ those terms can be replaced by $-C_a$ and (ii) the constraint $C^{\rm tot}$  can be solved for the dust momentum $P$ and $C_a^{\rm tot}$ for $P_j$. As a consequence one can write down an equivalent set of constraints given by
\bea
\label{eq:ConsDust2}
\widetilde{C}^{\rm tot}&=&P+h\quad{\rm with}\quad h(A,E):=\sqrt{C^2-q^{ab}C_aC_b}\nonumber\\
\widetilde{C}_j^{\rm tot}&=&P_j+h_j\quad{\rm with}\quad h_j(T,S^j,A,E)=S^a_j(C_a -hT_{,a})
\eea
where $q^{ab}$ is understood as a function of $E$ and we had to assume that the inverse of $S^j_{,a}$ denoted by $S^a_j$ exists. In regions of the phase space where $\det(S^j_{,a})=0$ those fields would not provide a good choice of spatial reference fields. In the case of the Hamiltonian constraint one obtains a quadratic equation in $P$ and chooses that solution of the possible two that yields a positive physical Hamiltonian, which has the correct flat spacetime limit.
Note that the function $h$ in contrast to $h_j$ in (\ref{eq:ConsDust2}) does not depend on the dust degrees of freedom anymore. If this happens one calls a constraint deparametrized, and given this, the construction of observables technically simplifies as well as deparametrization for the Hamiltonian constraint ensures that the final physical Hamiltonian will be time independent. The constraints in (\ref{eq:ConsDust2}) mutually commute up to SU(2) invariant combinations of Gauss constraints
\be
\{\widetilde{C}^{\rm tot}(x),\widetilde{C}^{\rm tot}(y)\}_{G_j=0}=\{\widetilde{C}_j^{\rm tot}(x),\widetilde{C}_k^{\rm tot}(y)\}_{G_j=0}=\{\widetilde{C}^{tot}(x),\widetilde{C}_j^{tot}(y)\}_{G_j=0}=0
\ee
which can be seen by direct computation or by the abstract argument that those constraints are linear in the dust momenta. Since their first class property has not changed by writing them in equivalent form, the Poisson bracket of any two constraints can only be a linear combination of those. But since they are linear in the momenta and the derivative in the Poisson brackets cancels the momentum, the only possible coefficients that are allowed on the righthand side are zero for $\widetilde{C}^{\rm tot}$ and $\widetilde{C}_j^{\rm tot}$ and non-zero coefficients for SU(2) invariant combinations of Gauss constraints\footnote{When working with ADM instead of Ashtekar variables the Gauss constraint is absent and in this case the constraints in (\ref{eq:ConsDust2}) mutually commute strongly.}.
In contrast the constraints in (\ref{eq:ConsDust1}) satisfy the Dirac algebra and thus by solving for the dust momenta we obtain a simplification of the constraint algebra. 
The construction of observables is performed in two steps now: First we reduce with respect to the spatial diffeomorphism constraint by using the fields $S^j$ as rods and obtain for the remaining degrees of freedom the following expressions \cite{Giesel:2007wi}:
\bea
\label{eq:Diffeo}
\tilde{A}^J_j(\sigma^k,t)&:=&\int_{\Sigma} d^3x \left|\det(S^j_{,a})\right|\delta(S^k(x),\sigma^k)A^J_aS^a_j(x) \nonumber\\
\tilde{E}^j_J(\sigma^k,t)&:=&\int_{\Sigma} d^3x \left|\det(S^j_{,a})\right|\delta(S^k(x),\sigma^k)E_J^aS^j_a(x)\nonumber\\
\tilde{T}(\sigma^k,t)&:=&\int_{\Sigma} d^3x \left|\det(S^j_{,a})\right|\delta(S^k(x),\sigma^k)T(x)
\eea
where we introduced a capital $J$ for the $su(2)$ index in order to distinguish between the Lie algebra index and the $j-index$ associated with the dust fields. The interpretation of the observables in (\ref{eq:Diffeo}) is that they give the value of the fields $A,E,T$ when the dust fields $S^j$ take the values $\sigma^j$. The abstract points $x$ in the spatial manifold $\Sigma$ over which the integration is performed above have therefore been labeled by the dust fields $S^j$. Analogous observables for the ADM variables where constructed in \cite{BK}, however the observabels with respect to $\widetilde{C}^{\rm tot}$ where not constructed in \cite{BK} but a formal Dirac quantization was used. This quantization is formal in the sense that similar to the Wheeler-DeWitt equation no representation for general relativity formulated in terms of ADM variables has been found so far in which the constraints can be promoted to well defined operators.
For the construction of observables with respect to the Hamiltonian and diffeomorphism constraint we take the  expression for $\tilde{A}^J_j(\sigma^k,t)$ and $\tilde{E}^j_J(\sigma^k,t)$ above and insert them into the standard formula for observables, that is a power series in powers of the clock field, in our case $T$ and multiple Poisson brackets between the Hamiltonian constraint and $\tilde{A}^J_j(\sigma^k,t)$ and $\tilde{E}^j_J(\sigma^k,t)$ respectively. Note that only the function $h$ in $\widetilde{C}^{\rm tot}$ will contribute in those Poisson brackets, because $P$ commutes with $\tilde{A}^J_j(\sigma^k,t)$ and $\tilde{E}^j_J(\sigma^k,t)$. Explicitly we obtain \cite{AQGIV}
\bea
\label{eq:FullObs}
{\bf A}^J_{j}(\sigma^k,\tau)&=&\exp(\chi_{h_\tau})=\sum\limits_{n=0}^{\infty}\frac{1}{n!}\{\tilde{h}(\tau),\tilde{A}^J_j(\sigma^k)\}_{(n)}\nonumber\\
{\bf E}^j_{J}(\sigma^k,\tau)&=&\exp(\chi_{h_\tau})=\sum\limits_{n=0}^{\infty}\frac{1}{n!}\{\tilde{h}(\tau),\tilde{E}^j_J(\sigma^k)\}_{(n)}
\eea
with
\be
\tilde{h}(\tau):=\int\limits_{\cal S}d^3\sigma(\tau -\tilde{T}(\sigma))\tilde{h}(\sigma)
\ee
here $\chi_{h_\tau}$ denotes the Hamiltonian vector field of $h_\tau$ , ${\cal S}$ denotes the range of $\sigma$ and is also called the \emph{dust space} and the iterative Poisson bracket is defined as 
$\{f,g\}_{(0)}=g\, ,\,$ $\{f,g\}_{(n)}:=\{f,\{f,g\}_{(n-1)}\}\}$. The interpretation of 
the quantities in (\ref{eq:FullObs}) is that they give the values of $A$ and $E$ respectively when the clock field $T$ takes the values $\tau$ and the rod fields $S^j$ take the values $\sigma^j$. $\tau$ and $\sigma^k$ can be understood as the physical time and space parameter. One can check by explicit computation that ${\bf A}^J_{j}(\sigma^k,\tau)$ and ${\bf E}^j_{J}(\sigma^k,\tau)$ indeed Poisson commute with $\widetilde{C}^{\rm tot}$ and $\widetilde{C}_j^{\rm tot}$
and are thus Dirac observables with respect to the Hamiltonian and spatial diffeomorphism constraint. 

The remaining Gauss constraint will be solved in the quantum theory via Dirac quantization.
Once the observables in (\ref{eq:FullObs}) are constructed we are interested in the generator of their dynamics. By construction they Poisson commute with $\widetilde{C}^{\rm tot}$ and 
$\widetilde{C}_j^{\rm tot}$ and thus the canonical Hamiltonian density $h_{\rm can}$ in (\ref{eq:AshAction}), that generates only gauge transformations, cannot be the generator as otherwise  the dynamics of the observables would be frozen. It turns out that the Hamiltonian density of the generator of the physical dynamics is given by the observables associated with the function $h$ in (\ref{eq:ConsDust2}). The observable of a function of $A,E$ can be easily constructed due to an automorphism property that the observable maps satisfy and hence we obtain
\be
{\bf h}(\sigma)=\sqrt{{\bf C}^2(\sigma,\tau) -{\bf q}^{jk}{\bf C}_j{\bf C}_k(\sigma,\tau)}
\ee
where ${\bf C}(\sigma,\tau)$ and ${\bf C}_j(\sigma,\tau)$ are the observables associated with the gravitational part of the Hamiltonian and spatial diffeomorphism constraint that can be obtained by replacing $A,E$ by their corresponding observables ${\bf A},{\bf E}$. Note that although  ${\bf C}_k(\sigma,\tau)$ and ${\bf C}(\sigma,\tau)$ depend on $\tau$ the final expression for ${\bf h}$ does not because on the Gauss constraint surface we have
\be
\{{\bf h}(\sigma),{\bf h}(\sigma^\prime)\}=0
\ee
This follows from the fact that $\widetilde{C}^{\rm tot}(x)$ Poisson commutes with itself and since $h$ does not include any dust degrees of freedom it immediately follows that $\{h(x),h(y)\}_{G_j=0}=0$.
The physical Hamiltonian is then given by
\be
\label{PhysHDust}
{\bf H}_{\rm phys}=\int\limits_{\cal S} d^3\sigma {\bf h}(\sigma)=\int\limits_{\cal S} d^3\sigma \sqrt{{\bf C}^2 -{\bf q}^{jk}{\bf C}_j{\bf C}_k}(\sigma)
\ee
where as before ${\bf q}^{ij}$ is understood as a function of ${\bf E}$. The equation of motion for the observables have the following form
\be
\label{eq:GIEOM}
\frac{d{\bf A}_j^J}{d\tau}=\{{\bf A}_j^J,{\bf H}_{\rm phys}\},\quad \frac{d{\bf E}_J^j}{d\tau}=\{{\bf E}_J^j,{\bf H}_{\rm phys}\}
\ee
and can be understood as a, with respect to the Hamiltonian and spatial diffeomorphism constraint, gauge invariant version of Einstein's equations for the reason that all terms in (\ref{eq:GIEOM}) are manifestly gauge invariant. In contrast to $H_{\rm can}$ in the ordinary Einstein equations ${\bf H}_{\rm phys}$ is nonvanishing on the constraint surface because only the total constraints including the dust degrees of freedom are vanishing there. Furthermore ${\bf H}_{\rm phys}$,  being itself an observable, has the following symmetries on the Gauss constraint surface
\be
\{{\bf H}_{\rm phys},{\bf C_j}(\sigma,\tau)\}_{G_j=0}=0\quad\rm{and}\quad \{{\bf H}_{\rm phys},{\bf h}(\sigma)\}_{G_j=0}=0
\ee
Those symmetries of ${\bf H}_{\rm phys}$ will become important when the quantization of the Brown-Kuchar model in section \ref{suse:qbkmod} is discussed.

%% file: ScalarMod.tex
In the last section we discussed the classical Brown-Kuchar model and saw that the four dust fields can be used as reference fields to construct observabels with respect to the Hamiltonian and diffeomorphism constraint.  The scalar field model, originally introduced  by Rovelli and Smolin \cite{RSscalar} and studied further by Kuchar and Romano \cite{KR}, considers gravity coupled to a massless scalar field whose action is of the form
\be
S_{\varphi}=-\frac{1}{2}\int\limits_M d^4X \sqrt{g}g^{\mu\nu}\phi_{,\mu}\phi_{,\nu}
\ee
The reason that one considers a massless scalar field is that otherwise the resulting Hamiltonian constraint would not be deparametrizable having the consequence that we end up with a time dependent physical Hamiltonian. Here we introduce only one additional matter field coupled to gravity\footnote{Note that also in the scalar field model the generalization to gravity plus arbitrary standard model matter is straightforward.} and thus we are only able to reduce one of the constraints at the classical level. That will be the Hamiltonian constraint. The remaining diffeomorphism and Gauss constraints will be solved in the quantum theory in this model. One of the motivations for this model is that in recent models in loop quantum cosmology (LQC) also a scalar field is chosen \cite{APSModel} as a clock, so this model here could be understood as the generalization of the APS model in \cite{APSModel} to full LQG. Further motivations will become clear in section \ref{suse:qbkmod} and \ref{suse:qscalar} when the quantization of the scalar field and the Brown-Kuchar model is discussed.  A 3+1-split of the action above yields the following total first class constraints
\bea
\label{eq:ConsScal1}
C^{\rm tot}&=&C+C^{\phi}, \quad\rm{with}\quad C^{\phi}=\frac{1}{2}\left(\frac{\pi^2}{\sqrt{\det(q)}} + \sqrt{\det(q)}q^{ab}\phi_{,a}\phi_{,b}\right)\nonumber\\
C_a^{\rm tot}&=&C_a+C_a^{\phi}\quad{\rm with}\quad C_a^{\rm dust}=\pi\phi_{,a}
\eea  
where as before $C,C_a$ denotes the gravitational parts of the Hamiltonian and diffeomorphism constraints shown in (\ref{eq:Constraints}) and $\pi$ denotes the momentum conjugate to $\phi$.
As before, the Gauss constraints is  not affected by the scalar field and thus it is still  the one shown in (\ref{eq:Constraints}). The Brown-Kuchar mechanism, that is replacing the terms that involve derivatives of $\phi$ by $C_a^\phi$ and this then again by $-C_a$ and then solving the Hamiltonian constraint for the scalar field momentum $\pi$ can be applied. One obtains a fourth order polynomial in $\pi$ and out of the four solutions only two are non trivial in the cosmological sector. The remaining $\pm$ sign ambiguity defines different regions in phase space and we choose $\pi=+h$ for the reason that only this solution includes homogenous and isotropic models.  Using this choice one obtains an equivalent Hamiltonian constraint of the form
\bea
\label{eq:ConsScal2}
\widetilde{C}^{\rm tot}&=&\pi-h\quad{\rm with}\quad h(A,E):=\sqrt{-\sqrt{q}C+\sqrt{q}\sqrt{C^2-q^{ab}C_aC_b}}
\eea
Now we are in an analogous situation to the Brown-Kuchar model in section \ref{suse:bkmod}: We realized that also the system of gravity coupled to a massless scalar field deparametrizes and we could proceed as before. This would mean to use the scalar field $\phi$ to construct Dirac observables with respect to the Hamiltonian constraint. What we would obtain are quantities $A_a^j(\tau,x)$ and $E_j^a(\tau,x)$ that commute with $\widetilde{C}^{\rm tot}$ in (\ref{eq:ConsScal2}) but not necessarily with the diffeomorphism or Gauss constraint. The latter two will be solved in the quantum theory via Dirac quantization. The algebra of $A_a^j(\tau,x)$ and $E_j^a(\tau,x)$ is isomorphic to the one of kinematical quantities and their dynamics, that is evolution with respect to the physical time parameter $\tau$, is generated by a physical Hamiltonian of the form
\be
\label{HphysScal}
H_{\rm phys}=\int\limits_\Sigma d^3x h(A,E)(x)
\ee
with $h(A,E)$ given in equation (\ref{eq:ConsScal2}). Alternatively, we cannot construct those observables with respect to the Hamiltonian constraint at the classical level but in the quantum theory and work with so called quantum Dirac observables, whose classical limit corresponds to those observables mentioned above. This will be explained more in detail in section \ref{suse:qscalar}. In this case all constraints will be solved by Dirac quantization in the quantum theory.
Likewise to the Brown-Kuchar model, also here the Hamiltonian densities strongly commute and $H_{\rm phys}$ is invariant under spatial diffeomorphisms, that is
\be
\{H_{\rm phys},h(x)\}_{G_j=0}=0\quad{\rm and}\quad \{H_{\rm phys},C_a(\tau,x)\}_{G_j=0}=0
\ee
In contrast to the Brown-Kuchar model a quantization of this partially reduced system will not yield the physical Hilbert space. 
The quantum theory of this model will be discussed in detail in section \ref{suse:qscalar}. Finally, let us mention that there exist another model introduced in \cite{Husain:2011tm} where only one instead of four dust fields are considered. This model seems to be a special case of the Brown-Kuchar model, where the momentum density of the dust has chosen to be equal to zero.

%% file: al.tex
We will now quantize gravity according to the algorithm for the quantization of constrained systems devised by Dirac (for the original account, see his \emph{Lectures on Quantum Mechanics}, for a modern treatment, see \cite{Thiemann:2007zz}). This means we proceed in two steps.
\begin{enumerate}
	\item Quantization of the canonical variables ("kinematic quantization")
	\item Impose the constraints as operator equations on states, and solve these equations to obtain \emph{physical states}.
\end{enumerate}  
The first step is what we will discuss in the present section. What we want is a representation of the canonical commutation relations
\begin{equation}
\left[A^i_a(x),E^b_j(y)\right]=\frac{\kappa}{2} \hbar\beta\delta_a^b\delta_j^i\delta(x,y)
\label{eq:ccr1}
\end{equation}
on a Hilbert space. Note that $\kappa\hbar=\ell_{\text{P}}^2$, the \emph{Planck area}. 
Fields evaluated at points are usually too singular to give good operators in the quantum theory. Thus one has to form suitably integrated ("smeared") quantities that correspond to well defined operators in the quantum theory. Poisson brackets then suggest commutation relations for these quantities, and one obtains an abstract algebra of operators. We will soon see that details can matter in this context. Different choices of smearing can lead to different algebras and hence to different quantum theories. In fact, in LQG we make a different choice of algebra than is customary in Yang-Mills theory \cite{Rovelli:1989za,Ashtekar:1991iw,Ashtekar:1993wf, Ashtekar:1994mh,Ashtekar:1994wa, Ashtekar:1995zh}. In the latter case, both the algebra and its representations used in the quantum theory make use of the metric as a classical background field in their construction. In general relativity, the metric is dynamic and hence can not be used as a background field. Moreover, a splitting of metric into background and dynamical part, while very useful in practical applications, is not very natural from a fundamental perspective. Hence, the algebra and its representation used in LQG does not make use of any background metric. This  makes LQG a very unusual quantum field theory. To illustrate this, we first consider the case of electromagnetism. The usual quantum theory is obtained by declaring  
\begin{equation}
\left[\int f^aA_a  \sqrt{\det q}\;\text{d}^3x, \int f'_bE^b \;\text{d}^3x\right]
=i\hbar \int f^a f'_a \sqrt{\det q}\;\text{d}^3x \; \id,
\end{equation}  
and by defining the ground state by 
\begin{equation}
a(f)\Omega=0.
\end{equation}
Here $f,f'$ are arbitrary smearing functions, and the definition of the annihilation operators $a$ makes use of the metric $q$ in various ways. But one could also define
\begin{equation}
E(S):=\int_S E^a\epsilon_{abc}\, \text{d}x^b\wedge\text{d}x^c
\end{equation} 
where $S$ is an oriented surface and $\epsilon_{abc}$ is the tensor density that is equal to the totally anti-symetric symbol in any coordinate system. We note that $E^a$ has density weight +1 whereas $\epsilon_{abc}$ has weight $-1$, so the integrand is a two-form and the integral, using the orientation of $S$, is hence coordinate independent. Similarly, defining
\begin{equation}
A(e)=\int_e A_a \,\text{d}x^a,
\end{equation} 
where $e$ is a curve, one finds (by a limiting procedure from the Poisson brackets of the point fields), 
\begin{equation}
[A(e),E(S)]=i\hbar I(e,S)\, \id, 
\end{equation}
where $I(e,S)$ is the signed intersection number for $S$ and $e$. The metric has thus dropped out of all definitions and relations. 

A similar thing can be done for gravity. For a surface $S$ and a Lie algebra valued smearing field $n$ on $S$ we define
\begin{equation}
E_n(S):=\int_S n^i E^a_i\epsilon_{abc}\, \text{d}x^b\wedge\text{d}x^c.
\end{equation} 
For $A$ we use the quantity analogous to $\exp(iA(e))$. We chose a local trivialization and define the \emph{holonomy} 
\begin{align}
h_e:=&\mathcal{P}\exp\int_e A\\
&=\um +\int_0^1A(e(t))\dot{e}^a(t)\,\text{d}t
+\int_0^1\text{d}t_1\int_{t_1}^1\text{d}t_2\, A_a(t_1)\dot{e}^a(t_1)A_a(t_2)\dot{e}^a(t_2) +\ldots
\end{align}
which is an element of SU(2), and gives the parallel transport map from the fiber over the beginning point $b(e)$ of the edge to the fiber over of its final $f(e)$, for the chosen trivialization. Under gauge transformations, i.e., changes of trivialization, $g:M\rightarrow SU(2)$
\begin{equation}
h_e \mapsto g(b(e))h_eg(f(e))^{-1}.
\end{equation}   
One finds
\begin{equation}
\label{eq:simple_comm}
[E_n(S),h_e]=\begin{cases}
\frac{\beta\ell_\text{P}^2}{2}
h_{e_1}\tau_jn^j(p)h_{e_2}& \text{ for a single transversal intersection } S\cap e=\{p\}\\
0 & \text{ if } S\cap e =\emptyset
\end{cases},
\end{equation}
where $\{\tau_j\}$ is a basis of su(2), and $\ell^2_\text{P}=\hbar\kappa$ is (a multiple of) the Plank length.
It is convenient to slightly generalize these variables. Given a \emph{graph} of paths $\gamma=\{e_1,e_2,\ldots e_n\}$ and a function $f:$SU(2)$^n\rightarrow \C$, one obtains the functional
\begin{equation}
f_\gamma[A]:=f(h_{e_1}[A],h_{e_2}[A], \ldots h_{e_n}[A])
\label{eq:cylin}
\end{equation}
A functional f is called \emph{cylindrical with respect to  $\gamma$} (written $f\in \cyl_\gamma$) if it is of the above form, and simply \emph{cylindrical} if it is of the above form for \emph{some} graph $\gamma$. We note:
\begin{itemize}
	\item A given cylindrical functional is cylindrical on \emph{many} graphs. Consider the example of a function $f_\gamma[A]=f(h_e[A])$, which is cylindrical w.r.t.\ the graph $\gamma=\{e\}$. Now consider a second graph $\gamma'={e_1,e_2,e_3}$, with $e_1\circ e_2=e$, and $e_3$ independent of $e$. Then $f_\gamma$ is also cylindrical w.r.t.\ $\gamma'$, as it can be written purely in terms of holonomies along edges in $\gamma'$,  $f_\gamma[A]=f(h_{e_1}[A]h_{e_2}[A])$.     
	\item For two cylindrical functions which are cylindrical on graphs with \emph{smooth} edges, one can not always find a \emph{finite} graph such that they are both cylindrical w.r.t.\ to that graph, because they can intersect infinitely many times. But for more regular edges, for example analytic or semi-analytic (roughly speaking: piecewise real analytic \cite{Lewandowski:2005jk}) ones this can not happen, and so one can always find such a graph. As a consequence, such cylindrical functions are closed under addition and multiplication and thus form an algebra, denoted $\cyl$. In the following, we will always assume edges (and also surfaces) to be semi-analytic.      
\end{itemize}   
We can use these observations to write the commutator between the canonical variables in a relatively concise form. For this, we assume without loss of generality that a surface $S$ and a graph $\gamma$ intersect only in vertices of $\gamma$. The commutator then reads:
\begin{equation}
[ f_\gamma, E_n(S)]\equiv X_n(S)( f_\gamma) =\frac{\beta \ell^2_\text{P}}{4}\sum\limits_{v\in V(\gamma)} n^i(v)\left[\sum\limits_{e \text{ at } v} \kappa(S,e,v) \widehat{J}_i^{(v,e)} f \right ](h_{e_1},h_{e_2}, \ldots)) 
\label{eq:ccr2}
\end{equation} 
where $V(\gamma)$ denotes the set of all vertices of $\gamma$
\begin{equation}
\kappa(S,e,v) = 
\begin{cases}
0 & \text { if $e$ intersects $S$ tangentially in $v$ or does not intersect $S$ at all}\\
1 &  \text { if $e$ intersects $S$ transversally and is above $S$}\\
-1&\text { if $e$ intersects $S$ transversally and is below $S$}\\
\end{cases}
\end{equation}
and 
\begin{equation}
\widehat{J}_k^{(v,e)}=\id\otimes\id\otimes\ldots \otimes 
\left\{
\begin{array}{c}
	L_k^e\\R^e_k
\end{array}
\right\}
\otimes\id \otimes \ldots, \qquad \text{ when }
\left\{
\begin{array}{c}
 e \text{ ingoing at } v \\ e\text{ outgoing at } v
\end{array}
\right\}. 
\end{equation}
$R$, $L$ denote the right/left invariant vector field on SU(2) associated with a basis $\tau_k$ of $su(2)$. The additional factor of 1/2 in \eqref{eq:ccr2} as compared to \eqref{eq:simple_comm} is due to our assumption that edges must end on the surface. An edge that continues on both sides of the surface as in  \eqref{eq:simple_comm} will thus count as two separate edges in \eqref{eq:ccr2}.
For general surfaces, the commutator above may not be a cylindrical function, again because edges and surfaces can interact each other infinitely often. Thus we also restrict the surfaces to be in a suitable regularity class. Then the operation $X_n(S)$ defined above is a derivation on the space $\cyl$ of cylindrical functions.  We note that the commutator has the Jacobi property, so
\begin{equation}
[f,[E_n(S),E_{n'}(S')]]=[X_n(S),X_{n'}(S')](f) 
\end{equation}
and the commutator on the right hand side is \emph{non-vanishing} in general. Thus we find that the operators corresponding to the spatial geometry do not commute.
 
The objects $E_n(S)$, together with the cylindrical functions $\cyl$ subject to the above commutator relations form the kinematic algebra $\mathfrak{A}$. Since it does not make reference to classical geometry on $\Sigma$, diffeomorphisms $\phi$ act in a simple way: 
\begin{equation}
\alpha_\phi(f)[A]:=f(\phi_*A), \qquad \alpha_\phi(E_n(S))=E_{\phi_*n}(\phi(S))
\end{equation}
are automorphisms of $\mathfrak{A}$. A similar statement holds for gauge transformations. 

To implement the constraints we need a representation of $\mathfrak{A}$, i.e., a mapping of $\mathfrak{A}$ into the operators of a Hilbert space. There are many representations of $\mathfrak{A}$, but one of them is special and therefore most important for LQG, the  \emph{Ashtekar-Lewandowski representation} of $\mathfrak{A}$ \cite{Ashtekar:1993wf, Ashtekar:1994mh,Ashtekar:1994wa,Ashtekar:1996eg}. To see how it works, note first that an inner product on $\cyl$ can be defined by 
\begin{equation}
\scpr{f_\gamma}{f'_\gamma}:=\int_{\text{SU(2)}^n} \text{d}\mu(g_1)\ldots \text{d}\mu(g_n)
\overline{f(g_1,g_2,\ldots g_n)}f'(g_1,g_2,\ldots g_n). 
\end{equation}
The measure $\text{d}\mu$ used above is the Haar measure on SU(2), and we have assumed without loss of generality that the two functions are cylindric w.r.t.\ the same graph, as discussed below \eqref{eq:cylin}. Closure with respect to the corresponding norm gives a Hilbert space $\mathcal{H}$. It can be shown that this space has a very suggestive structure, $\mathcal{H}=L^2(\abar,\text{d}\mu_{\text{AL}})$, the square integrable functions over a space of distributional connections, with respect to a certain measure.  

The action of the basic operators in this representation is analogous to that found in the Schr{\"o}dinger representation of quantum mechanics:
\begin{equation}
\pi(f)\Psi[A]=f[A]\Psi[A],\qquad \pi(E_n(S)) \Psi[A]= (X_n(S) \Psi)[A],
\label{eq:al_rep}
\end{equation}
where we have assumed that $\Psi$ is smooth enough for $X_n(S)$ to act. For example, $\Psi$  could be a cylindrical function based on a differentiable function on some power of SU(2). But the properties of this representation are very different from those of the Schr{\"o}dinger representation of quantum mechanics, and of the representations encountered in standard QFT. For example, eigenstates of the fluxes, i.e., the momentum variables, are normalizable, as we will see in a moment. Also, there is a precise analogue of this representation in the case of a scalar field, and it is unitarily inequivalent to the standard representation for a scalar in flat or curved background \cite{Ashtekar:2002vh}. 

The representation has several useful properties. It is irreducible and faithful. 
No background geometry was used in the definitions, so it carries a unitary representation of spatial diffeomorphisms and gauge transformations. It has an invariant and cyclic\footnote{\emph{Cyclic} means that $\{a\Omega| a\in\mathfrak{A}\}$ is dense in $\mathcal{H}$.} vector $\Omega$. 

The Hilbert space $\mathcal{H}$ has a very useful orthonormal basis. To explain, let us first consider a general compact Lie group $G$. Then there are two natural representations of $G$ on $\mathcal{H}_G=L^2(G, \text{d}\mu)$, the left- and right-regular representations
\begin{equation}
(\rho_\text{L}(g)f)(g')=f(g'g^{-1}), \qquad (\rho_\text{R}(g)f)(g')=f(gg'). 
\end{equation}
They both decompose into irreducibles, and since the two representations commute, there is a common basis of eigenvectors of the Casimir operators. Let $\pi$ be an irreducible representation of $G$, then 
\begin{align}
V(\pi,m)&:=\text{span}\{\pi_{mn}(\cdot)| n=1,2,\ldots \dim \pi \} \text{ is left invariant by } \rho_\text{L},\\
\overline{V}(\pi,n)&:=\text{span}\{\pi_{mn}(\cdot)| m=1,2,\ldots \dim \pi \} \text{ is left invariant by } \rho_\text{R}.
\end{align} 
The representation $V(\pi,m)$ induced by $\rho_\text{L}$ is $\pi$ itself. The one induced by 
$\rho_\text{R}$ on $\overline{V}(\pi,n)$ is its dual, $\overline{\pi}$, i.e. $\overline{\pi}(g)=\pi(g^{-1})^T$. The Peter-Weyl theorem now asserts that each irrep arises in the decomposition of the regular representations, and even more, that their matrix elements give a basis of
$\mathcal{H}_G$. Pick, for each equivalence class of irreps of $G$ a representative $\pi$, then the set of all $\sqrt{\dim \pi} \pi_{mn}$ for all equivalence classes form an orthonormal basis of $\mathcal{H}_G$.     

Now back to the LQG setting. Let $\mathcal{H}_\gamma=\overline{\cyl_\gamma}^{\norm{\cdot}}$. On the one hand, $\mathcal{H}_\gamma$ is a subspace of $\mathcal{H}$, on the other hand it is isomorphic to $L^2(\text{SU(2)}^n)$. Thus, an orthonormal basis for $\mathcal{H}_\gamma$ is given by 
\begin{equation}
\sqrt{(2j_1+1)\ldots (2j_n+1)}
\overset{j_1}{\pi}_{k_1l_1}(h_{e_1}[A])\cdot
\overset{j_2}{\pi}_{k_2l_2}(h_{e_2}[A])\cdot\ldots 
\overset{j_n}{\pi}_{k_n l_n}(h_{e_n}[A]).
\end{equation}
where the $j_1,j_2,\dots j_n$ label irreducible representations of SU(2).
We note, however, that in general $\mathcal{H}_\gamma\not\perp \mathcal{H_{\gamma'}}$. 
Given $\gamma$, there are functions in $\mathcal{H}_\gamma$ that do not depend on the holonomy along some edge of  $\gamma$, or that only depend on the product of two holonomies along adjacent edges in $\gamma$. Such functions are also elements of $\mathcal{H_{\gamma'}}$ for some $\gamma'$ that has less edges and vertices.  This makes linear dependencies among the elements of different graph Hilbert spaces rather generic. Take for example $\gamma=\{e\}$, $\gamma'=\{e_1,e_2\}$ with $e=e_1\circ e_2$. Then 
\begin{equation}
\pi_{mn}(h_e[A])=\sum_{m'}\pi_{mm'}(h_{e_1}[A])\pi_{m'n}(h_{e_2}[A]).
\end{equation}
Therefore we will introduce a family of slightly modified Hilbert spaces $\mathcal{H}'_\gamma$, which give a decomposition of $\mathcal{H}$ into orthogonal subspaces. But first we need to discuss the transformation properties of vectors under gauge transformations.

We start by considering just a single edge $e$. With respect to gauge transformations $g$, the vectors $\pi^j_{mn}(h_e)$ transform under the tensor product $j\otimes\overline{j}$, and can be visualized as the edge with representation $j$ sitting at its endpoint and representation $\overline{j}$ at its starting point.  
When several edges meet at a vertex $v$, contractions of the matrix indices of the representation matrices at that vertex can be done and correspond to vectors in the tensor product 
\begin{equation}
\mathcal{H}_v=\left(\bigotimes\limits_{e \text{ incoming } at v} j_e \right)\quad\bigotimes\quad\left( \bigotimes\limits_{e \text{ outgoing } at v} \overline{j_e}\right)
\label{eq:vertex}
\end{equation} 
To give an orthogonal basis of this space, one can simply decompose it into irreps, 
\begin{equation}
\mathcal{H}_v=\bigoplus_l c_l\, l,
\label{eq:vertex_decomp}
\end{equation} 
where $c_l$ counts the multiplicity of the spin $l$-representation.  When we apply this to the situation in LQG we obtain the following decomposition. Given a graph $\gamma$, 
\begin{equation}
\mathcal{H}_\gamma= \bigoplus_{\vec{j}}\mathcal{H}_{\gamma,\vec{j}}=\bigoplus_{\vec{j},\vec{l}}\mathcal{H}_{\gamma,\vec{j},\vec{l}}.  
\end{equation}
Here we have first decomposed into spaces in which the assignment of irreps to edges (labeled by $\vec{j}$) is fixed, giving essentially the tensor products of the spaces \eqref{eq:vertex}. Then we have further decomposed according to \eqref{eq:vertex_decomp}, labeling the irreducible subspace chosen at the vertices with $\vec{l}$.  

Now we can go back and remedy the problem that the decomposition into $\mathcal{H}_\gamma$ was not an orthogonal one (following \cite{Ashtekar:2004eh}). Given again a graph $\gamma$, we can call a labeling $\vec{j}$ of edges and $\vec{l}$ of vertices with irreps \emph{admissible} if no two-valent vertex has been assigned the trivial representation $l=0$, and none of the irreps assigned to the edges is trivial. Then we set
\begin{equation}
\mathcal{H}'_\gamma=\bigoplus_{\vec{j},\vec{l} \text{ admissible }}\mathcal{H}_{\gamma,\vec{j},\vec{l}}, 
\end{equation}
and obtain the desired orthogonal decomposition
\begin{equation}
\label{OrthDecomp}
\mathcal{H}=\bigoplus_\gamma \mathcal{H}'_\gamma.
\end{equation}

%% file: area_operator.tex
One of the special properties of the representation in LQG introduced in the last section is that one can define operators corresponding to geometrical objects such as volume area and length. Among those the most simple operator is the area operator from the point of view of the construction of the operator as well as with regards to the spectrum of the operators. 
The area operator was first introduced by Smolin \cite{Smolin:1992qz} and then further analyzed by Rovelli and Smolin in the loop representation \cite{VRS}. Ashtekar and Lewandowski discussed the spectrum of the area operator in the connection representation in \cite{Ashtekar:1996eg}.
In this section we want to discuss the implementation of the area operator as well as its spectrum in detail. At the end of the section we will briefly comment on the volume and length operator.
\\
\\
The classical area functional associated to a surface $S$ is given by the following expression
\be
Ar(S)=\int\limits_U d^2u\sqrt{\det(X^*q)}(u)
\ee
The ADM 3 metric is denoted by $q$, $X: U\to S$ is an embedding of the surface, where $U\subset \R^2$ and $X^*$ denotes the pull back of $X$. 
The coordinates on the embedded surface $S$ are given by the embedding functions $X^a$ with $a=1,2,3$ and let us denote the two coordinates by which the surface is parametrized by $u_1$ and $u_2$.  Given the embedding we can construct two tangent fields on $S$ 
\be
X^a_{,u_1}:=\frac{\partial X^a}{\partial u_1},\quad X^a_{,u_2}:=\frac{\partial X^a}{\partial u_2}
\ee
and also a co-normal vector field $n_a$ that is determined from the condition
\be
\label{conormal}
n_aX^a_{u_,i}=0\quad{\rm for}\quad i=1,2
\ee
The determinant in the area functional can be expressed as
\be
\det(X^*q)=q_{u_1u_1}q_{u_2u_2} - q_{u_1u_2}q_{u_2u_1}
=\left(X^a_{,u_1}X^b_{,u_1}X^c_{u_2}X^d_{,u_2} - X^a_{,u_1}X^b_{,u_2}X^c_{,u_2}X^d_{,u_1}\right)q_{ab}q_{cd}
\ee
In order to quantize the area functional we need to express it in terms of Ashtekar variables. For this purpose we consider the expression $\det(q)n_a n_b q^{ab}$ and use that we can express the inverse metric as 
\be
q^{ab}=\frac{1}{2}\frac{1}{\det(q)}\epsilon^{acd}\epsilon^{bef}q_{ce}q_{df}
\ee
Furthermore we see from (\ref{conormal}) that $n_a=\epsilon_{abc}X^{c}_{,u_1}X^d_{,u_2}$ yielding
\bea
\label{DetXq}
\det(q)n_an_bq^{ab}
&=&
\det(q)n_an_b\frac{1}{2}\frac{1}{\det(q)}\epsilon^{acd}\epsilon^{bef}q_{ce}q_{df}\nn \\
&=&
\epsilon_{ak\ell}X^k_{,u_1}X^\ell_{,u_2}\epsilon_{bmn}X^m_{,u_1}X^n_{,u_2}\frac{1}{2}\epsilon^{acd}\epsilon^{bef}q_{ce}q_{df}\nn \\
&=&
q_{u_1u_1}q_{u_2u_2} - q_{u_1u_2}q_{u_2u_1}
\eea
The inverse metric has a simple form in Ashtekar variables given by $q^{ab}=E^a_jE^b_k \delta^{jk}/ \det(E)$ and depends only on the densitized triad. From $E^a_j=\sqrt{q}e^a_j$ we get $\det(E)=\det(q)$  yielding
\be
\det(q)q^{ab}=E^a_jE^b_k \delta^{jk}
\ee
from which we can conclude using (\ref{DetXq}) that
\be
\sqrt{\det(X^*q)}=\sqrt{n_an_bE^a_jE^b_k \delta^{jk}}
\ee
Note that often one choses the basis $\tau_j:=-i\sigma_j/2$ in su(2) with $\sigma_j$ being the Pauli matrices for which the Cartan-Killing metric on su(2) $\eta_{jk}$ becomes 
$\eta_{jk}:=Tr(ad(\tau_j)ad(\tau_k))=-2\delta_{jk}$ and then one uses the Killing metric in the expression above and adjusts the pre-factors accordingly. 
The area functional is completely expressed in terms of Ashtekar variables now. In order to quantize the area functional we need to choose a regularization of the classical expression. We choose a partition of $U$ into closed set such that $U=\bigcup\limits_{i=1}^nU_i$. Then we can express the area functional as a sum of integrals over the individual sets $U_i$. Furthermore, we assume that the area of each $U_i$ is $\epsilon^2$ and we will denote a point centered in $U_i$ by $v$.
\bea
Ar(S)&=&\sum\limits_{i=1}^n\int\limits_{U_i}d^2u\sqrt{E^a_jE^b_k\delta^{jk}n_an_b}(X(u))\nn \\
&\approx&\sum\limits_{i=1}^n \epsilon^2 \sqrt{E^a_jE^b_k \delta^{jk}n_an_b}(v)\nn \\
&=&\sum\limits_{i=1}^n  \sqrt{(\epsilon^2E^a_jn_a)(\epsilon^2E^b_kn_b) \delta^{jk}}(v)\nn \\
&=&\sum\limits_{i=1}^n  \sqrt{E_j(S_{U_i})E_k(S_{U_i}) \delta^{jk}}(v)
\eea
In the limit $n\to\infty$ the Riemann sum above yields exactly the area functional. As can be seen in the last line we managed to express $Ar(S)$ as a function of the classical fluxes $E_j(S_{U_i})\equiv E_{\tau_j}(S_{U_i}) $ for which well defined operators exists. For this reason we obtain the quantum area operator simply by replacing the classical fluxes by their corresponding operators. Recall that the flux operators can be written in terms of right and left invariant vector fields $\rivf$ and $\livf$ associated to the edges of a given spin network function. 

\be
\flux\snf=\frac{\beta\ell_p^2}{4}\sum\limits_{\substack{v\in V(\gamma)\\ e\, \text{at}\, v}}\kappa(S,e,v)\hat{J}^{(e,v)}_j\snf
\ee
The term $\kappa(S,e,v)$ is as before $+1$ and $-1$ respectively for edges that intersect $S$ transversally and are above $S$ and below $S$ respectively at $v$ and zero in all other cases.
For the area operator we obtain
\be
\label{areaOP}
\hat{Ar}(S)\snf=\frac{\beta\ell_p^2}{4}\sum\limits_{i=1}^n\sqrt{\big(\sum\limits_{\substack{v\in V(\gamma)\\ e\, \text{at}\, v}}\kappa(S_{U_i},e,v)\hat{J}^{(e,v)}\big)^2}\snf
\ee
If the partition is chosen fine enough so that only one intersection point exists in $S_{U_i}$ instead of summing over all sets $U_i$ in the partition we can sum over all intersections points of an edge of type up or down. Note that in each intersection point more than one edge could meet. In this way we will loose all surfaces $S_{U_i}$, which do not contain an intersection point of an edge of type up or down. However, these would anyway not contribute since for them $\kappa(e,S_U)=0$. Thus we can reexpress the area operator as
\be
\label{areaOPYx}
\hat{Ar}(S)\snf=\frac{\beta\ell_p^2}{4}\sum\limits_{v\in P(S)}\sqrt{\big(\sum\limits_{e\text {at} v}\kappa(S_{U(v)},e,v)\hat{J}^{(v,e)}\big)^2}\snf
\ee
where  the set $P(S)$ of intersection points of edges up and down is given by
\be
P(S)=\{v\in e\cap S | \kappa(S,e,v)\not=0, \,\, e\in E(\gamma)\}
\ee
Let us now discuss the spectrum of the area operator. At each intersection point $x$ we have edges of type up, edges of type down and edges of type in that will not contribute to the spectrum. In order to write the expression under the square root in (\ref{areaOPYx}) in compact form we introduce the following operators:
\be
\hat{J}_j^{v,u}:=\sum\limits_{e\in E(v,u)} \hat{J}_j^{(v,e)}\quad \hat{J}_j^{v,d}:=\sum\limits_{e\in E(v,d)} \hat{J}_j^{(v,e)}
\ee
Here $E(v,u), E(v,d)$ denotes all edges of type up and down respectively that intersect each other in the point $v$. Then we have for each intersection point $v$
\bea
\label{Yud}
\left(\sum\limits_{\substack{e\in E(\gamma)\\ e\cap v\not=\emptyset}}\kappa(S,e,v) \hat{J}_j^{(v,e)}\right)^2
&=&
\left(\hat{J}^{v,u} - \hat{J}^{v,d}\right)^2 \nn \\
&=&(\hat{J}^{v,u})^2 +  (\hat{J}^{v,d})^2 - 2\hat{J}^{v,u}\hat{J}^{v,d} \nn \\
&=&2(\hat{J}^{v,u})^2 +  2(\hat{J}^{v,d})^2 - (\hat{J}^{v,u}+\hat{J}^{v,d})^2
\eea
We used in the second line that $[\hat{J}_j^{v,u},\hat{J}_k^{v,d}]=0$. Furthermore the operators $(\hat{J}^{v,u})^2$, $(\hat{J}^{v,d})^2$ and $ (\hat{J}^{v,u}+\hat{J}^{v,d})^2$ mutually commute. Moreover, we choose an explicit basis $\tau_j=-i\sigma_j/2$ for which the operators $\hat{J}^{(e,v)}$ satisfy the usual angular momentum algebra given by
$[\hat{J}^{(e,v)}_i,\hat{J}^{(e,v)}_j]=\epsilon_{ijk}\hat{J}^{(e,v)}_k$. Then we have that the operators $(\hat{J}^{(e,v)})^2\equiv \delta^{jk}\hat{J}^{(e,v)}_j\hat{J}^{(e,v)}_k$ locally act as
\begin{equation}
-\delta^{ij} R_i R_j=-\scpr{R}{R}\equiv -\Delta_{SU(2)}, \text{ or }-\delta^{ij} L_i L_j=-\scpr{L}{L}\equiv -\Delta_{SU(2)}
\end{equation}
where $-\Delta_{SU(2)}$ is the positive definite SU(2) Laplacian with spectrum $j(j+1)$, due to our choice of basis for su(2). Hence the same holds for the operators $(\hat{J}^{v,u})^2$, $(\hat{J}^{v,d})^2$, and $(\hat{J}^{v,u}+\hat{J}^{v,d})^2$, they act as Laplacians in the respective direct sum of representations. Therefore the spectrum of the operators involved in (\ref{Yud}) can be easily computed and we obtain 
 \be
{\rm Spec}(\hat{Ar}(S))=\frac{\beta\ell_p^2}{4}\sum\limits_{v\in P(S)}
\sqrt{2j_{u,v}(j_{u,v}+1) +  2j_{u,v}(j_{u,v}+1) -j_{u+d,v}(j_{u+d,v}+1)}
\ee
Here $j_{u,v},j_{d,v}$ denote the total angular momentum of the edges of type up (down respectively) at the intersection point $v$ and $j_{u+d,v}$ total coupled angular momentum of the up and down edges whose values range between $|j_{u,v}-j_{d,v}|\leq j_{u+d,v}\leq j_{u,v}+j_{d,v}$.
Let us consider the eigenvalue at one intersection point $v$. The smallest possible eigenvalue that we can get  occurs when either $j_{u,v}=0$ and $j_{d,v}=\frac{1}{2}$ or vice versa. The eigenvalue denoted by $\lambda_0$ is non vanishing and given by
\be
\lambda_0=\ell_p^2\beta\frac{\sqrt{3}}{8}
\ee
and is known as the area gap in LQG. The area gap plays an important role in the description of black hole physics within LQG and  black hole entropy calculations can be used the fix the value of the Immirzi parameter $\beta$.
\\
Finally let us say a few words about the other two geometrical operators, the volume and length operator in LQG. 
The volume operator enters crucially into the construction of the dynamics of LQG for the reason that the classical co-triad is expressed as the Poisson bracket between the connection and the  classical volume functional using the Thiemann trick (see section \ref{suse:ham}). The volume operator can be quantized in a similar manner than the area operator because the classical three dimensional volume of a given region $R$ is the integral over $R$ of $\sqrt{\det(q)}$ whose expression in Ashtekar variables is just $\sqrt{|\det(E^a_j)|}$. Thus it is again a function of the electric fields only and can be after a suitable regularization again be expressed in terms of flux operators. In the literature exist two different volume operators one introduced by Rovelli and Smolin (RS) \cite{VRS} and one introduced by Ashtekar and Lewandowski (AL) \cite{VAL}, which come out of a priori equally justified but different regularization techniques. Both volume operators act non-trivially only on vertices where at least three edges intersect. At a given vertex the operators have the following form
\bea
\label{eq:VolOp}
\hat{V}_{v,\rm RS}&=&C_{\rm RS}\sum\limits_{e_I\cap e_J\cap e_K=v}\sqrt{\big|\hat{Q}_{IJK}\big|}\nn \\
\hat{V}_{v,\rm AL}&=&C_{\rm AL}\sqrt{\Big|\sum\limits_{e_I\cap e_J\cap e_K=v}\epsilon(e_I,e_J,e_K)\hat{Q}_{IJK}\big|}
\eea
Here $\hat{Q}_{IJK}:=\epsilon_{ijk}\hat{J}^i_{e_I}\hat{J}^j_{e_J}\hat{J}^k_{e_K}$ is an operator involving only flux operators and thus right and left  invariant vector fields and $C_{\rm RS},C_{\rm AL}$ are regularization constants. The sum runs over all ordered triples of edges intersecting at the vertex $v$. The main differences between these two operators is that the RS-operator is not sensitive to the orientation of the triples of edges and therefore also planar triples of edges will contribute. The AL-operator has likewise to the $\kappa(S,e,v)$ in the area operator a similar sign factor $\epsilon(e_I,e_J,e_K)$ that vanishes whenever the triple of edges $e_i,e_j,e_k$ intersecting at a vertex $v$ are linearly dependent. Furthermore the sum over triples of edges involved in both operators, occurs outside the square root in case of the RS and inside the square root in case of the AL-operator.
\\ 
The spectral analysis of the volume operator is more complicated than for the area operator and can in general not be computed analytically.  A general formula for the computation of matrix elements of the AL-volume operator has been derived in \cite{JBVol}. Those techniques have been used to analyze the spectrum of the volume operator numerically up to a vertex valence of 7 in a series of papers \cite{JBDR}. Their work showed that the spectral properties of the volume operator depend on the embedding of the vertex that enters via the sign factors $\epsilon(e_i,e_j,e_k)$ into the construction of the AL-operator. Particularly the presence of a volume gap, that is a smallest non vanishing  eigenvalue, depends on the geometry of the vertex. For a more detailed discussion about the spectrum of the volume operator we refer the reader to the article of Brunnemann in these proceedings.
 A consistency check for both volume operators has been discussed in \cite{CCVol} where the Thiemann trick has been used to define an alternative flux operator. Those alternative flux operator is then compared to the usual flux operator and consistency of both operators could for instance fix the undetermined regularization constant $C_{\rm AL}=\ell_p^3/\sqrt{48}$ in the volume operator. Furthermore, the RS-operator did not  pass this consistency check and the reason that it worked for the AL-operator is exactly the presence of those sign factors $\epsilon(e_I,e_J,e_K)$ in the AL-operator. 
\\
\\
A length operator for LQG was introduced in \cite{Tlength}. The length operator is in some sense the most complicated one among the kinematical geometrical operators. Let us recall the the length of a curve $c: [0,1]\to\Sigma$ classically is given by
\be
\label{cllen}
\ell(c)=\int\limits_{0}^1\sqrt{q_{ab}(c(t))\dot{c}^a(t)\dot{c}^b(t)}dt=\int\limits_{0}^1\sqrt{e^i_{a}(c(t))e^j_b(c(t))\dot{c}^a(t)\dot{c}^b(t)\delta_{ij}}dt
\ee
here $\dot{c}^a$ denotes the components of the tangent vector associated to the curve.
When we express the metric $q_{ab}$ in terms of Ashtekar variables we obtain
\be
q_{ab}=\frac{1}{4}\epsilon_{acd}\epsilon_{bef}\epsilon_{ijk}\epsilon^{imn}\frac{E^c_jE^d_kE^m_eE^n_f}{\det(E)}
\ee
which is a non-polynomial function in terms of the electric fields and therefore a regularization in terms of flux operators similar to the area and volume operator does not exist.  Furthermore the denominator being the square of the volume density cannot be defined on a dense set in $\Hkin$ because it has a huge kernel.
One possibility to quantize the length used in \cite{Tlength} is to use for the co-triads that occur in (\ref{cllen}) the Thiemann trick and replace them by a Poisson bracket between the connection and the volume functional.  
This yields a length operator that involves a square root of two commutators between holonomy  operators along the curve $c$ and the volume operator. In this way the inverse volume density can be avoided and the volume occurs only linearly in the commutator. Also, the length operator does not change the graph or the spin labels of the edges likewise to the area and volume operator.
However, since the length operator becomes even a function of the volume operator its spectral analysis becomes even more complicated than for the volume operator itself and very little about the spectrum of the length operator is known except for low valence vertices.
\\
Another length operator was introduced in \cite{Blength} where the Thiemann trick was not used for the quantization. The regularization adapted in \cite{Blength} is motivated from the dual picture of quantum geometry and uses that the curve can be expressed as an intersection of two surfaces. This allows to express the tangent vector of the curve in terms of the normals of the surfaces. The inverse volume issue discussed above is circumvented by using a Tikhonov regularization for the inverse RS-volume-operator. For this length operator the spectral properties have only be analyzed for a vertex of valence 4, which is monochromatic, that is all spins are identical.
Recently an alternative length operator for LQG has been discussed in \cite{Mlength} where a different regularization has been chosen such that the final length operator can be expressed in terms of other geometrical objects the area, volume and flux operators. In this work the AL-operator is used and the inverse volume operator is also defined using a Tikhonov regularization similar to the one in \cite{Blength}.

%% file: gauss.tex
The classical expression for the smeared version of the Gauss constraint is given by
\be
G(\Lambda)=\int\limits_\Sigma d^3x \Lambda^j(D_aE^a_j)=-\int\limits_\Sigma d^3x (D_a\Lambda^j)E^a_j
\ee
This expression involves the densitized triad integrated over a 3-dimensional integral with smearing function $(D_a\Lambda^j)$. When the holonomy flux algebra is computed by means of a regularisation of holonomies $h^\epsilon_e(A)$ and fluxes  $E^{\epsilon^\prime}(S)$ both the holonomies and fluxes are regularized by expression that involve three dimensional integrals and $\epsilon$ and $\epsilon^\prime$ denote regulators. For this reason one can follow the same regularization strategy for the computation of $\{h_e(A),G(\Lambda)\}$ with the difference that we do not need a regulator for $G(\Lambda)$ since it involves already a 3 dimensional integral. Computing $\{h^\epsilon_e(A),G(\Lambda)\}$ and removing the regulator afterwards yields
\be
\label{HolGauss}
\{h_e(A),G(\Lambda)\}
=-\frac{\beta\kappa}{2}\int\limits_0^1 dt (D_a\Lambda^j)(e(t))\dot{e}^a(t)h_e(0,t)(A)\tau_jh_e(t,1)(A)
\ee
In the following we will choose an explicit basis for $su(2)$ given by $\tau_j:=-i\sigma_j/2$ with $\sigma_j$ being the Pauli matrices. In this case the structure functions $f_{jk\ell}$ are given by the Levi-Cevita symbol $f_{jk\ell}=\epsilon_{jk\ell}$. The expression $h_e(0,t)$ satisfies the differential equation
\be
\label{DGLHol}
\frac{dh_e(0,t)}{dt}=-h_e(0,t)A(e(t))\quad{\rm with}\quad A(e(t)):=A^j_a(e(t))\dot{e}^a(t)\tau_j
\ee
 The holonomy is defined as the unique solution $h_e(A):=h_e(0,1)$ with $h_e(0,0)=1_{SU(2)}$. The covariant derivative in  equation (\ref{HolGauss}) is phase space dependent since it involves the Ashtekar connection $A^j_a$. However, as we will show the expression under the integral can be expressed as a total time derivative of holonomies and $\Lambda:=\Lambda^j\tau_j$.  As a first step we consider
\bea
(D_a\Lambda)\dot{e}^a(t)&=&\frac{d\Lambda}{dt} + \epsilon^j_{k\ell}A^k_a\Lambda^\ell\tau_j\dot{e}^a(t)\nonumber\\
&=&\frac{d\Lambda}{dt} + A^k_a\Lambda^\ell[\tau_k,\tau_\ell]\dot{e}^a(t)\nonumber\\
&=&\frac{d\Lambda}{dt} +[A,\Lambda]
\eea
Furthermore we can use the product rule for the term $\frac{d}{dt}\left(h_e(0,t)\Lambda h_e(t,1)\right)$ and for the individual terms using the differential equation in (\ref{DGLHol})
\bea
\frac{d}{dt}h_e(0,t)&=&h_e(0,t)A(e(t))\nonumber\\
\frac{d}{dt}h_e(t,1)&=&-A(e(t))h_e(t,1)
\eea
to obtain
\be
\frac{d}{dt}\left(h_e(0,t)\Lambda h_e(t,1)\right)
=
h_e(0,t)\left(\frac{d\Lambda}{dt} +[A,\Lambda]\right)h_e(t,1)
\ee
Consequently, we obtain for the Poisson bracket the following expression
\be
\{h_e(A),G(\Lambda)\}
=\frac{\beta\kappa}{2}\left(\Lambda(e(0))h_e(A) - h_e(A)\Lambda(e(1))\right)
\ee
Knowing how $G(\Lambda)$ acts on holonomies we can generalize the action from holonomies to cylindrical functions and express it again in terms of right and left invariant vector fields. 
\bea
\{f_\gamma(A),G(\Lambda)\}
&=&
\frac{\beta\kappa}{2}\sum\limits_{e\in E(\gamma)}
\left(\Lambda^j(b(e))R_j^e - \Lambda^j(f(e))L_j^e\right)f_\gamma(A)\nonumber\\
&=&
\frac{\beta\kappa}{2}\sum\limits_{v\in V(\gamma)}\Lambda^j(v)\Big(\sum\limits_{e\in E(\gamma)}\kappa(S,e,v)\hat{J}_j^{(v,e)}f_\gamma\Big)(A)
\eea
where we rearranged the sum over edges as a sum over vertices and sums of the edges starting or ending at each vertex yielding $\kappa(S,e,v)=+1$ and $\kappa(S,e,v)=+1$ respectively. We realize that action of $G(\Lambda)$, likewise to the action of the flux, can be expressed as a vector field on $C^\infty(A)$. The quantization of $G(\Lambda)$ can then be obtained as an extension of $G(\Lambda)$ to cylindrical functions $\in C^{\infty}(\bar{\cal A})$ where $\bar{\cal A}$ is the quantum configuration space of generalized or also called distributional connections.
\bea
\widehat{G}(\Lambda)&=&i\hbar [ f_\gamma(A),G(\Lambda)]
=\frac{i\beta\ell_p^2}{2}\sum\limits_{v\in V(\gamma)}\Lambda^j(v)\Big(\sum\limits_{e\in E(\gamma)}\kappa(S,e,v)\hat{J}_j^{(v,e)}f_\gamma\Big)(A)
\nonumber\\
&=&
\frac{i\beta\ell_p^2}{2}\sum\limits_{v\in V(\gamma)}\Lambda^j(v)\left(\sum\limits_{\substack{e\in E(\gamma) \\
v=b(e)}} R_j^e -  \sum\limits_{\substack{e\in E(\gamma) \\
v=f(e)}} L_j^e\right) f_\gamma(A)
\eea
where we used in the last line the explicit expression for $\hat{J}^{(v,e)}_j$ and $\kappa(S,e,v)$.
Solutions to the Gauss constraint $\widehat{G}(\Lambda)f_\gamma(A)=0$ are those cylindrical functions for which the ingoing and outdoing edges at each vertex couple to the same resulting angular momentum $J_{\rm in}=J_{\rm out}$ so that the total angular momentum $J=J_{\rm in}\otimes J_{\rm out}=0$. In terms of the decomposition of the kinematical Hilbert space in terms of interwiners the gauge invariant Hilbert space is just the subspace where one associates to each each vertex an intertwiner that projects on the trivial representation. That this constructs gauge invariant spin networks functions follows from the fact that the holonomy transforms under a finite gauge transformation as $h_e(A^g)=g(b(e))h_e(A)g^{-1}(f(e))$, where $g\in SU(2)$ and $b(e)$ and $f(e)$ denote the beginning and final point of the edge, and hence the gauge transformations act at the vertices of the spin network function only. Furthermore, at the vertices the spin network functions transform in the resulting representation that the individual edges couple to, thus 
\be
{\cal H}_{\rm kin}^G=\bigoplus_{\gamma,\vec{j},} {\cal H}_{\gamma,\vec{j}, \vec{l}=0}
\ee 

%% file: RAQ.tex
In this section we briefly review  refined algebraic quantization that provides a framework to solve first class constraints in the quantum theory. The application of refined algebraic quantization in the context of Dirac quantization has been analyzed by Giulini and Marolf in \cite{Giulini:1998rk,Giulini:1998kf}. A review article about refined algebraic quantization and group averaging can be found in \cite{Giulini:1999kc}.
\\
In order to formulate the final quantum theory we are looking for solutions to the constraints in the quantum theory, hence we want to find states that are annihilated by the constraint operators, that is '$\hat{C}\psi_{\rm phys}=0$'. In general the constraint operators $\hat{C}$ are complicated functions of holonomies and fluxes and we are looking for the value zero in their spectra. 
In the simplest case the operators that we need to consider are bounded and their spectra is discrete. Then the operators are defined on the whole Hilbert space $\Hkin$ and the eigenvectors of the operators are elements of $\Hkin$. As we will see later on this is the case for the Gau\ss{} constraint operator. However, often the physically interesting operators are unbounded and in this case they are only defined on a dense subspace of $\Hkin$. Furthermore, in general, the spectrum of the operators will have a continuous part and the associated (generalized) eigenvectors will no longer belong to $\Hkin$. In our case this situation occurs whenever 
zero will not  lie in the discrete part of the spectrum of the constraint operators. A similar situation occurs already in standard quantum mechanics for the momentum  and position operators both being unbounded. The spectrum of the momentum operator $\hat{p}=-i\hbar \frac{d}{dx}$ is the real line and the (generalized) eigenfunctions are plane waves $\psi_k(x)=e^{ikx}$ that are not square integrable on $(\mathbb{R},dx)$ and thus no element of ${\cal H}_{QM}=L_2(\mathbb{R},dx)$. Likewise, the generalized eigenfunctions of the position operator are delta functions and therefore also no elements of ${\cal H}_{QM}$.
In those cases one has to look for solutions in a larger space than ${\cal H}_{QM}$ and, in the case of LQG,  ${\cal H}_{\rm kin}$ respectively. The mathematical framework that can be used here are so called rigged Hilbert spaces (also called Gelfand triples). They consists of a sequence of spaces
\be
 {\cal D}\subset {\cal H}\subset {\cal D}^\times
 \ee
where ${\cal D}$ is a dense subspace of $\cal H$ endowed with its own intrinsic topology, that is assumed to be stronger than the one induced from $\cal H$ and ${\cal D}^\times$ is the dual space of $\cal D$ containing all continuous anti-linear functionals on $\cal D$. The dense subspace is usually chosen by the requirement that it is the largest dense subspace of $\cal H$ such that it is an invariant domain for arbitrary powers of the elementary operators. For instance in the case of quantum mechanics the dense subspace is the ${\cal S}(\mathbb{R})$, the Schwartz space of smooth rapidly decreasing functions on $\mathbb{R}$. Associated with a given Gelfand triple
 there always exist a second rigged Hilbert space
\be
\label{Gelfand2}
 {\cal D}\subset {\cal H}\subset {\cal D}^\prime
 \ee
 which consists of $\cal D,\cal H$ and the dual space ${\cal D}^\prime$ of all linear and continuous functionals on $\cal D$ and it is the second rigged Hilbert space that we will discuss for LQG. In the case of quantum mechanics when the spectrum has a continuous part the associated (generalized) eigenvectors expressed as Dirac ket vectors live in $\cal {D}^\times$, whereas the home of the corresponding Dirac bra vectors is the space $\cal {D}^\prime$. For LQG we restrict our discussion to the latter Gelfand' triplet in (\ref{Gelfand2}).
Let us denote with $\Dkin$ the dense subspace of $\Hkin$ and as we saw before elements of $\Dkin$ are cylindrical functions.  Apart from the dense subspace $\Dkin$ itself the rigged Hilbert space framework requires that $\Dkin$ is endowed with an intrinsic topology stronger than the one induced from $\Hkin$ and as far as physics is concerned it does not yield a particular choice of a topology here. In order to avoid such a choice at this stage in LQG we do not consider the topological dual space $\Dkin^\prime$ but the algebraic dual instead denoted by $\Dkin^*$. Then we look for solutions in the algebraic dual ${\cal D}_{\rm kin}^*$ that is the space of all linear but not necessarily continuous functionals $\ell$ on ${\cal D}_{\rm kin}$. ${\cal D}_{\rm kin}^*$ is naturally equipped with the weak *-topology of pointwise convergence of nets\footnote{A net, that is a generalization of a sequence,  $(\ell^\alpha)$ converges in ${\cal D}_{\rm kin}^*$ if $(\ell_\alpha(f))$ converges to $\ell(f)$ for any $f$ in ${\cal D}_{\rm kin}$.}. As before we have the following topological inclusion 
\be
{\cal D}_{\rm kin}\subset{\cal H}_{\rm kin}\subset{\cal D}_{\rm kin}^*
\ee
since any functional that converges strongly in the norm of ${\cal H}_{\rm kin}$ will also converge pointwise. 
In order to formulate a requirement for solutions to the constraints in ${\cal D}_{\rm kin}^*$ we need to extend the action of the operators from ${\cal H}_{\rm kin}$ onto ${\cal D}_{\rm kin}^*$.
On those linear functionals $\ell$ that lie in ${\cal H}_{\rm kin}$ we want the action of the dual operator to agree with the usual one on ${\cal H}_{\rm kin}$. Let us denote the extension of an operator $\hat{O}$ by $\hat{O}^\prime$ then we define
\be
\label{DefOprime}
\left[\hat{O}^\prime\ell\right](f):=\ell(\hat{O}^\dagger f)
\ee
where $\dagger$ denotes the adjoint in $\Hkin$. Now suppose we have an $\ell\in\Hkin\subset\Dkin^*$ then using Riesz representation theorem we find a unique $f_\ell\in\Hkin$ such that $\ell$ can be expressed as $\ell=\langle f_\ell, .\rangle_{\Hkin}$ where $\langle .,.\rangle_{\Hkin}$ denotes the inner product on $\Hkin$. Then we obtain
\be
\label{lSol}
\left[\hat{O}^\prime\ell\right](f)=\ell(\hat{O}^\dagger f)=\langle f_\ell, \hat{O}^\dagger f\rangle_{\Hkin}=\langle \hat{O} f_\ell,  f\rangle_{\Hkin},
\ee
which explains the use of the adjoint in equation (\ref{DefOprime}). Looking for solutions in $\Dkin^*$ corresponds to finding linear functionals that  satisfy the following requirement
\be
\left[\hat{C}^\prime\ell\right](f)=\ell(\hat{C}^\dagger f)\stackrel{!}{=}0\quad{\rm for\,\, all}\,\, \hat{C},f\in\Dkin
\ee
Let us denote the space of solutions to the constraints by $\Dphys^*$. In general physical operators, those that commute with all constraint operators, will be unbounded and therefore only be defined on a dense subspace of the physical Hilbert space $\Hphys$ denoted by $\Dphys$, which is an invariant domain for the algebra of physical operators. The set $\Dphys^*$ is then the algebraic dual of $\Dphys$ and as before again one has also at the physical level a topological inclusion
\be
\Dphys\subset\Hphys\subset\Dphys^*
\ee
Given the situation that we have found $\Dphys^*$ there exists a systematic way to construct an inner product  by means of a so called rigging map $\eta$ 
\be
\eta :\Dkin\to\Dphys^*,\quad f\mapsto \eta(f)
\ee
that maps elements in $\Dkin$ into $\Dphys^*$. $\eta(f)$ being an element of $\Dphys^*$ is a linear functional on $\Dkin$ that additionally satisfies the condition in equation (\ref{lSol}), hence 
 $[\eta(f)](\tilde{f})$ is a complex number and if in addition 
 \bea
 \label{Cond1-3}
 (i) &  \overline{[\eta(f)](\tilde{f})}=[\eta(\tilde{f})](f)\\
 (ii) & [\eta(f)](f)\geq0\quad{\rm for\,\, all}\, f,\quad [\eta(f)](f)=0 \Rightarrow f=0 \\
 (iii) & [\eta(f)](\alpha g + \beta h)= \alpha[\eta(f)](g) + \beta[\eta(f)](h) 
 \eea
 are satisfied we can use $[\eta(f)](\tilde{f})$ do define a physical inner product on the image of $\eta$ given by $\eta(\Dkin)\in\Dphys^*$.  
 The bar in (i) denotes complex conjugation. Condition (iii) is trivially satisfied since $[\eta(f)]$ is a linear functional. Whether $(i,ii)$ are satisfied and thus $[\eta(.)](.)$ defines a positive semi definite sesquilinear form needs to be checked  once the explicit form of $\eta$ has been constructed. Apart from the minimal requirements (i-iii) we want the inner product to satisfy the property that adjoints with respect to the physical inner product $\langle .,.\rangle_{\rm phys}$ represent the adjoints in the corresponding kinematical case. That means that we can either first extend the operators to $\Dphys^*$ and then construct the adjoint or take the adjoint first in $\Dkin$ and then extend the adjoint operator to $\Dphys^*$ and we will obtain the same result. This yields the following condition for the physical inner product
 \be
 \label{AdjoinCond}
(iv)\quad  \langle (\hat{O}^\prime)^\dagger\psi,\tilde{\psi}\rangle_{\rm phys} = \langle (\hat{O}^\dagger)^\prime\psi,\tilde{\psi}\rangle_{\rm phys}
 \ee
 Here we denoted the adjoint with respect to the physical inner product with the same symbol $\dagger$ we used in the kinematical case. Furthermore, the rigging map needs to be constructed in such a way that the physical operators $\hat{O}$ defined on  $\Dphys$ respectively preserve the space of solutions $\Dphys^*$, that means
 \be
 \label{PresCond}
(v)\quad  \hat{O}^\prime\eta(f)=\eta(\hat{O}f)\quad{\rm for\,\, all}\,\, f\in{\Dkin}
 \ee
 A physical inner product on the image of $\eta$, that is $\eta(\Dkin)\subset\Dkin^*$ can then be defined  as
 \be
 \label{physInn}
 \langle\psi,\tilde{\psi}\rangle_{\rm phys} = \langle\eta(f),\eta(\tilde{f})\rangle_{\rm phys}:=[\eta(\tilde{f})](f)
 \ee
 The physical Hilbert space $\Hphys$ is then defined as the completion of $\eta$ with respect to $\langle . , .\rangle_{\rm phys}$
 With that definition of an inner product the condition (iv) in (\ref{AdjoinCond}) is automatically satisfied
 \bea
  \langle (\hat{O}^\prime)^\dagger\psi,\tilde{\psi}\rangle_{\rm phys} &=& \langle \psi,\hat{O}^\prime\tilde{\psi}\rangle_{\rm phys}\nn  \\
  &=&\langle\eta(f),\eta(\hat{O}\tilde{f})\rangle_{\rm phys}\nn \\
  &=&[\eta(\hat{O}\tilde{f})](f)=[\hat{O}^\prime\eta(\tilde{f})](f)=[\eta(\tilde{f})](\hat{O}^\dagger f)\nn \\
  &=&\langle\eta(\hat{O}^\dagger f),\eta(\tilde{f})\rangle_{\rm phys}\nn \\
  &=& \langle (\hat{O}^\dagger)^\prime\psi,\tilde{\psi}\rangle_{\rm phys}
  \eea
  An example how a rigging map can be explicitly constructed is the so called group averaging procedure. Consider a set of constraint operators ${\hat{C}_I}$ where $I$ labels the individual constraints that are self-adjoint and their first class algebra is a Lie algebra, then we can define unitary operators by using the exponential map
  \be
  \hat{U}(g)=\exp\left(i\sum\limits_I \theta^I\hat{C}_I\right)
  \ee
  with parameters $\theta^I\in T\subset\R$ yielding a unitary representation of the Lie group. The condition that a function $f$ is annihilated by the infinitesimal generators of the constraints carries over to the requirement that 
  a linear functional is a solution of the constraints if it is invariant under the action of those unitary operators, that is for $f\in\Dkin$
  \be
  [(\hat{U}(g))^\prime](f)=\ell((\hat{U}(g))^\dagger f) = \ell(f)\quad{\rm for\,\, all}\,\, g\in G
  \ee
Using the unitary operators we can define a projector on physical states given by
\be
\label{Proj}
\hat{P}=\int\limits_G d\mu_H(g)\hat{U}(g)
\ee
where $\mu_H$ denotes Haar measure on $G$. The property of the Haar measure that it is invariant under right and left translations are important for showing that the projector defined above indeed projects on physical states as can be seen below
\bea
\hat{U}(g)\hat{P}f&=&
\hat{U}(g)\int\limits_G d\mu_H(h)\hat{U}(h)f\nn \\
&=&\int\limits_G d\mu_H(h)\hat{U}(g)\hat{U}(h)f\nn \\
&=&\int\limits_G d\mu_H(h)\hat{U}(gh)f\nn \\
&=&\int\limits_G d\mu_H(g^{-1}\tilde{g})\hat{U}(\tilde{g})f\nn \\
&=&\hat{P}f
\eea
In the third line we used that $G$ is a group, in the fourth line we introduced the new integration variable $\tilde{g}:=gh$ and in the one before the last line we used that $\mu_h$ is invariant under translations. The rigging map $\eta$ can now be expressed in terms of the projector defined in  (\ref{Proj})
\be
\eta : \Dkin \to \Hphys \subset \Dphys^*\quad f\mapsto \eta(f):=\int\limits_G d\mu(g)_H\langle \hat{U}(g)f,.\rangle_{\rm kin}
\ee
and the physical inner product can then be defined as
\be
\langle\eta(f),\eta(\tilde{f})\rangle_{\rm phys}:=[\eta(\tilde{f})](f)
\ee
Likewise to the case of quantum mechanics where the Dirac bra and ket vectors in the spaces ${\cal D}^\times$ and ${\cal D}^\prime$  are distributions on $\cal D$ also here the linear functional $\eta(f)$ defined above is a distribution on $\Dkin$. In this sense the rigged Hilbert space framework combines Hilbert spaces with distribution theory and allows us to understand the case of unbounded operators with continuous spectra along the lines of the standard language used in quantum theory. \\
Before we will apply RAQ to LQG in order to solve the diffeomorphism constraint, we will discuss a simple example from quantum mechanics, where one basically could guess the physical Hilbert space immediately so that the individual steps of the RAQ program can be understood in a simple situation. We consider a two dimensional quantum mechanical system with kinematical Hilbert space $\Hkin=L_2(\R^2,dx^2)$ with one constraint given by $\hat{C}=\hat{p}_1$. In the first step of RAQ we need to choose a dense subspace $\Dkin$ of $\Hkin$ on which the elementary operators $\hat{x}_1,\hat{x}_2, \hat{p}_1, \hat{p}_2$ and arbitrary powers of them can be defined. As usual we choose ${\cal S}(\R^2)$, the Schwartz space of rapidly decreasing functions on $\R^2$ for $\Dkin$. Now we are looking for solutions to the constraint $\hat{C}=\hat{p}_1=-i\hbar\partial/\partial x_1$. These are functions that do not depend on the variable $x_1$ and since the inner product of $\Hkin$ involves an integral over $\R$ of $dx_1$ those functions are not normalizable and thus no elements of $\Hkin$. However, they can be mathematically defined as elements of $\Dkin^*$ and hence linear functionals $\ell$ on $\Dkin$ defined as
\be
\ell : \Dkin \to \C\quad f\mapsto\ell(f):=\langle \ell,\, f\rangle_{\rm kin}=\int\limits_{\R^2} d^2x \overline{\ell(x)}f(x) 
\ee
 The second step of RAQ consists of finding solutions in $\Dkin^*$ and in our case these are functionals that satisfy
\be
[\hat{C}^\prime\ell](f):=\ell(\hat{C}^\dagger f)=\int\limits_{\R^2} d^2x \bar{\ell}(x)\hat{p}_1^\dagger f(x)=i\hbar\int\limits_{\R^2} dx_1dx_2 \frac{\partial \overline{\ell}}{\partial x_1}(x_1,x_2) f(x_1,x_2)
\stackrel{!}{=}0
\ee  
Obviously, the elements of $\Dphys^*\subset\Dkin$ are those linear functionals that do not depend on $x_1$.Therefore, physical operators defined on $\Hphys$ or a dense subspace $\Dphys$ respectively are operators that depend only on $\hat{x}_2,\hat{p_1},\hat{p_2}$ because those commute with the constraint $\hat{C}=\hat{p}_1$. Now, since $\hat{p}_1$ is a constraint operator involving $\hat{p}_1$ correspond to classical objects that vanish on the constraint surface and are thus rather uninteresting physical operators. For this reason the relevant physical operators will be functions of $\hat{x}_2,\hat{p}_2$ only. Hence, the physical Hilbert space in this simple example is just $\Hphys=L_2(\R,dx_2)$. Let us show that we end up with the same result when we use the rigging map to construct the physical Hilbert space.
The projector for  group averaging in this example has the form
\be
\hat{P}:=\int\limits_\R dt e^{it\hat{p}_1}
\ee
and using $\hat{P}$ to define the corresponding rigging map $\eta:\Dkin\to\Dphys^*$ we obtain
\be
\eta(f):=\int\limits_\R dt \langle e^{it\hat{p}_1} f, . \rangle_{kin}=\int\limits_\R dt \int\limits_{\R^2} d^2x \overline{f(x_1+t,x_2)}
\ee
where we used that the unitary operator $\exp(it\hat{p}_1)$ generates translations. The physical inner product can then be defined as
\be
\label{QMphys}
\langle \eta(f),\, \eta(\tilde{f})\rangle_{\rm phys}
:=[\eta(\tilde{f})](f)
=
\int\limits_\R dt \int\limits_{\R^2} dx_1 dx_2 \overline{\tilde{f}(x_1+t,x_2)}f(x_1,x_2)
\ee
Let us introduce the variable $t^\prime:=t+x_1$ then we rewrite the expression in (\ref{QMphys}) as
\be
\langle \eta(f),\, \eta(\tilde{f})\rangle_{\rm phys}
=\int\limits_\R dx_2 \left(\int\limits_\R dt^\prime \overline{\tilde{f}(t^\prime,x_2)}\right)\left(\int\limits_\R dx_1 f(x_1,x_2)\right)
=\langle \tilde{\psi},\, \psi \rangle_{L_2(\R, dx_2)}
\ee
showing that the physical Hilbert space we end up with is indeed $\Hphys=L_2(\R,dx_2)$.
\\
\\
Finally, let us say a few words about alternative methods to solve the constraints in cases where the refined algebraic quantization (RAQ) framework cannot be applied. This is for instance the case for the Hamiltonian constraints since their algebra involves structure functions instead of structure constants. A way to circumvent this problem is to replace the Hamiltonian constraint by the corresponding master constraint $\bf M$ defined in (\ref{Master}) \cite{TMCP}
, that (i) satisfies a trivial algebra and is constructed in such a way that one obtains a self-adjoint operator \cite{Thiemann:2005zg}. Now one could on the one hand apply RAQ to find solutions to the Master constraint. On the other hand one can also use the direct integral decomposition (DID) method in order to construct the physical Hilbert space.  In the latter case one uses the fact that for self adjoint operators $\hat{O}$ on separable Hilbert spaces the Hilbert space can be written as a direct integral 
\be
{\cal H}\simeq\int\limits_\R^\oplus d\mu(\lambda) {\cal H}_\lambda
\ee
whereby $\mu$ is the spectral measure and ${\cal H}_\lambda$ are again separable Hilbert spaces with an inner product induced from ${\cal H}$. On each of this Hilbert spaces the operator $\hat{O}$ acts by multiplication. So assuming that $\hat{O}$ is a constraint operator the physical Hilbert space is just the Hilbert space ${\cal H}_\lambda$ associated with the eigenvalue $\lambda=0$, that is $\Hphys={\cal H}_0$. For LQG the kinematical Hilbert space on which the master constraint operator $\bf\hat{M}$ will be defined is non separable. However, $\Hkin$ decomposes into a direct sum of separable Hilbert spaces each of which are left invariant by the action of  $\bf\hat{M}$ and therefore the DID method can be applied to each of the individual Hilbert spaces in the direct sum, see \cite{Han:2009ay} for an application of this method.

%% file: diffeo.tex
When looking for solutions of the diffeomorphism constraint, we would like to proceed in a similar way as for the Gauss constraint. Doing so, the first step consists in defining the infinitesimal version of the spatial diffeomorphism operator which is classically of the form
\be
\vec{C}(\vec{N})=\int\limits_\Sigma d^3x N^aF^j_{ab}E_j^b(x)
\ee
where $N^a$ is the shift vector, $F_{ab}=F^j_{ab}\tau_j/2$ the curvature associated to $A^j_a$ and $E_j^a$ the densitized triad, where we neglected the terms involving the Gauss constraint $G_j$. Considering the term involving the Gauss constraint it is possible to rewrite $\vec{C}(\vec{N})$ as an integral over $\Sigma$ of $E_j^a$ and $({\cal L}_{\vec{N}}A^j)_a$ yielding, similar to the Gauss constraint, to an expression that involves the densitized triad smeared over a 3 dimensional integral with a phase space dependent smearing function. Now, when  one tries to proceed and expresses $\vec{C}(\vec{N})$ in terms of holonomies and fluxes one realizes that the operator corresponding to $\vec{C}(\vec{N})$ does not exist on ${\cal H}_{\rm kin}$. Note that recently in \cite{Laddha:2011mk} using the habitat introduced in \cite{Lewandowski:1997ba} a quantization of the infinitesimal diffeomorphism operators was performed using a densitized shift vector of the form $N^a=E^a_j/\sqrt{q}$ yielding an expression that is less singular and can be quantized. As we will see in section \ref{suse:qbkmod} the same is also true for the classical expression $q^{ab}C_aC_b$.
\\
However, that $\hat{C}_a$ itself does not exist is not a problem as far as the Dirac quantization program is concerned because the requirement that the representation should be diffeomorphism covariant ensures that finite spatial diffeomorphisms are implemented unitarily (see also section \ref{suse:reps} for a more detailed discussion). As a consequence we can work with the operators denoted by $\hat{U}(\varphi)$ for $\varphi\in$ Diff($\Sigma$) instead when looking for solutions of the diffeomorphism constraint.
\\
In order to define the action of $\hat{U}(\varphi)$ on a given spin network function we define the multilabel $s=\{\gamma,\vec{j},\vec{I}\}$. Then we have for $T_s=\snf$
\be
\hat{U}(\varphi) T_s = T_{\varphi\cdot s}\quad{\rm f.a.}\,\, \varphi\in Diff(\Sigma)
\ee
with $\varphi\cdot s=\{\varphi \cdot\gamma,\varphi\cdot\vec{j},\varphi\cdot\vec{I}\}$ and
\begin{eqnarray}
\varphi\cdot\gamma&:=&\left(\varphi\cdot e =\varphi(e)\right)_{e\in E(\gamma(s))} \nonumber\\
\varphi\cdot\vec{j}&:=&\left((\varphi\cdot\vec{j})_{\varphi(e)}=j_e\right)_{e\in E(\gamma(s))} \nonumber\\
\varphi\cdot\vec{I}&:=&\left((\varphi\cdot\vec{I})_{\varphi(v)}=I_v\right)_{v\in V(\gamma(s))}
\end{eqnarray}
where $E(\gamma(s))$ denotes the set of edges of the underlying graph $\gamma(s)$ of $T_s$ and $V(\gamma(s))$ denotes the set of vertices of $\gamma(s)$.
We see that the action of $\hat{U}(\varphi)$ maps the graph $\gamma(s)$ on which $T_s$ is defined onto its image under the diffeomorphism $\varphi$. Note that definitions above are understood in the way that the order in which the edges at the vertices are coupled to obtain the corresponding intertwiners is not changed under the action of the diffeomorphism operator.
That the infinitesimal constraints do not exist as operators can be also seen from the fact that the family of operators $\hat{U}(\varphi)$ are not weakly continuous. A family of unitary operators $\hat{U}(\varphi_t)$ is said to be weakly continuous if we have that
\be
\lim_{t\to 0}\langle T_s\, , \hat{U}(\varphi_t) T_{s^\prime}\rangle = \langle T_s\, ,\, T_{s^\prime}\rangle\quad{\rm f.a.}\quad T_s,T_{s^\prime} \in \Hkin
\ee
Let $\varphi_t^V$ be a one-parameter family of diffeomorphisms generated by a vector field $V\not=0$. If we choose $\gamma$ in the support of $V$ then there exists $\epsilon>0$ such that for all $t\in (0,\epsilon)$ $\varphi_t^V(\gamma)\not=\gamma$. Now we choose $T_s=T_{s^\prime}$ and obtain
\be
\lim_{t\to 0}\langle T_s\, , \hat{U}(\varphi^V_t) T_{s}\rangle=
\lim_{t\to 0}\langle T_s\, ,  T_{\varphi^V_t(s)}\rangle=0\not=1=\langle T_s\, ,\, T_s\rangle
\ee
which shows that the finite diffeomorphisms are not weakly continuous.
For the reason that any cylindrical function $f_\gamma$ can be written as a linear combination of spin network functions, the action of $\hat{U}(\varphi)$ can be easily extended to $f_\gamma$. In contrast to the Gauss constraint, solutions to the spatial diffeomorphism constraint will not live in a subspace of $\Hkin$ and we need to apply the RAQ program here, which has been done in \cite{Ashtekar:1995zh}. An application to 2+1 Euclidian gravity can be found in \cite{Marolf:1997eb} and in \cite{Sahlmann:2006qs} the RAQ quantization program was applied to a scalar field toy model.
\\
 As a first step in RAQ we need to choose a dense subspace of $\Hkin$ on which arbitrary finite powers of the elementary operators are defined. In the case of LQG $\Dkin=Cyl$ the space of smooth cylindrical functions. Now we are looking for solutions in the algebraic dual $\Dkin^*=Cyl^*$ and for this purpose the action of operators and particularly of $\hat{U}(\varphi)$ on $\Hkin$ needs to be extended to functionals $\ell$ in  $\Dkin^*$ and is given by
\be
[\hat{U}^\prime(\varphi)\ell](f)=\ell(\hat{U}^{-1}(\varphi)f)\quad{\rm f.a.}\quad f\in\Dkin, \varphi\in Diff(\Sigma)
\ee
where we denoted the extension of the operator with a prime as above. This allows us to formulate the requirement for solutions $\ell$
\be
\label{SolCond}
[\hat{U}^\prime(\varphi)\ell](f)=\ell(\hat{U}^{-1}(\varphi)f)\stackrel{!}{=}\ell(f) \quad{\rm f.a.}\quad f\in\Dkin, \varphi\in Diff(\Sigma)
\ee
Those solutions are elements of $\Dphys^*\subset\Dkin^*$ defined as the set of those $\ell\in\Dkin^*$ that satisfy the condition in (\ref{SolCond}). Since the spin network functions lie dense in $\Dkin=Cyl$ we can restrict the construction of solutions to the diffeomorphism constraint to them and then express diffeomorphism invariant cylindrical functions as linear combinations of the solutions associated with spin network functions.
 \be
\label{SolCondTs}
[\hat{U}^\prime(\varphi)\ell](\snf)=\ell(\hat{U}^{-1}(\varphi)\snf)=\ell(T_{\varphi^{-1}(\gamma),\vec{j},\vec{I}})
\stackrel{!}{=}\ell(T_s) \quad{\rm f.a.}\quad f\in\Dkin, \varphi\in Diff(\Sigma)
\ee
Let us introduce the orbits of $s$ associated with $Diff(\Sigma)$ denoted by $[s]$ 
\be
[s]:=\{\varphi\cdot s,\,\, \phi\in Diff(\Sigma)\}
\ee
For all orbits $[s]$ a diffeomorphism invariant distribution can then be constructed as
\be
\label{Ls}
\ell_{[s]}=\sum\limits_{s^\prime\in [s]} \langle T_{s^\prime},\, .\rangle_{\rm kin}
\ee
We need to ensure that the sum in (\ref{Ls}) converges when applied to some spin network function $T_s$ as otherwise $\ell_{[s]}$ would not be an element of $\Dphys^*$.
Fortunately, due to the orthogonality property of spin network functions that follows from the Peter \& Weyl theorem, already mentioned in section \ref{suse:al}, in the expression
\be
\label{Ls2}
\ell_{[s]}(T_{\tilde{s}})=\sum\limits_{s^\prime\in [s]} \langle T_{s^\prime},\, T_{\tilde{s}}\rangle_{\rm kin}={\chi}_{[s]}(\tilde{s})
\ee
only one term will be non-vanishing where $\chi$ denotes the characteristic function. First, we only obtain a nonvanishing contribution if $\tilde{s}\in[s]$, meaning the graph $T_{\tilde{s}}$ is defined on must be diffeomorphic to the graph associated with the multi-label $s$. Furthermore the righthand side of (\ref{Ls2}) is only nonvanishing if additionally the representations associated with the edges agree. 
Note that in the context of group averaging (\ref{Ls2}) can be understood as group averaging over the orbits of $[s]$ under Diff($\Sigma$) using a counting measure on the orbit.
That those $\ell_{[s]}$ are invariant under the action $\hat{U}(\varphi)$ can be easily seen
\bea
\hat{U}^\prime(\varphi)\ell_{[s]}(T_{\tilde{s}})&=&
\sum\limits_{s^\prime\in [s]} \langle T_{s^\prime},\, \hat{U}^\dagger(\varphi)T_{\tilde{s}}\rangle_{\rm kin}\nonumber\\
&=&\sum\limits_{s^\prime\in [s]} \langle \hat{U}(\varphi)T_{s^\prime},\, T_{\tilde{s}}\rangle_{\rm kin}\nonumber\\
&=&\sum\limits_{s^\prime\in [s]} \langle T_{\varphi(s^\prime)},\, T_{\tilde{s}}\rangle_{\rm kin}\nonumber\\
&=&\sum\limits_{s^{\prime\prime}\in [s]} \langle T_{s^{\prime\prime}},\, T_{\tilde{s}}\rangle_{\rm kin}\nonumber\\
&=&\ell_{[s]}(T_{\tilde{s}})
\eea
As far as the operators corresponding to strong Dirac observables with respect to spatial diffeomorphism are concerned, these are operators that strongly commute with $\hat{U}(\varphi)$, one can show \cite{Ashtekar:1995zh} that $\Hkin$ splits into mutually orthogonal super selection sectors, that is, it decomposes into a direct sum of Hilbert spaces associated with the individual orbits $[s]$
\be
\Hkin=\bigoplus\limits_{[s]} {\cal H}_{\rm kin}^{[s]}\quad{\rm with}\quad {\cal H}^{[s]}_{\rm kin}=\bigoplus\limits_{s^\prime\in[s]}{\cal H}_{\rm kin}^{s^\prime}
\ee
and (strongly) diffeomorphism invariant operators preserve those individual Hilbert spaces ${\cal H}^{[s]}_{\rm kin}$. The rigging map $\eta$ is then constructed for each individual ${\cal H}^{[s]}_{\rm kin}$ separately. For this purpose let us choose some dense subspace ${\cal D}_{\rm kin}^{[s]}\subset{\cal H}^{[s]}_{\rm kin}$ and consider the algebraic dual 
${({\cal D}_{\rm kin}^{[s]})}^*$. A family of rigging maps $\eta_{a}:{\cal D}_{\rm kin}^{[s]}\to{({\cal D}_{\rm diff}^{[s]})}^*$ can then be defined as
\be
\label{etadiff1}
\eta_{a}(T_{s}):=a_{[s]}\ell_{[s]}(T_{s})
\ee
where $a_{[s]}>0\in\R$ is some up to now unspecified number labeling the family of maps. However, in general $a_{[s]}$ depends on the orbit of the multilabel $s$ and not on the orbit of the graph $\gamma(s)$ only. As a consequence we will not be able to define a symmetric inner product by means of $\eta_a$ unless we will choose the coefficients $a_{[s]}$ in such a way that they depend on the equivalence class of the graphs only.
\\
\\  
In order to discuss how the coefficients $a_{[s]}$ need to be modified and also make the connection to the group averaging more transparent we will rewrite $\eta_{a}$ as sum over diffeomorphisms in Diff($\Sigma$). Note that the naive ansatz where one averages over the total group Diff($\Sigma$) does not work for the reason that there exist uncountably infinitely many diffeomorphisms that leave a given spin network $T_s$ invariant. Thus the trivial diffeomorphisms need to be factored out in order to obtain a well defined element of $({\cal D}_{\rm diff}^{[s]})^*\subset{({\cal D}_{\rm kin}^{[s]})}^*$. We consider the diffeomorphisms that leave a given spin network function $T_s$ invariant and distinguish between two cases:
\bea
TDiff_{[s]}&:=&\{\varphi\in Diff(\Sigma) | \varphi\cdot s=s\}\nonumber\\
TDiff_{[\gamma]}&:=&\{\varphi\in Diff(\Sigma) | \varphi\cdot \gamma=\gamma,\,\, \varphi(e)=e\,\,\, f.a. e\in E(\gamma(s))\}
\eea
where $\varphi(e)=e$ should be interpreted as saying that the edge including its orientation are preserved. We used a notation analogous to the one introduced in \cite{Ashtekar:2004eh}. However, in \cite{Ashtekar:2004eh} the group averaging and the rigging map is defined in terms of equivalence classes of graphs and not of spin network functions. 
Both $TDiff_{[s]}$ and $TDiff_{[\gamma]}$ are subgroups of $Diff(\Sigma)$. $TDiff_{[s]}$ includes all diffeomorphisms that preserve the spin network $T_s$ whereas elements of $TDiff_{[\gamma]}$ additionally preserve all edges of the underlying graph $\gamma(s)$ and hence do not depend on the multilabel $s$ anymore but only on the graph. The quotient of the two groups turns out to be a finite group, the group of spin network symmetries of $T_s$  denoted by
\be
S_{[s]}:=TDiff_{[s]}/TDiff_{[\gamma]}
\ee
The group averaging can then be expressed as an averaging over the group of spin network symmetries and an averaging over the remaining diffeomorphisms that fill the orbit $[s]$  of $s$ which is $Diff(\Sigma)/TDiff_{[s]}=Diff_{[s]}(\Sigma)$. The projector associated to the averaging over $S_{[s]}$ can be expressed as
\be
\label{ProjGS}
P_{S_{[s]}}:=\frac{1}{|S_{[s]}|}\sum\limits_{\varphi\in S_{[s]}} \hat{U}(\varphi)
\ee
Note, that in general the size of  $S_{[s]}$ and $S_{[s^\prime]}$ will be be different even when the graphs associated with the multilabels $s$ and $s^\prime$ are diffeomorphic. 
As a consequence, we need to consider this in the definition of the prefactor that enters into the rigging map $\eta_a$ to ensure that the prefactor depends on the equivalence class of the graphs only.
Hence, when we consider also the averaging with respect to those diffeomorphisms that move the spin network we obtain for the rigging map
\be
\eta_{a}(T_{s})=a_{[\gamma]}{|S_{[s]}|}\sum\limits_{\varphi\in Diff_{[s]}(\Sigma)}\langle \hat{U}(\varphi)P_{S_{[s]}} T_{s}\, , \, .\rangle_{\rm kin}
\ee
Here we introduced $a_{[\gamma]}$  as the real number that one obtains from defining the group averaging in terms of graphs and not spin network functions in order to make a comparison with the notation used in \cite{Ashtekar:2004eh} possible. We have defined the rigging map above in a way that the norm of the most unsymmetric spin network function is given by $a_{[\gamma]}$ which can be chosen to be one. 
\\
A diffeomorphism invariant inner product can the be constructed on the image of $\eta_{a}$ denoted by $\eta_{a}(\Dkin^{[s]})\subset {({\cal D}_{\rm diff}^{[s]})}^*$ and the diffinvariant Hilbert space ${\cal H}_{\rm diff}^{[s]}$ is then the completion of ${\cal D}_{\rm diff}^{[s]}$ with respect to $\langle . , . \rangle_{\rm diff}$
\be
\label{innerProdDiff}
\langle \eta_{a}(T_{s^\prime}), \eta_{a}(T_s)\rangle_{\rm diff}
:= [\eta_{a}(T_s)](T_{s^\prime})
=a_{[\gamma]}{|S_{[s]}|}
\sum\limits_{\varphi\in Diff_{[s]}(\Sigma)}\langle \hat{U}(\varphi)P_{S_{[s]}} T_{s}\, , \, T_{s^\prime}\rangle_{\rm kin}
\ee
It can be shown that the so defined inner product satisfies all five requirements mentioned in section \ref{se:RAQ} in equations (\ref{Cond1-3}), (\ref{AdjoinCond}) and (\ref{PresCond}). The Hilbert space ${\cal H}_{\rm diff}$ can then be constructed as the direct sum of the individual Hilbert spaces ${\cal H}^{[\gamma]}_{\rm diff}$.
Those requirements in (\ref{Cond1-3}), (\ref{AdjoinCond}) and (\ref{PresCond}) do not yield any further restrictions on the factor $a_{[\gamma]}$ in (\ref{innerProdDiff}) except the already 
implemented condition that $a_{[\gamma]}$ is a real positive number. 
Consequently, as far as operators are considered that correspond to strong Dirac observables with respect to spatial diffeomorphisms the inner product of the solution space is not unique because we have a freedom to normalize the inner products associated with the mutually orthogonal spaces ${\cal H}^{[\gamma]}_{\rm diff}$. However, one expects that operators corresponding to weak observables, that is, operators that commute with the spatial diffeomorphism constraint only modulo constraint operators, will mix super selection sectors and could therefore yield additional requirements for the normalization constants and hence further restrict the ambiguity of the inner product on ${\cal H}_{\rm diff}$.
\\
\\
Finally, let us briefly mention that also the Gauss constraint can be solved using group averaging. Let us denote the unitary operators corresponding to finite SU(2) gauge transformations by $\hat{U}(g)$, then the associated rigging map is given by
\be
\eta(T_s)=\int\limits_{SU(2)^\Sigma} \prod\limits_{x\in\Sigma}d\mu_H((g(x))\langle \hat{U}(g) T_s\, , \, .\rangle_{kin}
\ee
Using $\eta$ above to construct the gauge invariant Hilbert space ${\cal H}_{\rm kin}^G$ yields to exactly the same result as we obtained in section \ref{se:gauss}.
Note, that solving the Gauss constraint can equivalently be done either before solving the diffeomorphism constraints or afterwards. The resulting Hilbert space ${\cal H}_{\rm kin}^G$ is, similar to the kinematical Hilbert space $\Hkin$ still non-separable. In \cite{Fairbairn:2004qe} a proposal was introduced to obtain a separable Hilbert space for $\Hdiff$ by allowing fields to have isolated point of non-differentiability. The associated gauge symmetry is an extension of Diff($\Sigma$) and includes homeomorphisms that, together with their inverse are smooth everywhere except at at most finitely many points. An application of the RAQ program with respect to this extension of Diff($\Sigma$) yields a separable $\Hdiff$. 

%% file: reps.tex
We have already discussed one representation of the canonical commutation relations \eqref{eq:ccr1} (or rather their integrated version \eqref{eq:ccr2}). This was the diffeomorphisms covariant representation \eqref{eq:al_rep} on the Ashtekar-Lewandowski Hilbert space $\Hkin$.  In this section we would like to discuss other representations. Why would it be interesting to do so? There are some aspects of the standard representation that are peculiar, and one might wonder whether there are other representations that do not have these properties. Examples are: In the standard representation,
\begin{itemize}
	\item there are no operators representing the connection $A$ directly, 
  \item the generator of the unitary maps implementing diffeomorphisms does not exist,
  \item the spectrum of geometric operators is purely discrete.
\end{itemize}
Natural requirements for a ``fundamental'' representation include:
\begin{enumerate}
	\item The representation is irreducible, i.e., each state in the representation Hilbert space is cyclic.
	\item The representation is diffeomorphism covariant, i.e., there exist unitary operators implementing the diffeomorphisms.
	\item There is at least one state that is \emph{invariant} under diffeomorphisms.	
\end{enumerate}
A representation with all these characteristics is, however, equivalent to the the Ashtekar-Lewandowski representation \eqref{eq:al_rep}. This is a consequence of the uniqueness theorem
\begin{prop}[\cite{Fleischhack:2004jc,Lewandowski:2005jk}] 
There is only one cyclic representation of $\mathfrak{A}$ with diffeomorphism invariant cyclic vector -- the Ashtekar-Lewandowski representation \eqref{eq:al_rep}. 
\end{prop}
We note that to really prove the above proposition, the algebra $\mathfrak{A}$ and the class of diffeomorphisms has to be defined in great detail. We also note that there are interesting representations that violate one or more premises of the above proposition.\footnote{To be precise, the examples given in the literature are for the somewhat simpler case of the structure group SU(2) replaced by U(1), but it is likely that they generalize to the SU(2) case.} For example, \cite{Varadarajan:2007dk} exposes a diffeomorphism invariant representation that is reducible. In \cite{Dziendzikowski:2009rv}, a cyclic and diffeomorphism invariant representatiopn of an algebra slightly larger than the one from the above proposition are described. 

In the following, we will however describe a simpler class of representations due to Koslowski, \cite{Koslowski:2007kh,Sahlmann:2010hn,Koslowski:2011vn}. These representations have been called \emph{representations with classical spatial background geometry}. The basic observation is that the Ashtekar-Lewandowski representation is a member of a family of very similar representations. The AL ground state is peaked on a degenerate triad $\bg{E}=0$. But it is easy to construct similar representations which, however are peaked on different classical triad fields $\bg{E}$. A rough sketch of the situation is as follows:
\begin{align}
\text{AL ground state } \Omega \qquad &\widehat{=} \qquad \delta_{0}(E) \text{ in momentum rep.}\\
\text{Ground states w.\ background } \Omega_{\bg{E}} \qquad &\widehat{=} \qquad \delta_{\bg{E}}(E) \text{ in momentum rep.}
\end{align}
How are these representations defined precisely? The Hilbert space, and the representation of the holonomies stays the same, 
\begin{equation}
\mathcal{H}_{\bg{E}}=\Hkin, \qquad \pi_{\bg{E}}(h)=h, 
\label{eq:bgrep1}
\end{equation}
but the representation of the fluxes is changed by adding a c-number term, 
\begin{equation}
\pi_{\bg{E}}(E_{n}(S))=X_n(S)+\bg{E}_n(S)\,\id, \qquad \bg{E}_n(S)=\int_S *\bg{E}_i n^i\, \text{d}^2x.
\label{eq:bgrep2}
\end{equation}
The new representations have the following properties. 
\begin{itemize}
	\item The representations $\pi_{\bg{E}}$ are cyclic.
	\item One can regularize the geometric operators in the representations $\pi_{\bg{E}}$  with exactly the same methods as in the standard representation (described in sec. \ref{suse:geo}). The resulting operators are 
		\begin{equation}
		\widehat{Ar}(S) = \widehat{Ar}_\text{AL}(S) + Ar(\bg{E},S)\, \id, \qquad 
		\widehat{V}(R) = \widehat{V}_\text{AL}(R) + V(\bg{E},R)\, \id,
		\label{eq:geo_background}
		\end{equation}
		where $S$ is a surface, $R$ a region, the subscript AL denotes operator in the standard representation\footnote{$\widehat{V}_\text{AL}$ can be either, the Ashtekar-Lewandowski, or the Rovelli-Smolin version, depending on which regularization procedure is chosen.}, and the c-number terms are given by the classical value in the respective background. This result is very simple to state, but the proof is non-trivial \cite{Sahlmann:2010hn}.
	\item Only the symmetries of $\bg{E}$ can be implemented unitarily in the new representations.  
\end{itemize} 
We also note that for $\bg{E}=0$, we recover the standard representation from \eqref{eq:bgrep1},\eqref{eq:bgrep2}.
Equation \eqref{eq:geo_background} shows that the new representations can be interpreted as containing a ``geometric condensate''.

If we want to use the more general representations in the place of the standard representation when implementing constraints, the failure of the diffeomorphisms to be implemented unitarily is of concern. For example, for the operators $U(\phi)$ implementing the diffeomorphisms in the standard representation one finds that generically
\begin{equation}
U(\phi)\pi_{\bg{E}}(E_n(S))U(\phi)^\dagger\neq \pi_{\bg{E}}(E_{\phi_*n}(\phi(S)))
\end{equation} 
for diffeomorphisms $\phi$. 
But one can easily show that one can also not find other unitaries that will do the job. The reason is that the geometrical background $\bg{E}$ is fixed can can not transform under any operation on $\mathcal{H}_{\bg{E}}\equiv\Hkin$. There is a way to remedy this problem, at the price of going over to a much larger Hilbert space and a reducible representation. Let us use the notation 
\begin{equation}
\ket{T}_{\bg{E}} \equiv \ket{T,\bg{E}} 
\end{equation}
for a spin network function $T$ interpreted as an element of $\mathcal{H}_{\bg{E}}$. Then define 
\begin{equation}
\mathcal{H}_{[\bg{E}]}=\bigoplus_{\overline{\bg{E}}\in [\bg{E}]} \mathcal{H}_{\overline{\bg{E}}}.
\label{eq:bigh}
\end{equation}
Here the direct sum is over all background triads in one gauge and diffeomorphism equivalence class, 
\begin{equation}
\overline{\bg{E}}\in [\bg{E}]\quad \Longleftrightarrow\quad \exists g, \phi:\quad  \overline{\bg{E}}=\text{Ad}_g(\phi_*\bg{E}),
\end{equation}
where $g$ denotes a gauge transformation and $\phi$ a diffeomorphism. We note that the Hilbert space \eqref{eq:bigh} is thus labeled by a spatial metric modulo diffeomormphisms, that is, a spatial geometry, or a point in superspace. $\mathfrak{A}$ is represented on $\mathcal{H}_{[\bg{E}]}$ through the direct sum of the representations $\pi_{\overline{\bg{E}}}$ for $\overline{\bg{E}}\in [\bg{E}]$, and gauge transformations, diffeomorphisms, as well as their semidirect product can be implemented unitarily. The representation on $\mathcal{H}_{[\bg{E}]}$ is not cyclic, however.  

Starting from $\mathcal{H}_{[\bg{E}]}$, it is possible to obtain states that solve the diffeomorphism and Gauss constraint by group averaging, just as in the case of the standard representation discussed in detail in sections \ref{se:RAQ}. Let us consider the diffeomorphisms as an example. We make the definitions
\begin{align}
\text{TDiff}_{([s],\bg{E})}&:= \{ \text{Diffeos }\phi: \phi_*\bg{E}=\bg{E}  \text{ and } \phi\cdot s=s\}\\
\text{TDiff}_{([\gamma],\bg{E})}&:= \{ \text{Diffeos }\phi: \phi_*\bg{E}=\bg{E}  \text{ and } \phi(e)=e \text{ for all edges $e$ of $\gamma$}\}
\end{align}
Note that these definitions exactly parallel those used in the detailed treatment of the group averaging procedure in section \ref{se:Diffeo}.  Like in the standard case both of these sets actually form groups. Moreover their quotient , the group of spin network symmetries given by
\begin{equation}
\text{S}_{([s],\bg{E})}= \text{TDiff}_{([s],\bg{E})}/\text{TDiff}_{(\gamma,\bg{E})}
\end{equation}
can be shown to be a finite group. Here, it is understood that $T_s$ is cylindrical on the graph $\gamma$.  
Then group averaging for a spin net $T_\gamma$ in the $\bg{E}$-sector of $\mathcal{H}_{[\bg{E}]}$ effectively reduces to averaging over diffeomorphisms modulo $\text{TDiff}_{([s],\bg{E})}$ denoted by $\text{Diff}_{[s]}(\Sigma)=\text{Diff}(\Sigma)/\text{TDiff}_{([s],\bg{E})}$ and over $\text{S}_{([s],\bg{E})}$. More precisely, let
\begin{equation}
\label{eq_diff}
(T_s,\bg{E}|T_{s'}, \bg{E'}\rangle
:= a_{[\gamma]}|S[s]|
\sum_{G(\phi)\in\text{Diff}_{[s]}(\Sigma)} 
\sscpr{T_s,\bg{E}}
{P_{S_{([s],\bg{E})}} U^\dagger_\phi}
{T_{s'}, \bg{E'}},
\end{equation}
where the projection $P_{S_{([s],\bg{E})}}$ is defined as
\begin{equation}
\label{eq_pro}
P_{S_{([s],\bg{E})}}
\ket{T_s,\bg{E}}:=\frac{1}{\betr{\text{S}_{[s],\bg{E})}}} \sum_{G(\phi)\in
\text{S}_{[s],\bg{E})}} U_\phi\ket{T_s,\bg{E}}.
\end{equation}
Now it is easy to show that 
\begin{prop}
The linear functionals $(T_s,\bg{E}|$ are well defined, finite, and diffeomorphism invariant, 
\begin{equation}
(T_s,\bg{E}|\circ U_\phi=(T_s,\bg{E}|\qquad \text{ for all } \phi\in\text{Diff}.  
\end{equation}
\end{prop}
Again, these definitions reduce, for the case $\bg{E}=0$, to the diffeomorphism averaging map defined in section \ref{se:Diffeo}. 
Similar results can be obtained for gauge transformations and, taking the semidirect product of diffeomorphisms and gauge transformations, for bundle automorphisms. This means that the quantum kinematics can be developed to the same point for the new representations as for the standard representation. Among other things, this shows nicely that diffeomorphism invariance is not antithetical to being peaked on a fixed geometry. 

As an example consider the operator $\widehat{Vol}$ for the volume of the entire spatial slice. Also in the new representations, it commutes with all automorphisms. It thus defines an operator on $\mathcal{H}_\text{aut}$. Moreover, this operator acts in precisely the way one would expect. If $f$ is an eigenstate of the volume operator in the standard representation, with eigenvalue $\lambda$, then  $|f,\bg{E})$ is an eigenvector of $\widehat{Vol}$ with eigenvalue $\lambda+\bg{Vol}$.

We also note that the invariant states resulting from the vacuum representation are again contained in these constructions as the special case $\bg{E}=0$. Finally, while $H_{[\bg{E}]}$ is large, this is partially remedied by the group averaging. For example, vectors 
$\ket{1,\bg{E}}$, $\ket{1,\bg{E'}}$ $\in H_{[\bg{E}]}$ are mapped onto \emph{the same} vector in $\mathcal{H}_\text{aut}$. More generally   
\begin{equation}
(f,\phi_*\bg{E}|=(\phi^{-1}_* f,\bg{E}|.
\end{equation}

%% file: ham.tex
As we have seen in section \ref{suse:canon}, the Hamilton constraint of the classical theory is given by
\begin{equation}
C=\frac{\beta}{2}
\underbrace{
\frac{1}{\sqrt{q}}E^a_iE^b_j\epsilon^{ij}{}_kF^k_{ab}
}_{=:C_\text{E}}
-\beta(1+\beta^2)
\underbrace{
\frac{1}{\sqrt{q}}E^a_iE^b_jK_{[a}^iK_{b]}^j
}_{=:T}.
\label{eq:hamil}
\end{equation}
In the present section we will discuss how to turn this classical expression into a well defined operator. The general difficulty with this is obviously that $C$ is a complicated nonlinear function in the phase space variables, hence ordering problems present themselves. There are also some specific difficulties with the expression: 
\begin{itemize}
	\item \eqref{eq:hamil} contains the inverse volume element. The volume element itself has a large kernel when quantized, see the discussion in section \ref{suse:geo}, so its inverse is ill defined. 

\item The expression \eqref{eq:hamil} contains the curvature $F$ of $A$, as well as the extrinsic curvature $K$. For neither of them there is a simple operator in the quantum theory.  

\end{itemize}
A guiding principle in the quantization process can be the Dirac algebra \eqref{eq:dirac1}--\eqref{eq:dirac3}. In particular, the quantum Hamiltonian constraint should be invariant under gauge transformations, covariant under diffeomorphisms, and the commutator of two Hamilton constraints should give a diffeomorphism constraint. 

We should say that the knowledge about the quantization and implementation of the Hamilton constraint is not complete. Many things remain to be understood. But we will show that at least there is a strategy that leads to well defined constraint operators. Given the difficulties outlined above this is highly nontrivial in itself.

The quantization strategy we will describe in the following is due to Thiemann \cite{Thiemann:1996aw,Thiemann:1996av,Thiemann:1997rv}, but draws on important earlier work and ideas by Rovelli, Smolin, Lewandowski and others. Our presentation is in part based on \cite{Ashtekar:2004eh}.
\subsubsection*{Thiemann's tricks}
The quantization is based on two key ideas. The first one is to use various ingenious classical identities to express parts of the Hamilton constraint in terms of Poisson brackets before quantization. The second one is to express curvature in terms of holonomies. Let us explain them in turn. 

Let 
\begin{equation}
V=\int_\Sigma\text{d}^3x\, \sqrt{\det q}, \qquad 
\overline{K}=\int_\Sigma\text{d}^3x\,K^i_aE^a_i
\end{equation} 
be the total volume of the spatial slice, and the integrated extrinsic curvature. 
Then
\begin{equation}
\frac{E^a_iE^b_j\epsilon^{ijk}}{\sqrt{\det q}}=\frac{4}{\kappa}\epsilon^{abc} \{V, A^k_c\},\qquad
K^j_a=\frac{2}{\kappa}\{\overline{K},A^j_a\}.
\label{eq:ttrick}
\end{equation}
These identities can be used to write
\begin{align}
C_\text{E}(N)&=c\int_\Sigma\text{d}^3x\, N \epsilon^{abc}\tr \left(F_{ab}\{A_c,V\}\right),\\
T(N)&=c'\int_\Sigma\text{d}^3x\, N \epsilon^{abc}\tr \left(\{A_a,\overline{K}\}\{A_b,\overline{K}\}\{A_c,\overline{K}\}\right),
\label{eq:tpb}
\end{align}
where we have used the notation for the two parts of the Hamilton constraint introduced in \eqref{eq:hamil}. The constants are $c=8/\kappa$ and $c'=16/\kappa^3$. The idea behind these reformulations is that it is natural to replace Poisson brackets by commutators in the quantization process, 
\begin{equation}
\{\;\cdot\; , \; \cdot\;\} \longrightarrow \frac{1}{i\hbar}[\;\cdot\; , \; \cdot\;]. 
\end{equation} 
This means that the quantization would be greatly simplified if operators existed for the quantities $V,\overline{K}$. Indeed we have already seen in section \ref{suse:geo} that an operator exists for $V$. With respect to $\overline{K}$, the identity
\begin{equation}
\overline{K}=\{V,C_\text{E}\}
\label{eq:extr_curv}
\end{equation}
suggests to first quantize $C_\text{E}$, and then use the commutator with the volume operator to define the operator for $\overline{K}$. Thus we have already dealt with two of the difficulties regarding the quantization of $C$: The inverse volume element is gone, and the extrinsic curvature is dealt with. What remains is the quantization of the curvature $F$ of $A$. Here we use the well known fact that holonomies encode information about curvature. Let $S$ be an oriented surface such that the integral $\int_S F$ is small, and let $\alpha$ be the (oriented) boundary of $S$.  
Then the first term on the right hand side of 
\begin{equation}
\int_S F=\frac{1}{2}\left(h_\alpha-h_\alpha^{-1}\right) +O\left(\left(\int_S F\right)^2\right)
\end{equation}
is a good approximation to the left hand side. Let $e$ be an edge starting at a point $s(e)$. A similar approximation plus a second Taylor expansion gives 
\begin{equation}
\epsilon \dot{e}^a(s(e)) \{A(s(e))_a,V\}\approx h^{-1}_e\{h_e,V\}
\end{equation}
where $\dot{e}$ is the tangent to $e$ in a chosen parametrization $e(t)$ , and $\epsilon$ is the coordinate length $\epsilon=\int_e dt$ of the edge in the given parametrization.   
In this way, we can express curvatures and connections by holonomies. Putting everything together we can get a Riemann sum approximation of the Euclidean part of the constraint, 
\begin{equation}
{C}_\text{E}(N)\approx C_\text{E}^{(\Box)}
:=\frac{c}{2}\sum_\Box N(v_\Box) \sum_{I=1}^3\left[
\left(h_{\alpha_I(\Box)}^{-1}-h_{\alpha_I(\Box)}\right)
h^{-1}_{s_I(\Box)}
\left\{h_{s_I(\Box)},V\right\}
\right].
\label{eq:e_reg}
\end{equation}
Here $\{\Box\}$ is a decomposition of $\Sigma$ into 3-dimensional cells, and for each cell a point $v_\Box$ has been fixed.  $\{\alpha_I(\Box)\}$ is a set of loops and $\{s_I(\Box)\}$ a set of edges such that their tangents span the tangent space in the point $v_\Box$ in the following sense: there is a basis $\{b_I(\Box)\}$ of the tangent space at $v(\Box)$, such that $b_I(\Box)$ is tangent to both $\alpha_I(\Box)$ and $s_I(\Box)$, and compatible with their orientations.  We call the data ($\{\Box\}$, $\{v_\Box\}$, $\{s_I(\Box)\}$, $\{\alpha_I(\Box)\}$) a \emph{regulator} of $C_\text{E}$, and sometimes denote it simply by $\Box$. The exact shape of these cells, loops and edges does not matter. The approximation is good as long as the cells are much smaller than the scale on which the fields $A,E$ vary, and the loops and edges stay within the cell.

Finally, we can consider families of regulators such that the cells shrink to points. Then the corresponding approximations will converge to the exact result for a wide variety of such families. 

The same kind of arguments can also be made for the second part of the Hamiltonian constraint $T(N)$. The connection components $A_a$  in \eqref{eq:tpb} can be replaced by holonomies along edges with suitable tangents, and the integrated exterior curvature $\overline{K}$ by Poisson brackets of $V$ with the \emph{regulated} Euclidean part \eqref{eq:e_reg}, as per \eqref{eq:extr_curv}. The resulting expression is quite complicated and contains ambiguities, but the correct refinement limit is obtained for a large class of regulators. 

\subsubsection*{Quantization}
We will now come to the quantization. The general idea is clear: Pick a family of regulators which converge to the continuum result. Replace Poisson brackets by commutators, and holonomies and volume operators by their operator counterparts, and obtain operators 
\begin{equation}
\widehat{C}_\text{E}^{(\Box)}(N)
=\frac{c}{2}\sum_\Box N(v_\Box) \sum_{I=1}^3\left(
\left(h_{\alpha_I(\Box)}^{-1}-h_{\alpha_I(\Box)}\right)
h^{-1}_{s_I(\Box)}
\left[h_{s_I(\Box)},\widehat{V}\right]
\right).
\end{equation}
on the kinematic Hilbert space. Now take the refinement limit $\Box\rightarrow \Sigma$ to obtain an operator $\widehat{C}_\text{E}$. There are, however, several difficulties when putting this program into practice:
\begin{enumerate}
\item In the limit of infinite refinement, the operator is in danger of creating infinitely many loops and edges. Hence the limit may be ill defined.
\item Even if problem 1.\ can be overcome, the operator will generically not converge, since typically $\widehat{C}_\text{E}^{(\Box)}\Psi\perp \widehat{C}_\text{E}^{(\Box')}\Psi$ for regulators $\Box\neq\Box'$. 
\item Since $h$ and $\widehat{V}$ do not commute, there are ordering ambiguities. 
\item There is a lot of ambiguity in the choice of regulators since now there is no guarantee that different families of regulators will converge to the same operator, if they converge at all. 
\end{enumerate}
The first problem can be solved by a suitable ordering. Let us consider the action on a spin network. The volume operator acts only at the vertices, hence ordering it to the right will force the loops and edges that are created by $\widehat{C}_\text{E}$ to be attached to the vertices of the spin network only. Thus, for a given spin network, only finitely many new edges and loops can be created. This also partially solves problem 3. To deal with the rest of the difficulties, we will be less ambitious, and not demand convergence in the kinematic Hilbert space. Rather, we consider the matrix elements of  $\widehat{C}_\text{E}^{(\Box)}$ between one kinematic state and one diffeomorphism invariant one. It turns out that due to the diffeomorphism invariance of the one state, many of the ambiguities in the attachment of the loops and edges do not change the matrix elements. What is more, for several types of regulators it is known that the matrix elements converge, 
\begin{equation}
\lim_{\Box\rightarrow \Sigma} (\Psi|\widehat{C}_\text{E}^{(\Box)}|f_\gamma\rangle \quad \text{ is well defined.}
\end{equation}    
Typically, the matrix elements already become constant at a finite refinement, namely when the decomposition of $\Sigma$ into cells is already so fine that there is at most one vertex of $\gamma$ per cell. 

Now we have to be careful. Convergence of the above matrix elements does not imply that there exists a limit operator on the kinematic Hilbert space. Rather, we can interpret $(\Psi|\widehat{C}_\text{E}^{(\Box)}|$ as an element in the (algebraic) dual space of $\cyl$, and hence conclude that there is an operator 
\begin{equation}
\widehat{C}_\text{E}^\dagger: \Hdiff \longrightarrow \cyl^*.  
\label{eq:map}
\end{equation} 
The detailed features of this operator depend on the chosen family of regulators. But the generic features do not:
\begin{itemize}
	\item $\widehat{C}_\text{E}^\dagger$ acts locally at the vertices. 
	\item It acts by creating and annihilating edges and loops. 
\end{itemize}
One can proceed in the same way with the quantization of $T(N)$, but since the quantized expression contains double commutators with $\widehat{C}_\text{E}^{(\Box)}$, the operator action becomes extremely complicated. Nevertheless it is well defined and finite. 
\subsubsection*{Solutions}
Given the definition of the Hamilton constraints we sketched above, what are the solutions? They are states $\Psi$ in $\Hdiff$ such that 
\begin{equation}
(\Psi| C(N) f\rangle =0 \qquad \text{for all } f\in \cyl \text{ and all } N.
\end{equation}
One simple solution is the LQG vacuum $\ket{}$, which can also be interpreted as a state in $\Hdiff$. But more complicated solutions exist. For working out the set of solutions in some detail, details of the regularization used in the quantization of the constraints have to be fixed, since they do matter. Suffice it to say that so called \emph{exceptional edges} play an important role in the construction of solutions. Exceptional edges are edges of the type created by the quantum constraint itself. We will not discuss this in detail, but refer to \cite{Thiemann:1996aw,Thiemann:1996av,Thiemann:1997rv,Ashtekar:2004eh} for more detailed accounts. 

Solutions lie in the intersection of the kernels of all Hamilton constraints. Formally, the projector on this space can be expressed and approximated as follows \cite{Reisenberger:1996pu}:
\begin{equation}
\begin{split}
P_C &=\delta(\widehat{C})
=\int\text{D}N e^{i\widehat{C}(N)}\\
&=1+i\int\text{D}N\int N(x)\widehat{C}(x)+\frac{i^2}{2} \int\text{D}N\iint N(x_1)N(x_2)\widehat{C}(x_1)\widehat{C}(x_2) +\ldots.
\end{split}
\end{equation}
$\widehat{C}(x)$ denotes the local action of the constraint, which is zero unless $x$ is the position of a vertex of the state acted upon. 
The path integral over $N$ gives an infinite result, but by requiring diffromorphism invariance, it can be split into a divergent term that can be normalized away, and a finite remainder \cite{Rovelli:1996qc}.

The matrix elements of the projector can then be expanded into a series
\begin{equation}
(T_{\gamma_1} | P_C T_{\gamma_2})=\sum_{N=0}^\infty\; \sum_{v_1}\ldots \sum_{v_n} c_{v_1\ldots v_N}(T_{\gamma_1}| \widehat{C}(v_1) \widehat{C}(v_2) \ldots \widehat{C}(v_N) |T_{\gamma_2})
\label{eq:sf}
\end{equation}
where the finite sums are over all vertices of $\gamma_2$ and $c_{v_1\ldots v_N}$ is the finite remainder of the integral  over the lapse function. It only depends on the diffeomorphism equivalence class of the vertex set $\{v_1,v_2,\ldots v_N \}$. 
We note that a priori the multiple applications of the local constraint in \eqref{eq:sf} do not make sense, since we have up to now only defined the constraint operators in such a way that domain and range are disjoint, see \eqref{eq:map}. But it is possible to enlarge the domain of definition in such a way that multiple applications of the constraints become possible \cite{Rovelli:1996qc,Lewandowski:1997ba}. We will sketch how this is done when we discuss the question of anomalies below.

These matrix elements are interesting, because in principle they contain all the information about the inner product on the Hilbert space of physical states,
\begin{equation}
(T_{\gamma_1}| P_C T_{\gamma_2}) =\scpr{P_C\, T_{\gamma_1}}{P_C\, T_{\gamma_2}}_\text{phys}.
\end{equation}
The expansion \eqref{eq:sf} can be interpreted as a kind of Feynman expansion, organized in terms of how many times the constraint acts. The individual terms can be nonzero only if the action of the constraint operators on $T_{\gamma_2}$ produces exactly $T_{\gamma_1}$. Thus the non-zero diagrams can be thought of as terms coming from the evolution of one spin network state into another. More precisely, they can be labeled by a two-complex, whose faces carry representations and whose edges carry intertwiners. The complex has the graphs $\gamma_1, \gamma_2$ as boundaries, and the internal vertices correspond to the cation of the constraints. These diagrams are called \emph{spin foams}, and they show up independently in approaches that discretize the covariant path integral for gravity, compare the contribution by Rovelli. That they show up in an expression for the physical inner product of the canonical theory is a very encouraging link between canonical and covariant picture.
In fact, in the light of the recent developments that are treated in Rovelli's contribution to this volume, we are getting close to actually having a precise correspondence
\begin{equation}
\text{quantum Hamilton constraint} \qquad \longleftrightarrow \qquad \text{spinfoam model}.
\end{equation}
We will now  discuss some further aspects of the Hamilton constraint quantization.
\subsubsection*{Symmetry, anomaly freeness, ambiguities} 
In principle, it would be desirable to produce a symmetric, or even selfadjoint Hamiltonian constraint, 
\begin{equation}
C^\dagger(N)=C(N). 
\end{equation}
But this turns out to be hard in practice, and there are even some no-go theorems \cite{Lewandowski:1997ba}. Interestingly, there are heuristic arguments to the effect that one can not have both, symmetric constraints and a constraint algebra that is anomaly free.

We have seen that the constraints classically close to form an algebra with respect to the Poisson bracket. The same should happen on the quantum level, now with respect to the commutators. Otherwise the gauge symmetries may have been broken when quantizing the theory. Such an anomaly in the gauge symmetries would strongly suggest the quantum theory to be unphysical. In particular, we are interested in the commutators 
\begin{equation}
[C(M),C(N)]
\end{equation}	
since by the above construction, we can already see that the Hamilton constraints transform correctly under gauge transformations and diffeomorphisms. Classically the above commutator is proportional to a diffeomorphism constraint, hence at minimum one requires that the commutator should vanish states of $\Hdiff$. 
The problem is that the constraints map $\Hdiff$ to a certain subspace of $\cyl^*$ which is strictly larger than $\Hdiff$. So the above commutator is not well defined, as it stands. There are two proposed solutions to this problem. The first, by Thiemann \cite{Thiemann:1997rv}, is to look at the commutator on $\Hkin$, before removing the regulator. He finds
\begin{equation}
[C^{(\Box)}(M), C^{(\Box)}(N)]= \text{ something }\neq 0, \qquad (\Psi| \text{ something }=0 \text{ for }\ |\Psi) \in \Hdiff. 
\end{equation}
In this sense, 
\begin{equation}
[C(M),C(N)]\rvert_{\Hdiff} =0, 
\end{equation}	
and the quantization is anomaly free. The other solution to defining the commutator is by Le\-wan\-dows\-ki and Marolf \cite{Lewandowski:1997ba}. They introduce a certain class of elements of $\cyl^*$ that is slightly larger than $\Hdiff$. Without going into technical details, a \emph{vertex-smooth state} $|\Psi)$ is a state
\begin{equation}
|\Psi)\in\cyl^*\;:\; (\Psi | U_\phi f_\gamma) \text{ is a function of } V(\phi(\gamma)),  
\end{equation}
i.e., of the set of vertices of the graph $\phi(\gamma)$, for \emph{any} diffeomorphism $\phi$. 
Trivial examples of vertex smooth states are given by diffeomorphism invariant states. A less trivial example is the linear functional given by 
\begin{equation}
\Psi' \mapsto (\Psi|\int_\Sigma N \widehat{\sqrt{\det q}}| |\Psi'\rangle
\end{equation}
for a lapse function $N$ and $|\Psi)$ in $\Hdiff$. 

Lewandowski and Marolf observe that $(\Psi| C(N)$ is vertex-smooth for a large class of regulators, and that its action can be extended to vertex-smooth states. Moreover, they find 
\begin{equation}
(\Psi_\text{vs}|[C(M),C(N)]=0, 
\end{equation}
where $\Psi_\text{vs}$ is vertex-smooth. As far as diffeomorphism invariant states are concerned, this result would be expected for an anomaly free representation. But since it holds for all vertex-smooth states, it is surprising and a little worrisome, since the term in the Dirac algebra that results from the Poisson bracket of two Hamiltonian constraints, a diffeomorphism constraint, would be expected to act non-trivial on most vertex smooth states. But this has to be checked explicitly, and it may be possible to find quantizations of this term that indeed vanish on vertex-smooth states. New light on this question may be shed by new results of Laddha and Varadarajan \cite{Laddha:2010hp,Laddha:2010wp,Laddha:2011mk}, who employ new techniques to define constraints and their commutator algebra. 

We should not finish without pointing out that there are various ambiguities in the above procedure that are poorly understood, for example regarding the loop attachment and the representation of the newly created links (see however \cite{Perez:2005fn}). Overall, it is however very encouraging that we can find a family of well defined constraint operators that are anomaly free in a certain sense, and that lead to a convergence of the canonical and the spin-foam picture. Given the complexity of the Hamiltonian constraints of general relativity, these results are highly non-trivial. 
\\
Some of the techniques discussed in this section will be used in the following two sections where the quantization of the two classical models introduced in section \ref{suse:bkmod} and \ref{suse:scalar}.

%% file: qbkmod.tex
In this section we will discuss the quantization of the Brown-Kuchar model introduced in section \ref{suse:bkmod}. We want to quantize the reduced phase space whose elementary variables are given by the observables ${\bf A}^J_j$ and ${\bf E}^j_J$ shown in equation (\ref{eq:FullObs}). Hence we need to look at the algebra of these elementary observables in order to know what kind of representations are possible for the corresponding quantum theory. A property of those models where deparametrization occurs is that the algebra of the elementary observables is isomorphic to the kinematical one, that is
\be
\{ {\bf A}^J_j(\sigma),{\bf E}^k_K(\sigma^\prime)\}=\frac{\kappa}{2}\delta^{J}_K\delta_j^k\delta^3(\sigma,\sigma^\prime)
\ee
In general the algebra of observables can be more complicated and is given by the expression \cite{Thiemann:2004wk}
\be
\{O_{A,T},O_{E,T}\}\simeq O_{\{A,E\}^*,T}
\ee
which involves the Dirac bracket denoted by $\{.,.\}^*$. Here we denoted the general observables associated with $A,E$ with respect to a reference field $T$ by $O_{A,T}$ and $O_{E,T}$ respectively. The reason why the Dirac bracket occurs on the righthand side is that the originally first class constraint of a given system together with the gauge-fixing constraints for the clock fields $C_I:=T_I-\tau_I$ , where $I$ labels the individual reference fields of a given model, form a system of second class constraints. Applied to general relativity, we would start with the first class system given by the Hamiltonian $C$ and spatial diffeomorphism constraint $C_a$ . Then for the Brown-Kuchar-model we choose four reference fields $T,S^j$ with $j=1,2,3$ and obtain four gauge-fixing constraints $C_0=T-\tau=0$, $C_j=S^j-\sigma^j=0$. In the framework of gauge unfixing introduced in \cite{Vytheeswaran:1994ra} the construction of the observables in section \ref{suse:bkmod} corresponds to transforming the second class constraints $C$ and $C_a$ again into first class constraints and using the gauge-fixing constraints to construct a projector that maps $A$ and $E$ onto their corresponding observables with respect to the now first class constraints $C$ and $C_a$. Thus, the way how observables are constructed in the relational framework is a particular case of the gauge unfixing procedure.
\\
It seems that for the Brown-Kuchar-model the quantization of the reduced phase space seems to be a trivial task and considering only the algebra it looks like even a Fock quantization would be possible. However, this is not the case because likewise to the Dirac quantization where one requires that the kinematical representation needs to allow to implement the constraints as well defined operators, here we are only interested in those representations in which the physical Hamiltonian ${\bf H}_{\rm phys}$ in (\ref{PhysHDust}) can be implemented as a well defined operator. Since ${\bf H}_{\rm phys}$ consists of terms that involve the gravitational contribution of the Hamiltonian and diffeomorphism constraint, Fock quantization is excluded. However, a possible representation would be the one of the (gauge invariant) kinematical Hilbert space in LQG on which the constraint operators can be defined. Note that this representation becomes physically in this model since we are quantizing the reduced phase space here. Hence, as a first possible representation for $\Hphys$ let us choose $\Hphys=L_2(\bar{\cal A},d\mu_{AL})$, restricted to its gauge invariant subspace. Now, our task is to quantize the generator of the dynamics, that is ${\bf  H}_{\rm phys}$. On the classical reduced phase space the expression ${\bf h}^2(\sigma)={\bf C}^2-{\bf q}^{jk}{\bf C}_j{\bf C}_k$ is constrained to be positive. Implementing this in the quantum theory would correspond to defining self-adjoint operators for ${\bf h}^2(\sigma)$ and restrict for each $\sigma$  the spectral resolution of the Hilbert space to the positive part of the spectrum. Since this is technically impossible at the moment because of the complexity of the operators in the full theory, we use the absolute value under the square root and instead and quantize
\be
{\bf H}_{\rm phys}=\int\limits_{\cal S} d^3\sigma \sqrt{\left|{\bf C}^2 -{\bf q}^{jk}{\bf C}_j{\bf C}_k\right|}(\sigma)
\ee
 As a first step, likewise to the construction in section \ref{suse:ham},  we need to regularize the classical expression. For this purpose we choose a partition of the spatial dust manifold ${\cal S}$ into 3 dimensional cells $\Box$ such that ${\cal S}=\bigcup \Box$. Hence, ${\bf H}_{\rm phys}$ can be written as
 \be
 {\bf H}_{\rm phys}=\sum\limits_{\Box}\int\limits_{\Box} d^3\sigma \sqrt{\left|{\bf C}^2 -{\bf q}^{jk}{\bf C}_j{\bf C}_k\right|}(\sigma)
\ee
Let us denote the refinement limit in which the partition becomes the continuum  by $\Box\to{\cal S}$, the volume of the cells by $V_0(\Box)$ and a point inside $\Box$ by $\sigma(\Box)$, then we can rewrite ${\bf H}_{\rm phys}$ as a limit of a Riemann sum
\be
{\bf H}_{\rm phys}=\lim_{\Box\to\Sigma}\sum\limits_{\Box} V_0(\Box) \sqrt{\left|{\bf C}^2 -{\bf q}^{jk}{\bf C}_j{\bf C}_k\right|}(\sigma(\Box))
\ee
As a second step we will reformulate the expression under the square root above so that we are able to use quantization techniques that have been successfully applied to the case of the Hamiltonian constraint and have been discussed in section \ref{suse:ham}. For simplicity we will restrict our discussion to the euclidean part of the Hamiltonian constraint $C_E$. As has been explained in section \ref{suse:ham} once this part has been quantized the remaining part can be quantized using the operator for $C_E$.
We introduce the (rescaled) magnetic field $B^j_J$ and its contraction with a co-triad given by
\be
{\bf B}^j_J:=2\epsilon^{jk\ell}{\bf F}^{J}_{k\ell},\quad {\bf B}:={\bf B}^j_J\tau_J{\bf e}_j
\ee
where $\tau_J=-i\sigma_J/2$ are, as before, a choice of a basis for su(2), $\sigma_j$ are the Pauli matrices and ${\bf F}^J_{jk}$ is the curvature associated to ${\bf A}^J_j$. Note that the index position of the capital indices is not important here since these are the su(2) Lie algebra indices are pulled with $\delta^{JK}$. Using that $Tr(\tau_J\tau_K)=-\frac{1}{2}\delta_{JK}$ we obtain
\be
Tr({\bf B})=-\epsilon^{jk\ell}{\bf F}^J_{k\ell}e^J_j
\ee
The co-triad can be expressed in terms of triads by the following formula
\be
{\bf e}^J_j=\frac{1}{2}\frac{1}{\det({\bf e}^j_J)}\epsilon^{JMN}\epsilon_{jmn}{\bf e}^m_M{\bf e}^n_N
\ee
together with the identities ${\bf E}^j_J=\sqrt{\det({\bf q})}{\bf e}^j_J$ and $\det({\bf e}^J_j)={\rm sgn}(\det({\bf e}))\sqrt{\det({\bf q})}$ we obtain
\be
Tr({\bf B})=-{\rm sgn}(\det({\bf e}))\frac{\epsilon^{JKL}{\bf F}^J_{k\ell}{\bf E}^k_K{\bf E}^\ell_L}{\sqrt{\det({\bf q})}}
\ee
Consequently we have $[Tr({\bf B})]^2={\bf C}^2_E$. For the second term under the square root ${\bf q}^{jk}{\bf C}_j{\bf C}_k$ using ${\bf q}^{jk}=E^j_JE^k_K\delta^{JK}/\det({\bf q})$  and ${\bf C}_j={\bf F}^K_{jk}{\bf E}^k_K$ we obtain
\be
{\bf q}^{jk}{\bf C}_j{\bf C}_k=\frac{{\bf F}^L_{\ell j}{\bf E}^\ell_L{\bf E}^j_J {\bf F}^M_{mk}{\bf E}^m_M{ \bf E}^k_K\delta^{JK}}{\det({\bf q})}
\ee
On the other hand  when we use that $Tr(\tau_I\tau_J\tau_K)=\frac{1}{4}\epsilon_{IJK}$ and consider the term $4\rm{Tr}({\bf B}\tau_K)$ we obtain
\be
4\rm{Tr}({\bf B}\tau_K)=-{\rm sgn}(\det({\bf e}))\frac{{\bf F}^I_{k\ell}{\bf E}^k_I{\bf E}^\ell_K}{\sqrt{\det({\bf q})}}
\ee
Thus we have 
\be
{\bf q}^{jk}{\bf C}_j{\bf C}_k=16\rm{Tr}({\bf B}\tau_J)Tr({\bf B}\tau_K)\delta^{JK}=:\delta^{JK}C_JC_K
\ee
Let us introduce the quantities
\be
{\bf C}(\Box):=\int\limits_{\Box} d^3\sigma {\bf C}(\sigma)\quad {\bf C}_J(\Box):=\int\limits_{\Box} d^3\sigma {\bf C}_J(\sigma)
\ee
then in the refinement limit we can rewrite ${\bf H}_{\rm phys}$ as
\be
{\bf H}_{\rm phys}=\lim\limits_{\Box\to{\cal S}}\sum\limits_{\Box} \sqrt{\left| {\bf C}^2(\Box)- \delta^{JK}{\bf C}_J(\Box){\bf C}_K(\Box)\right|}
\ee
and this finishes the regularization of the classical expression. 
Using the notation $\tilde{\tau}_\mu:=(-\mathbb{I}_2,4\tau_J)$ with $\mu=0,1,2,3$ and the classical identity in (\ref{eq:ttrick}) we can rewrite the regularized expressions as
\be 
{\bf C}_{\mu}(\Box)=\int\limits_{\Box} d^3\sigma Tr({\bf B}\tilde{\tau}_\mu)=\frac{4}{\kappa}\int\limits_{\Box}Tr({\bf F}\wedge \{{\bf V}({\Box}),A\}\tilde{\tau}_\mu)
\ee
where  ${\bf }V({\Box})$ is the volume of $\Box$ given by
\be
V({\Box})=\int\limits_\Box d^3\sigma \sqrt{\det({\bf q})}
\ee
The corresponding quantum operator is then defined as
\be
\widehat{\bf H}_{\rm phys}=\lim\limits_{\Box\to{\cal S}}\sum\limits_{\Box} \sqrt{\left| \widehat{\bf C}^\dagger_0(\Box)\widehat{\bf C}_0(\Box)- \delta^{JK}\widehat{\bf C}^\dagger_J(\Box)\widehat{\bf C}_K(\Box)\right|}
\ee
and one needs to show that the limit yields a well defined operator on $\Hphys$. For the operator $\widehat{\bf C}_0(\Box)$ this has been shown in \cite{TQSD} and is briefly discussed in section \ref{suse:ham}. The operator $\widehat{\bf C}_I(\Box)$ can be quantized using similar techniques since in its definition also enters the $Tr({\bf B})$ term with an additional $\tilde{\tau}_J$ matrix inside the trace. At this point the symmetries of the classical physical Hamiltonian become important. As mentioned in section \ref{suse:bkmod} the physical Hamiltonian is invariant under (active) diffeomorphisms on the dust manifold ${\cal S}$ and we would like to preserve this symmetry also in the quantum theory having the consequence that $\widehat{\bf H}_{\rm phys}$ needs to be implemented as a spatially diffeomorphism invariant operator. The representation we choose for $\Hphys$ is the gauge invariant sector of the usual kinematical representation of LQG, namely $L_2(\bar{\cal A},d\mu_{AL})$. For this representation it was shown in \cite{Ashtekar:1995zh} that spatially diffeomorphism invariant  operators, need to be quantized in a graph preserving way, meaning that those operators do not modify the underlying graph that a spin network function is defined on. However, in its usual quantization discussed in section \ref{suse:ham} the operator $\widehat{\bf C}_0$ is quantized in a graph changing way. Thus, choosing the usual kinematical representation of LQG for $\Hphys$ together with the requirement that the classical symmetries of ${\bf H}_{\rm phys}$ carry over to the quantum theory forces us to quantize the operators $\widehat{\bf C}_\mu$ in a graph preserving way. Since the Hilbert space in the chosen representation decomposes into an orthogonal sum of the Hilbert spaces associated with each individual graph, this means that each of these graph Hilbert spaces needs to be preserved separately. In order to implement  this graph preserving property we introduce the notion of a minimal loop:
Given a graph $\gamma$, consider a vertex $v\in V(\gamma)$ and a pair of edges $e,\tilde{e}\in E(\gamma)$ of edges starting at the vertex $v$. A loop $\alpha_{\gamma,e,\tilde{e}}$ in $\gamma$ starting at $v$ going a long the edge $e$ and ending at $v$ along the edge $\tilde{e}^{-1}$ is said to be minimal provided that there exist no other loop in $\gamma$ with these properties and fewer edges transversed. Using the notion of a minimal loop we can define an operator for each graph at a given vertex $v$ 
\be
\label{Cmugv}
\widehat{\bf C}_{\mu,\gamma,v}=\frac{1}{\ell_p^2 |T_v(\gamma)|}\sum\limits_{(e_1,e_2,e_3)\in T_v(\gamma)}\epsilon^{IJK}\frac{1}{|L_{\gamma,v,e_I,e_J}|} 
\sum\limits_{\alpha\in L_{\gamma,v,e_I,e_J}}Tr\big(\tilde{\tau}_\mu \hat{h}_\alpha \hat{h}_{e_K}[\hat{h}^{-1}_{e_K}, \hat{V}_{\gamma,v}]\big)
\ee
here $T_v(\gamma)$ denotes the set of ordered triples of edges at the vertex $v$, whose tangent vectors at $v$ are linearly independent. and $L_{\gamma,v,e_I,e_J}$ is the set of minimal loops\footnote{Note that the introduction of the minimal loop here causes no further complication when the semiclassical limit of this operator is considered. This can for instance be seen in \cite{Giesel:2006uk} where the semiclassical limit in the context of AQG is discussed and also the concept of the minimal loop is used in the quantization.} Furthermore, $\hat{V}_{\gamma,v}$ is the Ashtekar-Lewandowski volume operator shown in equation (\ref{eq:VolOp}). The operator for the physical Hamiltonian for each graph $\gamma$ is then defined as
\be
\label{Hphysg}
\widehat{\bf H}_{{\rm phys},\gamma}
=\sum\limits_{v\in V(\gamma)}\sqrt{\big|P_\gamma\left(\widehat{\bf C}^\dagger_{\gamma,v}\widehat{\bf C}_{\gamma,v}-\delta^{JK}\widehat{C}^\dagger_{J,\gamma,v}\widehat{C}_{K,\gamma,v}\right)P_\gamma\big| }
\ee
whereby $P_\gamma: \Hphys\to{\cal H}^\prime_{\rm phys,\gamma}$ is an orthogonal projection operator defined analogously to the orthogonal decomposition shown in (\ref{OrthDecomp}), that needs to be introduced in order to ensure that ${\bf H}_{{\rm phys},\gamma}$ is graph preserving. Although the loop is attached along already existing edges of the graph it can still be the case that when a holonomy operator is acting the resulting product of representation includes the trivial one. The final operator is then defined as
\be
\widehat{\bf H}_{{\rm phys}}=\bigoplus\limits_\gamma \widehat{\bf H}_{{\rm phys},\gamma}
\ee
The fact that  $\widehat{\bf H}_{{\rm phys}}$ has to be quantized in a graph preserving way comes from the spatial diffeomorphism invariant corresponding classical expression of the physical Hamiltonian. However, one might take the point of view that including those projection operators into the operator $\widehat{\bf H}_{{\rm phys}}$  to enforce the graph preserving property of the operator looks slightly artificial. A way to avoid this issue and thus also the projection operators in (\ref{Hphysg}) is to change the representation for $\Hphys$. One possible other representation introduced in the framework of Algebraic Quantum Gravity (AQG) \cite{Giesel:2006uj} is von Neumann's infinite tensor product representation (ITP). In the context of AQG one does not work with the embedded graphs used in LQG but considers one (fundamental) abstract combinatorial graph on which the quantum dynamics is defined. The embedding of the graph into a given spatial manifold happens only in the semiclassical sector of the theory and how the abstract graph is embedded is encoded in semiclassical states. To each edge of the abstract graph one associates an $L_2(SU(2),d\mu_H)$ Hilbert space and one considers a graph with countable infinitely many edges. One of the motivations to introduce the AQG model was that semiclassical computations of dynamical operators technically simplify in this setup. This is due to the fact that for graph preserving operators the current existing semiclassical states can be used and those operators can be defined more naturally in the AQG framework. In \cite{Giesel:2006uk} a combinatorial graph of cubic topology was chosen and considering this graph we can define an (algebraic) operator for the physical Hamiltonian. The algebraic version of the operator in (\ref{Cmugv}) is given by
\be
\label{Cmuv}
\widehat{\bf C}_{\mu,v}=\sum\limits_{s_1,s_2,s_3=\pm 1}s_1s_2s_3\epsilon^{I_1I_2 I_3} 
Tr\big(\tilde{\tau}_\mu \hat{h}_{\alpha_{I_1s_1,I_2s_2}} \hat{h}_{e_{v,I_3s_3}}[\hat{h}^{-1}_{e_{v,I_3s_3}}, \hat{V}_{v}]\big)
\ee
here $e_{v,Is}$ denotes the edge starting at $v$ and going in positive (s=+1) or negative (s=-1) I-direction and ${\alpha_{Is,J\tilde{s}}}$ is the unique minimal loop in the algebraic graph of cubic topology, that starts at $v$ goes along the edge $e_{Is}$ and comes along the edge $e^{-1}_{J\tilde{s}}$ before ending in $v$ again. In analogy one can define an algebraic volume operator, that is for a cubic graph of the form
\be
\widehat{V}_{v}:=\ell_p^3\sqrt{\left| \frac{1}{48}\sum\limits_{s_1,s_2,s_3=\pm 1}\epsilon^{IJK}\epsilon_{LMN}\hat{J}^L_{e_v,Is_1}\hat{J}^M_{e_v,Js_2}\hat{J}^N_{e_v,Ks_3}\right|}
\ee
The final algebraic operator is then given by
\be
\widehat{\bf H}_{\rm phys}
=\sum\limits_{v \in V(\gamma)}\sqrt{\left| \widehat{\bf C}_v^\dagger\widehat{C}_v - \delta^{JK}\widehat{\bf C}^\dagger_{J,v}\widehat{\bf C}_{K,v}\right|}
\ee
where the sum runs over the countable infinitely many vertices of the algebraic graph. In contrast to the LQG framework in AQG trivial representations associated to the edges are allowed. The picture of the dynamics is then that dynamical operators do not change the underlying infinite abstract algebraic graph but only representations associated to the edges. However, since trivial representations are allowed subgraphs of the fundamental algebraic graph can and will be modified so that the quantum dynamics in the algebraic framework looks similar to the graph modifying one in LQG. In both formulations the usual LQG and the AQG model geometrical operators are defined on the physical Hilbert space and are thus observables. Hence, these models are two examples where the discrete spectra of those operators is carried over to the physical sector.
\\
A further possibility to formulate a model for LQG, in which operators can be defined in a graph changing way is the scalar field model discussed in section \ref{suse:scalar} whose quantization will be discussed in the next section.

%% file: qscalar.tex
In this section we discuss the quantization of the scalar field model whose classical theory was introduced in section \ref{suse:scalar} and we will closely follow the presentation from \cite{Domagala:2010bm}. As already mentioned at the end of section \ref{suse:scalar} the physical Hilbert space $\Hphys$ of this model will be constructed from the gauge invariant subspace of the diffeomorphism invariant Hilbert space denoted by ${\cal H}_{\rm diff}^G$ for the reason that the diffeomorphism as well as the Gauss constraint are solved by means of Dirac quantization. As discussed in section \ref{se:RAQ} and \ref{se:Diffeo} the diffeomorphism invariant Hilbert space can be constructed by using a rigging map $\eta_{\rm diff}:\Dkin\to {\cal D}_{\rm diff}^*\subset \Dkin^*$. In addition in order to construct the operator corresponding to the physical Hamiltonian $H_{\rm phys}$ of this model, we need another Hilbert space denoted by ${\cal H}_{\rm diff,x}$  associated to a subgroup denoted by $Diff(\Sigma,x)$ of $Diff(\Sigma)$ including those diffeomorphisms, which preserve a given  point $x\in\Sigma$. The construction of the rigging map for $Diff(\Sigma,x)$ works analogously to that of $Diff(\Sigma)$ when using the techniques introduced in section \ref{se:RAQ}. Next the gauge invariant subspace of these two Hilbert spaces can be easily obtained because for each gauge invariant cylindrical function $\tilde{f}\in$ Cyl,  the linear functional $\eta_{\rm diff}(\tilde{f})$ is not affected by gauge transformations that act on $f\in$ Cyl in the sense that the expression $[\eta_{\rm diff}(\tilde{f})](f)$ is invariant under gauge transformation. Consequently, we obtain ${\cal H}_{\rm diff}^G$  and ${\cal H}_{\rm diff,x}^G$ respectively by restricting $\Dkin=$Cyl to the subspace of gauge invariant cylindrical functions.
\\
\\
When we decide to do not reduce with respect to the Hamiltonian constraint at the classical level, we need to construct solutions to the Hamiltonian constraint in the quantum theory, this yields an equation of the form
\be
\left(\hat{\pi}(x)-\hat{h}(x)\right)\Psi=0
\ee
Taking into account that $\hat{\pi}$ is quantized as $-i\delta/\delta\phi(x)$ (setting $\hbar=1$) we obtain as a (formal) general solution
\be
\label{SolC}
\Psi(\phi,A)=e^{\int_\Sigma d^3x\hat{\phi}(x)\hat{h}(x)}\psi(A)
\ee
where $\psi(A)$ is an SU(2) gauge and spatially diffeomorphism invariant function.
\\
\\
Physical operators, these are operators that correspond to classical Dirac observables, will be defined on $\Hphys$ or a dense subspace $\Dphys$ of it. We explained in section \ref{se:RAQ} that the action of operators on $\Hkin$ can be extended to $\Dkin^*$. Therefore, in our case, we have for symmetric diffeomorphism and gauge invariant operators  $\hat{L}$ defined originally on $\Hkin$ a natural action on ${\cal H}_{\rm diff}$, where we denote the extended operator by $\hat{L}^\prime$, given by
\be
[\hat{L}^\prime\eta_{\rm diff}(f)](\tilde{f}):=[\eta_{\rm diff}(f)](\hat{L}^\dagger\tilde{f})=[\eta_{\rm diff}(f)](\hat{L}\tilde{f})
=\langle \eta_{\rm diff}(\hat{L}f)\, ,\, \eta_{\rm diff}(\tilde{f})\rangle_{\rm diff}
\ee
So far, the operators $\hat{L}$ are only observables with respect to the diffeomorphism and Gauss constraint but not with respect to the Hamiltonian constraint. In this model those observables are not constructed at the classical level, like in the Brown-Kuchar model of section \ref{suse:bkmod} but are constructed as Dirac observables directly in the quantum theory on $\Hphys$. A quantum Dirac observable $\hat{O}$ is defined as an operator on $\Hphys$ (or a dense subspace $\Dphys$) with the following properties
\begin{itemize}
\item $\hat{O}$ is SU(2)-gauge and spatially diffeomorphism invariant.
\item The operator $\hat{O}$ commutes with the Hamiltonian constraints, that is $[\hat{C}^{\rm tot}(x),\hat{O}]=0$ for all $x\in\Sigma$.
\end{itemize}
Inspired by the relational framework \cite{RRF,DObs} for the classical theory, one can (formally) define a family of Dirac observables. Let $\hat{L}$ be an SU(2) gauge and diffeomorphism invariant linear operators, then the operator $\hat{O}(L)$ defined as
\be
{O}_{\tau}(\hat{L}):=e^{i\int\limits_\Sigma d^3x \left( \hat{\phi}(x)-\tau(x)\right)\hat{h}(x)}\,\hat{L}\,e^{-i\int\limits_\Sigma d^3x\left(\hat{\phi}(x)-\tau(x)\right)\hat{h}(x)}
\ee
is a quantum Dirac observable. Here $\tau$ is the value that the reference field $\phi$ takes while being transformed along its gauge orbit. We need to choose $\tau(x):=\tau$ with $\tau$ being a constant real number in order to ensure that the resulting operator is spatially diffeomorphism invariant. 
Note that the expression above can also be obtained from the requirement that the (formal) solutions shown in (\ref{SolC}) should be mapped into solutions by Dirac observables.
The classical interpretation of these Dirac observables is precisely the analogue of those formal power series used to construct the observables ${\bf A}^J_j$ and ${\bf E}^j_J$ in the Brown-Kuchar model shown in equation (\ref{eq:FullObs}). Explicitly, we have in the quantum theory
\be
\label{OpEvol}
O_\tau(\hat{L})=\sum\limits_{n=0}^\infty\frac{i^n}{n!}\left[\hat{L},\int\limits_{\Sigma}(\hat{\phi}(x)-\tau)\hat{h}(x))\right]_{(n)}
\ee
which is the quantization of the classical expression
\be
O_\tau(L)=\sum\limits_{n=0}^\infty\frac{1}{n!}\{L,\int\limits_{\Sigma}(\phi(x)-\tau)h(x))\}_{(n)}
\ee
\\
The quantum dynamics of the observables is given by the following equation
\be
-i\frac{d}{d\tau}{O}_\tau(\hat{L})=\left[{O}(\hat{L}),\hat{H}_{\rm phys}\right]
\ee
and those equations are the analogue of the Heisenberg picture in quantum mechanics.
Our final task is to implement the operator $\hat{H}_{\rm phys}$ that is a quantization of the classical expression in (\ref{HphysScal}). Promoting the individual terms under the square root in $H_{\rm phys}$ to operators we obtain the heuristic expression
\be
\hat{H}_{\rm phys}=\int\limits_\Sigma d^3x \sqrt{-\sqrt{\hat{q}}\hat{C} +\sqrt{\hat{q}}\sqrt{\hat{C}^2-\widehat{q^{ab}C_aC_b}}}
\ee
where $\hat{q}:=\widehat{\det(q)}$. This operator will be defined on a suitable domain of ${\cal H}_{\rm diff}^G$ and thus act only on states that are spatially diffeomorphism invariant as long as we restrict to the operator evolution defined by (\ref{OpEvol}). The operators corresponding to the classical expression $q^{ab}C_aC_b$ should annihilate diffeomorphism invariant states. Therefore, assuming a suitable operator ordering for $H_{\rm phys}$ we assume that we can work with the simplified operator
\be
\hat{H}_{\rm phys}=\int\limits_\Sigma d^3x \sqrt{-2\sqrt{\hat{q}}\hat{C}}:=\int\limits_\Sigma d^3x \hat{h}(x)
\ee
and this is also the physical Hamiltonian suggested by Rovelli and Smolin in \cite{RSscalar}. Although we can use some of the already existing quantization techniques in the literature for $\hat{C}$ and $\sqrt{\hat{q}}$ respectively, what we need to define is an operator for $\sqrt{-2\sqrt{q}C}$ and the already existing operators are the Hamiltonian constraint 
$\hat{C}$ smeared against arbitrary lapse functions (graph-modifying) in \cite{TQSD}, the master constraint mentioned in section \ref{suse:Cldynamics} \cite{TMCP} (graph-modifying and graph-preserving) and the physical Hamiltonian of the Brown-Kuchar model (graph preserving) in \cite{AQGIV}. The physical Hamiltonian $\hat{H}_{\rm phys}$ for the scalar field model will be defined on a suitable domain of ${\cal H}_{\rm diff}^G$ in a graph changing way. Let us briefly sketch the quantization procedure.  The regularization of the classical expression for $C$ will be that of \cite{Ashtekar:2004eh}, where the operator valued distribution for $\hat{C}$ is defined as
\be
\int\limits_\Sigma d^3x N(x)\hat{C}(x)=\sum\limits_{x\in\Sigma} N(x)\hat{C}^\prime_{x}
\ee
One of the differences to the regularization chosen by Thiemann in \cite{TQSD} is the way how the loop is attached to the graph on which the spin network functions are defined on. In Thiemann's proposal the loop runs along the edges $e_I$ and $e_J$ that belong to the graph $\gamma$ (see section \ref{suse:ham}) whereas in the regularization in \cite{Ashtekar:2004eh} the loop lies in the plane spanned by the edges $e_I$ and $e_J$ but is only connected to the graph at the vertex $v$. 
Note that neither the operator $\widehat{\sqrt{q}}\hat{C}(x)$ nor the operator $\hat{h}(x)$ are defined on ${\cal H}_{\rm diff}^G$ due to their dependence on $x$ that breaks diffeomorphism invariance. Therefore, we need the Hilbert spaces ${\cal H}_{\rm diff,x}^G$ in order to implement those operators in the quantum theory.
Each individual operator $\hat{C}^\prime_x$ maps its domain from ${\cal H}_{\rm diff}$ to ${\cal H}_{\rm diff,x}$ and as discussed in \cite{Domagala:2010bm} defines naturally an operator on  ${\cal H}_{\rm diff,x}$. The operators $\hat{C}^\prime_x$ from \cite{Ashtekar:2004eh} are non-symmetric, which is no problem as long as we are working with constraint operators. However,  here we are quantizing a physical Hamiltonian, that involves a square root and we will need to perform a spectral decomposition in order project on the positive part of the spectrum of the operator under the square root. Therefore, we consider the operators 
\be
\hat{C}_x:=\frac{1}{2}\left(\hat{C}^{\prime}_x+\hat{C}^{\prime\dagger}_x\right)
\ee
The operator valued distribution, that we consider for implementing the Hamiltonian density $h(x)$ of ${H}_{\rm phys}$ is then of the form
\be
\label{Cdistr}
\hat{C}(x):=\sum\limits_{x^\prime\in\Sigma}\delta(x,x^\prime)\hat{C}_{x^\prime}
\ee
Likewise one can define an operator valued distribution for the classical expression $\sqrt{q}$ given by
\be
\label{qdistr}
\widehat{\sqrt{q}}(x)=\sum\limits_{x^\prime\in\Sigma}\delta(x,x^\prime)\widehat{\sqrt{q}}_{x^\prime}
\ee
Note that both operator values distributions above are well defined for the reason that when they are smeared against an arbitrary smearing function $F$ and applied to some cylindrical function  only a finite number of non zero terms occur in the sum
\be
\int\limits_\Sigma d^3x F(x)\widehat{\sqrt{q}}(x)f_{\gamma}=\sum\limits_{k=1}^N F(v_k)\widehat{\sqrt{q}}_{v_k}f_{\gamma}
\ee
where $v_1,...,v_N$ are the vertices of the graph $\gamma$ and similar for the operator $\hat{C}(x)$. The expressions in (\ref{Cdistr}) and (\ref{qdistr}) can then be used to define an operator for $H_{\rm phys}=\int_\Sigma h(x)$ on (a subspace of) ${\cal H}_{\rm diff,x}^G$
\be
\hat{h}(x):=\sum\limits_{x^\prime\in\Sigma}\delta(x,x^\prime)\sqrt{-2\widehat{\sqrt{q}}_{x^\prime}^{\frac{1}{2}}\hat{C}_{x^\prime}\widehat{\sqrt{q}}_{x^\prime}^{\frac{1}{2}}}
\ee
Due to the square root in the equation above $\hat{h}(x)$ is only well defined on the subspace of ${\cal H}_{\rm diff,x}^G$ where the spectrum of $\widehat{\sqrt{q}}_{x^\prime}^{\frac{1}{2}}\hat{C}_{x^\prime}\widehat{\sqrt{q}}_{x^\prime}^{\frac{1}{2}}$ is positive. In order to be able to consider only the positive part of the spectrum we need to  choose a selfadjoint extension for $\hat{h}(x)$ and in general this choice might not be unique. Let us denote the subspace of ${\cal H}_{\rm diff,x}$ corresponding to the positive part of the spectrum by ${\cal H}_{\rm diff,x,+}$. As discussed in \cite{Domagala:2010bm} there exists a natural map $\eta_\Sigma:{\cal H}_{\rm diff,x}^G\to{\cal H}_{\rm diff}^G$ with $\eta_{\rm diff(\Sigma,x)}(f)\mapsto\eta_{\rm diff(\Sigma)}(f)$. The domain of the physical Hamiltonian operator is then the image of $\eta_\Sigma$ on ${\cal H}_{\rm diff,x,+}$ and we obtain as the physical Hilbert space ${\cal H}_{\rm phys}=\eta_\Sigma({\cal H}_{\rm diff,x,+})$ with inner product
\be
\langle e^{\int_\Sigma d^3x\hat{\phi}\hat{h}}\psi,e^{\int_\Sigma d^3x\hat{\phi}\hat{h}}\psi^\prime\rangle_{\rm phys}:=\langle \psi,\psi^\prime\rangle_{\rm diff}
\ee
with $\psi,\psi^\prime\in{\cal H}_{\rm diff}^G.$ The final form of the operator for $\hat{H}_{\rm phys}$ is then given by
\be
\hat{H}_{\rm phys}=\int\limits_{\Sigma}\hat{h}(x)=\sum\limits_{x\in\Sigma}\sqrt{-2\widehat{\sqrt{q}}_{x}^{\frac{1}{2}}\hat{C}_{x}\widehat{\sqrt{q}}_{x}^{\frac{1}{2}}}
\ee
Finally, let us comment on the classical symmetries of the Hamiltonian in this model. Likewise to the Brown-Kuchar model $H_{\rm phys}$ is invariant under spatial diffeomorphisms. However, for the reason that here a different representation than the kinematical representation of LQG was chosen, the requirement that $H_{\rm phys}$ needs to be quantized in a graph preserving way is absent.  Furthermore, also the classical Hamiltonian densities $\{h(x),h(y)\}=0$ commute a general property of those deparametrized models. In the framework of the habitat, being the home of so called vertex smooth states, briefly mentioned at the end of section \ref{suse:ham}, it was shown in \cite{Lewandowski:1997ba} that the commutator of two Hamiltonian constraints smeared against arbitrary lapse functions vanishes on the habitat. Thus, one would expect that the Hamiltonian densities in the scalar field model will also vanish on the habitat $[\hat{h}(x),\hat{h}(y)]=0$.
\\
Although the graph modifying property of the physical Hamiltonian in the scalar field model has some advantages as far as the rather artificial infinitely many conservation laws for graph preserving operators in the Brown-Kuchar model are concerned, when the semiclassical sector of the model is considered the analysis becomes more complicated. The reason for this is that with the current existing semiclassical techniques for the full theory, graph modifying operators cannot be analyzed and those techniques would need a strong improvement in order to be able to analyze the semiclassical sector of the scalar field model in full generality. 
\\
Summarizing, both quantum models, the Brown-Kuchar as well as the scalar field model could be a step in the right direction and help in the future to extract some physical information out of full LQG but of course this is a long range project and on the way there are still many technical issues to solve.

%% file: concl.tex
In the present lectures we have given an introduction to loop quantum gravity, in particular its canonical quantization techniques. We also saw how it makes contact with the path integral formulation developed in spin foam gravity. The latter topic, and its connection to what we presented here is covered in detail in the contribution by Rovelli. But also in what we did cover, we have left out many details, and did not even touch on any applications, such as to the quantum theory of black hole horizons (covered in the contribution by Barbero, Lewandowski and Villase\~nor), or to cosmology (covered in the contribution by Singh).  A good starting point to get an overview over all of these developments is the contribution by Ashtekar in these proceedings. 

However, we hope that we have explained at least some of the other big achievements of loop quantum gravity, namely its description of quantum geometry and the corresponding dynamics. The quantum theory of (extrinsic and intrinsic) geometry, as described in section \ref{se:qukin} comprises in particular geometric operators with a discrete spectrum, the scale of which is set by Planck lenght, and diffeomorphism invariant states. We have furthermore seen that, based on this, well defined Hamiltonian constraints, and in the case of the matter models we considered, well defined Hamiltonian operators can be obtained. This is a highly non-trivial result, given the complicated nature of the classical dynamics. Moreover, there is a clear connection to the spin foam approach to loop quantum gravity.  

Although many of the structures that we have described have already been investigated for some time, there are still lots of new developments. Some of those have already been mentioned in the main text, but there are many more that we could not cover in these lectures, among them new connections between the full theory and symmetry reduced models \cite{Brunnemann:2010qk,Fleischhack:2010zt}, coherent states for quantum geometry \cite{Thiemann:2000bw,Thiemann:2002vj,Bahr:2007xn,Flori:2008nw,Flori:2009rw,Freidel:2010tt}, an interpretation of quantum geometry in terms of polyhedra in flat space  \cite{gr-qc/0501075,Freidel:2009ck,Freidel:2010aq,Freidel:2010tt,Rovelli:2010km,Bianchi:2010gc}, and corresponding stunning results about the quantum volume \cite{Bianchi:2011ub} and the use of spinor techniques \cite{Freidel:2010tt,Borja:2010rc,Livine:2011gp,Livine:2011vk,Livine:2011zz}. 

Let us finally list some important questions that are the subject of ongoing investigation in loop quantum gravity: 
\begin{itemize}
\item Barbero-Immirzi parameter: What role does it ultimately play in loop quantum gravity with and without matter?
\item Controlled approximations: Loop quantum gravity is a non-perturbative approach to the quantization of gravity, but approximations will be vital to do physics. How can we find controlled approximations to situations with symmetries from the full theory, or approximately solve the Hamilton constraints? In the context of the reduced models the Hamiltonian constraint is already solved and what we end up is an evolution equation in the physical Hilbert space. However, also here one needs approximation techniques for the reason that the evolution equations are similar complicated to the solution equations of the Hamiltonian constraint.	
	\item Loop quantum gravity and matter: Which types of matter can be consistently coupled to loop quantum gravity? What are the implications of quantized space-time geometry to the propagation of matter?    
\item Physics from Hamilton constraints and Hamiltonians: How does one extract physics from the solutions to the constraints? In particular one should be able to understand how ordinary quantum field theory and classical general relativity are embedded into loop quantum gravity. The first should correspond to a sector of quantum gravity where quantum fluctuations of the geometry are small but matter is still treated as a quantum object, whereas for general relativity both the matter and geometry quantum fluctuations are expected to be negligible. Furthermore, it is important to analyze how ambiguities in the quantization of constraints and physical Hamiltonians do reflect in physical properties of the theory.
\item Connection to spin foam gravity: What is the precise relation between scattering amplitudes and the physical inner product? Which quantization of the Hamilton constraint corresponds to which vertex amplitude? 
\end{itemize}  
For some of these, there are already some insights. Answers to these questions will be crucial for the path that loop quantum gravity takes in the future.  